\documentclass[acmsmall,screen]{acmart}
\acmJournal{TOPLAS}

\usepackage{hyperref}
\usepackage{listings}
\usepackage{subcaption}
\usepackage{mathenv}

\let\RHD=\ltimes

\def\O#1/{\ensuremath{\mathinner{\mathop{\mathsf{#1}}}}}
\newcommand\Id{\O id/}
\newcommand\Def{\triangleq}
\newcommand{\quant}[2]{\mathopen{\text{\divide\thinmuskip3 \divide\medmuskip3 \divide\thickmuskip3 $#1#2.$}\mskip1mu}}
\newcommand{\lam}{\quant\lambda}
\newcommand{\coind}{\mathinner{\mathbf{coind}}}
\newcommand{\ccat}{\mathinner{\mathrel{\mathtt{+\!\!+}}}}

\definecolor{amber}{rgb}{1.0, 0.75, 0.0}
\definecolor{applegreen}{rgb}{0.55, 0.71, 0.0}
\definecolor{citrine}{rgb}{0.89, 0.82, 0.04}
\definecolor{beaublue}{rgb}{0.74, 0.83, 0.9}

\definecolor{nord9}{HTML}{81A1C1}
\definecolor{nord10}{HTML}{5E81AC}
\definecolor{background}{HTML}{eceff4} %nord6
\newcommand{\aside}[1]{\ignorespaces}

\newcommand\scala{Scala}
\newcommand\strymonas{\textsc{{strymonas}}}

\newenvironment{bullets}%
{\begin{list}{$\bullet$}{\setlength{\leftmargin}{1.5ex}%
      \setlength{\itemindent}{.5ex}%
      \setlength\itemsep{0.2mm}}%
\setlength{\parindent}{2ex}%
}%
{\end{list}}

\lstdefinestyle{sOcaml}{language=[Objective]Caml,
  	literate={+}{{$+$}}1 {/}{{$/$}}1
           {=}{{$=$}}1
           {>}{{$>$}}1 {<}{{$<$}}1
           {<>}{$\not=$}1
           {->}{{$\rightarrow$}}2
           {>=}{{$\geq$}}2 {<-}{{$\leftarrow$}}2
           {<=}{{$\leq$}}2
           {|}{{$\mid$}}1
           {|>}{{$\triangleright$}}2
           {>>}{{$\RHD$}}1
           {|>>}{{$\ominus$}}2
           {'a}{$\alpha$}1
           {'b}{$\beta$}1
           {'c}{$\gamma$}1
           {'s}{$\sigma$}1
           {'w}{$\omega$}1
           {'z}{$\zeta$}1
           {'z1}{$\zeta_1$}1
           {'z2}{$\zeta_2$}1
           {list}{list}1
           {...}{\ldots}2
           {forall}{$\forall\ignorespaces$}2
           {exists}{$\exists\ignorespaces$}1
           {\#\#\#}{{$\leadsto$}}3
           {===}{{$\equiv$}}1
           {\#\#+}{\color{red}}1
           {\#\#-}{\color{black}}1
}

\lstnewenvironment{code}{}{}

\lstMakeShortInline[columns=fullflexible]|

\makeatletter
\let\LS=\lst@UseLostSpace
\makeatother

\begin{document}

\title{Complete Fusion for Stateful Streams}
\subtitle{Equational Theory of Stateful Streams and Fusion as
  Normalization-by-Evaluation}

\author{Oleg Kiselyov}
\orcid{0000-0002-2570-2186}
\affiliation{%
  \institution{Tohoku University}
  \country{Japan}}
\email{oleg@okmij.org}

%% \author{Aggelos Biboudis}
%% \affiliation{%
%%   \country{Switzerland}}
%% \email{biboudis@gmail.com}

\author{Tomoaki Kobayashi}
\orcid{https://orcid.org/0009-0006-7147-9997}
\affiliation{%
  \country{Japan}}
\email{tomoaki.kobayashi.t3@alumni.tohoku.ac.jp}

\author{Nick Palladinos}
\affiliation{%
  \country{Greece}}
\email{nick.palladinos@gmail.com}

\begin{abstract}
Processing large amounts of data fast, in constant and small space is
the point of stream processing and the reason for its increasing
use. Alas, the most performant, imperative processing code tends to be
almost impossible to read, let alone modify, reuse~-- or write
correctly.

We present both a stream compilation theory and its implementation as a
portable stream processing library \strymonas\ that lets us assemble complex
stream pipelines just by plugging in simple combinators, and yet attain
the performance of hand-written imperative loops and state
machines. The library supports finite and infinite streams and
offers a rich set of combinators: from \textsf{map},
\textsf{filter}, take(while) to flat-map (nesting), zip,
map-accumulate and sliding windowing. The combinators may be
freely composed, and yet the resulting convoluted imperative code
contains no traces of combinator abstractions: no closures,
intermediate objects or tuples.  The high-performance is portable and
statically \emph{guaranteed}, without relying on compiler or
black-box optimizations.  We greatly exceed in performance the available
stream processing libraries in OCaml.
The library exists in two versions, OCaml and Scala 3, and supports
pluggable backends for code generation (currently: C, OCaml and
Scala).

The declaratively built \strymonas\ pipelines are all stateful. The
stream state introduced in the library is not directly
observable, however. Therefore, the \strymonas\ API looks like the familiar
interface of `pure functional' combinators. Nevertheless, programmers
may introduce their own stream state and share it across the pipeline.

\strymonas\ has been developed in tandem with the equational theory of
stateful streams. Our theoretical model represents all desired
pipelines and guarantees the existence of unique normal forms,
which are mappable to (fused) state machines. We describe the
normalization algorithm, as a form of normalization-by-evaluation.
Stream pipeline compilation and optimization are represented as
normalization, and are hence
deterministic and terminating, with the
guaranteed outcome.  The equational theory lets us state and prove the
correctness of the complete fusion optimization.
\end{abstract}

%% \category{D.3.2}{Programming Languages}{Language Classifications}[Applicative (functional) languages]
%% \category{D.3.4}{Programming Languages}{Processors}[Code Generation]
%% \category{D.3.4}{Programming Languages}{Processors}[Optimization]

%% \terms{Languages, Performance}


\maketitle

\section{Introduction}

Stream processing is transforming, combining, correlating or
reducing possibly unbounded sequence(s) of data, which are
accessed strictly linearly rather than randomly and
repeatedly~-- and processed uniformly. The upside of the limited
expressiveness is the opportunity to process large amount of data
efficiently, in constant and small space. With more data (`big data')
coming every day, stream processing is coming to forefront, as
evidenced by high-speed automated trading, software-defined radio and
processing data from radio telescopes
or accelerators.

Stream processing also exhibits the painful \emph{abstraction
  vs. performance trade-off}.  Manually written imperative
state-machines offer the highest performance and the least memory
overhead, but are not reusable or extensible, and difficult to write
correctly and debug. Libraries of freely composable declarative stream
components let us quickly assemble an obviously correct pipeline, but
suffer from the high overhead of abstractions, mainly due to repeated
creation and disposal of closures, buffers, objects, etc. For some
stream operations such as zip, the overhead may be unbearable and so
the operation is not offered by the library, limiting its
expressiveness: Java Streams is a notable example.\footnote{\relax
  ``This set of operations is quite versatile but, as
  can be expected, it does not cover all the operations a programmer
  may require. One such operation is zip, as we can observe in one of
  the most visited posts about Java Streams in Stackoverfow: Zipping
  streams using JDK8 with lambda (java.util.stream.Streams.zip). Even
  7 years later, we are now on Java 14 release and there is no zip
  operation for Streams yet.'' \emph{Bridge the Gap of Zip
    Operation}\cite{java8_7years}.}
Code generation, or compilation in general, offers the
resolution (see, e.g. StreamIt discussed in \S\ref{s:related}): giving
the user a declarative DSL to be
compiled to an imperative state machine.

Transforming the composition of stream operations into a state machine
is thus the central problem. Although its simple instances (involving
mapping and filtering and accumulation) are straightforward,
difficulties snowball as we add zip and flatmap (parallel and
sequential concatenation), prompting the introduction of various
restrictions (e.g., on the occurrences of flatmap in a pipeline; see
\S\ref{s:related} for details), which ultimately limits the
expressiveness. The most complex are pipelines which involve zipping
of pipelines both of which contain flatmap combinators.  Such
pipelines inevitably occur when correlating or mixing\footnote{\relax
e.g., overlaying two video or mixing of two audio streams, both
compressed. An example is the `amix' operation of the popular video/audio
processing application \textsf{ffmpeg}. See \S\ref{s:evaluation} for
more discussion of this example.}
compressed
streams. Compression is common, e.g., in sensor data streams;
the decompression involves flat-mapping.
  Such, sometimes called ``dynamic rate'' pipelines, although not common
  among StreamIt's (mostly signal processing) benchmarks, are
  ``absolutely necessary'' \cite[\S5.3]{Thies10}
  to express the benchmarks where they occur:
  MPEG and JPEG encoders and decoders, and also rasterizers in graphics
  pipelines. The (de)compressor often needs to
maintain internal state.
However, none of the streaming DSLs to
our knowledge support this general case, with the exception of
\cite{strymonas-2017,staged-rewriting}. However the former failed to
compile such pipelines fully down to state machines (some closures
remained) and the latter relied on an unusually deep dependence on the
Scala 2 compiler and hence not available in the latest Scala 3
compiler. Neither provided any argument about correctness of such
compilation.

Thus the challenges of the compilation approach:
\begin{enumerate}
\item How to support both flatmap and zip, in any combination without
  limitation? How to support stateful filtering and flatmapping,
  typical of compression and decompression?
\item How to guarantee the compilation quality: compiling down to
  an imperative state machine?
  How to go beyond the `best-effort' optimization?
\item How to verify the optimizations correct?
\item How to give some guarantees about the correctness of the
  compiler implementation itself?
\item How to let the user add further facilities, e.g., windowing, and
  related optimizations (such as specializations to small or very large window
  sizes), while preserving all other guarantees? That is, how to add a
  degree of extensibility to the system without voiding the guarantees?
\end{enumerate}

This paper presents the compilation model \emph{and} its implementation that
meets all the challenges. Its novel contributions are:

\subsubsection*{Theoretical Contributions}
\begin{enumerate}
\item a formal calculus of stateful coinductive streams and their
  equational laws, covering zip and flatmap, e.a.;
\item unique stream normal forms that are mappable to state
  machines;
\item an algorithm to calculate the normal forms as a form of
normalization-by-evaluation;
\item the notion of linear stream and linearization;
\item zip-conversion: converting out, and hence eliminating zip from pipelines;
\item mechanized proofs of the equational laws, zip-conversion and
  linearization.
\end{enumerate}

\subsubsection*{Implementation Contributions}
\begin{enumerate}
\item the implementation of the normalization algorithm, using staging;
\item staged closure conversion, when linearizing flatmapped streams;
\item the high-performance 
  stream processing library \strymonas, supporting zip and
  flatmap in any combination, as well as accumulating map, windowing and
  compression/decompression;
\item static guarantees of \emph{complete fusion} 
  (Defn.~\ref{d:complete-fusion}), as well as type and scope safety,
  ensured by construction.
\end{enumerate}

The present paper is, hence, both theoretical and
engineering: in fact, theory and its implementation have
been evolving in tandem.

Below we make our claims precise (\S\ref{s:claims}), 
after reminding the terminology and
circumscribing the scope; \S\ref{s:overview} shows many examples. We
use the name \strymonas\ to refer both to the library and to its
underlying compilation theory~-- in its current version, which differs
significantly from the original version \cite{strymonas-2017} (see
\S\ref{s:related} for detail).

\subsection{Streaming Model}

\strymonas\ is a DSL to build streaming applications, also called
pipelines, which are
represented as connected directed graphs whose nodes are called `operators'
and edges are communication channels (using the terminology of
\cite{catalog-optimizations}). In \strymonas:
\begin{itemize}
\item graphs are acyclic;
\item an operator may have no more than one output edge: splits, or
  forks, are not directly supported (but see below). An operator
  without output channels is called `consumer': typically |fold| or
  |iter|. A complete \strymonas\ application has exactly one
  consumer. Abusing the terminology,
  we call a not yet consumer-terminated pipeline 
 (i.e., the pipeline that has an output edge and still can be extended)
  a (\strymonas) stream.
\item an operator may have any number of input channels. An operator with
no input channels is called `producer', such as |of_arr| that presents
an array as a finite stream, or |iota| that generates an infinite
streams of integers. File descriptors may also act as \strymonas's
producers. A complete \strymonas\ application has one or more
producers.
\item most operators have one input and one output channel~--
  so-called stream transformers: e.g., |map| and |map_accum|; |filter|
  and map-filter; |take| and |take_while|; |drop| and |drop_while|; as
  well as sliding windowing combinators.

Here is a characteristic example of a linear pipeline. To be explained
in all detail in \S\ref{s:taste}, it is shown here to give a general
feeling. 
\begin{code}
iota C.(int 1) |> map C.(fun e -> e * e) |> filter C.(fun e -> e mod (int 17) > int 7) 
            $\LS$|> take C.(int 10) |> fold C.(+) C.(int 0)
\end{code}
\S\ref{s:taste} also shows the OCaml and C code generated by
\strymonas\ for this example: an imperative while-loop running in
strictly constant space.

\item there are operators with two input and one output edge:
|zip| and |zip_with|, to process streams in lockstep.
More advanced merging of streams (e.g., sorted merge) requires an
extension of the core and the theory, and left
for future work.

As an example, the operator |take| mentioned above is actually implemented in
terms of |zip_with|:
\begin{code}
let take n s = zip_with (fun _ x -> x) (from_to C.(int 1) n) s
\end{code}
It seems wasteful: generating the stream
of numbers |1..n| just to throw them away. As to be explained
in detail in \S\ref{s:tupling}, there is actually no waste, and the
implementation is as good as written by hand.

\item the operator |flat_map|, superficially looking as a transformer, realizes
stream nesting, or concatenation of arbitrary number of
streams. The structure of the inner pipeline is statically known. For example,
\begin{code}
iota C.(int 1) |> flat_map (fun x -> from_to x C.(x + int 5)) |> take C.(int 10)
\end{code}
produces the stream 1,2,3,4,5,6,2,3,4,5.
\item operators may keep an internal state and share the state. The state
  has to be declared. An example is computing the stream of
  differences between two consecutive input samples:
\begin{code}
let diff : int cstream -> int cstream = fun st ->
  let- z = initializing_ref C.(int 0) in
  st |> map_raw C.(fun e k -> let- v = letl (e - dref z) in (z := e) @. k v)
\end{code}
It is harder to understand without the explanations (which are given,
in \S\ref{s:stateful}). One may get an impression that \strymonas\ is
powerful but also not that simple.
\item data items~--- the units of communication~-- are not merely
  integers or floating-point numbers. Tuples, records, etc. are also
  supported. For example, here is the FIR (finite impulse response)
  filtering, taken from \cite{strymonas-PEPM24}:
\begin{code}
let fir_filter ?(decimation=0) (weights:float array) : F32.t cstream -> F32.t cstream =
  let ntaps = Array.length weights in
  let (module Win) = make_window F32.tbase ntaps (decimation+1) in
  Win.make_stream >> map_raw F32.(Win.dot tbase lit weights ( +. ) ( *. ))
\end{code}
The function |Win.make_stream| converts the stream of floats to the
stream of windows of the size |ntaps|, advancing by |decimation+1| steps.
Window is an abstract data type;
the operator `|>>|' is the left-to-right function composition.
The following |map_raw| maps each window to a number: the result of
the dot-product of the window samples with the |weights|. The generated C
code runs in strictly constant-space; moreover, it is vectorizable.

  Tupling may compensate to an extent for the lack of
  splitters (i.e., operators with two or more output channels), as we
  describe below.
\item
  \strymonas\ is typed: the type of an operator describes its input
  and output channels and the type of data items in them. The |diff|
  example showed the stream types in the type annotation (which is not
  necessary and given only for clarity).
\end{itemize}

\strymonas\ applications are built by connecting the input channel
of one operator with the output of another (using `\lstinline{|>}',
which is a left-to-right function application in OCaml), 
provided the types of the
data items match. Beside the type matching, there are no
restrictions on connecting the operators. In particular, |flat_map|
and |zip| may occur repeatedly and in any combination.

\begin{figure}[ht!]
\centering
\begin{subfigure}{\textwidth}
 \centering
 \includegraphics[width=0.43\linewidth]{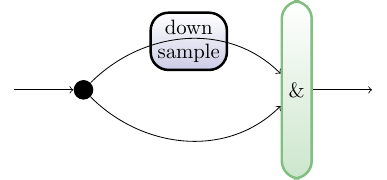}
 \hskip 2em
 \includegraphics[width=0.43\linewidth]{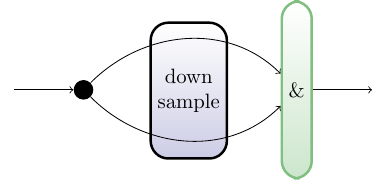}
\end{subfigure}
\caption{Unsupported (left) and supported (right) split-join processing.}
\label{f:pipelines}
\end{figure}

\strymonas\ deliberately does not provide |unzip|, or any such
splitting operator. Splitting admits pathological pipelines
such as the one in Fig.~\ref{f:pipelines}(left), which requires
unlimited buffering. As we stress below, \strymonas\ emphasizes
constant-space processing. Ensuring constant space processing in the
presence of splitters is possible, with so-called synchronicity
constraints \cite{Caspi-TCS,Paspi-Pouzet-SyncKahn}; StreamIt, reviewed
in \S\ref{s:related} accomplishes them with the static rate specification.
\strymonas, however, is built around dynamic scheduling.

Therefore, \strymonas\ does not provide for splitting, in general. A
limited form of split-join processing (such as the one in
Fig.~\ref{f:pipelines}(right))~-- where split streams are processed in
lockstep, at the same rate~-- is possible with tupling.

\subsection{Scope}
Stream processing is an immensely vast area \cite{stephens-survey},
spanning from digital signal processing and (synchronous) dataflow to
MapReduce, databases and complex event processing. Paper
\cite{catalog-optimizations} gives a good recent overview.

Our place is what \cite{catalog-optimizations} calls
fusion,\footnote{\relax Among other optimizations catalogued in
\cite{catalog-optimizations},
  we also use operator reordering and algorithm
  selection.} which, broadly speaking, means merely
running stream operators within the same operating system process. The
stream operators may still run in separate threads and may still use
an arbitrary amount of auxiliary storage
\cite{catalog-optimizations}. A stricter view of fusion has become
common in programming language community (see, e.g.,
\cite{coutts_stream_2010,biboudis_phd} for surveys), which we adopt in
this paper: fusion means running the entire streaming application
within a single thread and without unbounded
intermediate storage. Furthermore, we call the fusion \emph{complete} when
there are no intermediate data structures or function calls at all.
\begin{definition}[Complete Fusion]\label{d:complete-fusion}
  A streaming application is called
  completely fused when it runs without function calls and memory
  allocations however temporary, provided each individual
  operator may be executed without calls and allocations.
\end{definition}

Our direct competitors are the streaming libraries offered for
virtually all modern programming languages and epitomized by the
industrially widely used Java8 Streams and its descendants. All mature
libraries solve the stream fusion problem~-- \emph{with
  qualifications}~\cite{coutts_stream_2007,farmer_hermit_2014,svenningsson_shortuct_2002}:
expressivity is restricted or the complete fusion is not guaranteed.
A good example is Java, whose streaming, since the roll-out of Java 8
Streams, is often taken as the standard. Streaming from an array
amounts to an ordinary loop, well-optimized by a Java JIT
compiler~\cite{biboudis_streams_2015}. However, |java.util.Streams|
are still much slower than hand-optimized loops for non-trivial
pipelines~\cite{biboudis_clash_2014,
  strymonas-2017,streams_staticanalysis}.  Furthermore, the library
still cannot zip several streams, let alone nested or infinite
streams.

\strymonas's goal is to be library like Java8 streams, applicable to
the same industrial use cases. However, we aim: (i) for larger expressivity
(in particular, supporting |zip| and |flat_map| with no restrictions,
as well as infinite streams), plus (ii) the guaranteed complete fusion:
producing the sequential code that an (ideal) human programmer could
write for the given pipeline. A good illustration of both aims is
software-defined radio, specifically, FM reception
\cite{strymonas-PEPM24}~-- a relative large and realistic streaming application
demanding high performance. \strymonas\ proved successful in
implementing it, attaining and exceeding the performance of GNU Radio.

Since we restrict our attention to fusion, the parallel execution is
out of scope. Although \strymonas\ applies to an extent to
event processing (e.g., it provides sliding
windows), complex event processing \cite{Cugola_2012}
is the topic for another time.
The sequential execution is by no means the only way to run streaming
applications. However, it is one of the main ones~-- the one supported by
our competition (streaming libraries in many programming languages).
Sequential execution is also what a single worker in a ``big data''
processing application does.

Furthermore, \cite{Thies10} demonstrates
that the opportunities for pipeline parallelism are not that many in
realistic signal-processing--related 
benchmarks. Also, \cite{Sermulins05} show on their
benchmarks that on modern superscalar CPUs simply fusing all operators
(and removing any pipeline parallelism) achieves the performance of
most or all of the sophisticated optimizations presented in that
paper. The FM radio reception
\cite{strymonas-PEPM24} is another illustration: a single-core single-thread
\strymonas\ application attains and exceeds the performance of GNU
Radio, which relies on multiprocessing.

\subsection{Precise Claims}
\label{s:claims}

\paragraph{Guarantees}
It is guaranteed, \emph{by the very construction}, that the generated
target code (i) is well-typed and hygienic; (ii) contains no function
calls; no closures, tuples, records and other such intermediate data
structures~-- provided that individual operators do not, on their own,
call functions or allocate temporarily. The complete fusion is thus
statically assured for all pipelines.

The complete fusion is, mainly, a space guarantee: the ability to run
the main processing loop without any GC or even stack allocations.
The absence of function calls and temporary data structures also
improves locality and enables further compiler optimizations. We also
save time otherwise needed to construct/deconstruct/communicate
temporary data structures. Thus complete fusion is expected to improve
performance. Our benchmarks bear it out. Comparing our generated
code against related libraries in OCaml
%% and Java/Scala,
which
use closures/objects extensively, shows the complete fusion is
correlated with improved execution time, often to significant extent
(see \S\ref{s:evaluation}).

Both guarantees come from the tagless-final code generation
\cite{carette-finally-jfp,tagless-final-oxford}: the target code is
produced via an abstraction, a simple typed first-order API, which
provides no facilities to create closures, records, etc. in the first
place. The fact that such a simple backend API supports all pipelines
expressible in \strymonas\ (which is all Java8-supportable
pipelines, and many, many more) is surprising, even to us (see
`principal difficulties' below).

\paragraph{Portability}
\strymonas\ is portable, demonstrated by providing two independent
implementations, in OCaml and Scala 3. The target code, in which user
actions are expressed and the result is delivered, is not fixed
either. Just by swapping backends, all implementing the same API, one
generates different code -- \emph{all completely fused}. Current backends:
OCaml (using MetaOCaml), C (plain OCaml or Scala).
%%, Scala (using Scala 3 quotations~\cite{Stucki:2018,Dotty}).
More backends may be added, such as
flow charts (i.e., circuit descriptions), MLIR and WASM (there is a
preliminary support for WASM already).

\paragraph{Evaluation}
We evaluate \strymonas\ on a series of benchmarks collected from the
literature, see \S\ref{s:evaluation}. \strymonas\ reproduces the
performance of the original strymonas \cite{strymonas-2017} on then published
benchmarks but significantly exceeds it on the new complex zipping
benchmarks.
Thus \strymonas\ not just approaches but \emph{matches} closely the
performance of the hand-written code in all cases (where hand-written
code was feasible).
\aside{
Not all libraries support all combinators: e.g., Java
Streams does not run our benchmarks involving |zip|. For the remaining
benchmarks, we offer speed-up up to
$32\times$ over Java Streams. (Although |zip| is offered by
Guava, it is not specialized for |long|s used in our benchmarks.)
According to \cite{java8_7years} the fastest |zip| support, as
a standalone dependency, is in
JAYield.\footnote{\url{https://github.com/jayield/jayield}} \strymonas\ is
faster up to $22\times$ ($5\times$-$17\times$ for the zipped filters and
zipped flatMaps resp., see
Fig.\ref{f:microbenchmarks2}).}  Compared to the state-of-the art OCaml
streaming
library\footnote{\url{https://github.com/odis-labs/streaming}} we
offer $8\times$-$35\times$ speed-up (see
Fig.\ref{f:microbenchmarks1}).
In addition, the paper
\cite{strymonas-PEPM24} evaluates \strymonas\ on a relatively large,
realistic, performance-demanding application: FM Radio reception.

\paragraph{Correctness}
The entire process of transforming a high-level pipeline description to a
low-level target code representation is typed, all throughout,
eliminating a wide class of errors (which, albeit simple, tend to be
most insidious, especially scoping errors). Furthermore, the
transformations are based on explicitly stated equational laws (see
\S\ref{s:derivation}), which are verified in Agda, see \S\ref{s:Agda}.
Some transformations, such as eliminating zipping, are very
subtle; errors are present even in published work.

\paragraph{Principal difficulties}
The source of much trouble is zipping of the pipelines which both contain
flatmap (nesting), i.e., producing data items at statically
unpredictable rates. Our earlier work \cite{strymonas-2017} failed to
achieve complete fusion in these cases, resorting to closures, and
conjectured that such failure is inevitable. We were wrong.

A common misconception is that just by applying a sufficiently
powerful inliner (built into a compiler, or as a separate tool~--
partial evaluator) one may achieve complete
fusion. The paper \cite{strymonas-2017} already showed this not to be the
case. Although inlining certainly helps, by enabling other compiler
optimizations (\cite{catalog-optimizations}), completely avoiding
intermediate data structures requires a creative step, the right
representation of streams with the set of transformations on them.
% The situation is similar to that of FFT. One cannot get to FFTW
% quality code by PE alone.

\paragraph{Key ideas}
A high-level pipeline description is transformed to an intermediary
language (IR), designed to easily map to a target
language. This separation of the host and target languages also separates
overheads: the overhead of transformations and abstractions remains at
the code generation time and does not leak into the generated code.

Commonly, transformations and optimizations are based on repeated
application of re-writing rules~-- which are often non-confluent and
non-terminating. Therefore, the existence or uniqueness of a normal
form is not assured and improvements are hard to
predict.\footnote{See \cite{strymonas-2017} for examples of
  unpredictability of such rule-based optimizations for stream fusion.}
What sets \strymonas\ apart is that reductions to IR and accompanying
optimizations are performed as \emph{normalization-by-evaluation}
(NbE): deterministic evaluation to the normal form, existing
by construction. The evaluation is certain to terminate with the
predictable result, which gives performance assurances.

The key to the complete fusion even for complex pipelines is the
particular closure conversion (see \S\ref{s:closure-conv}), performed
without inspection of the code and hence maintaining all hygiene and
typability guarantees without exotic type systems.

\subsection{Structure of the paper}

The next section introduces \strymonas\ and illustrates the complete
fusion, from straightforward pipelines to tupling to stateful stream
processing. Section \ref{s:derivation} is the theoretical part of the
paper. It develops the equational theory of stateful streams;
introduces the key notions of a linear stream, linearization and
zip-conversion; and defines the normal forms of stream pipelines,
proves their existence and uniqueness, and introduces the
normalization algorithm showing its correctness. The implementation
section~\ref{s:implementation} describes how the theory maps to
practice: how normal forms are converted to code, and how, hence, the
compilation of a stream pipeline to a state machine is
performed. \S\ref{s:evaluation} describes the performance evaluation,
and \S\ref{s:related} overviews and compares to the related work. 
Appendices demonstrate more interesting applications: run-length
compression/decompression and nested grouping/aggregation
(essentially, the composition of three Mealey machines).

There are two implementations of the library, in OCaml and Scala, with
the same functionality. Throughout the paper we use the OCaml version.
The complete code~-- of the library, the examples including all examples
in the present paper, and the benchmarks~-- is freely available
in the strymonas repository
\url{http://strymonas.github.io/}. The current version is 2.1.

\section{What \strymonas\ can do}
\label{s:overview}
This section is to introduce \strymonas, and, together with
\S\ref{s:evaluation}, justify our expressivity claims.  Although the
library interface in Fig.\ref{f:raw-API} concisely and completely
specifies the capability of \strymonas, it may not be
easy to grasp the full variety of stream pipelines expressible
with the interface. This section helps on examples, usually taken from
related/competing libraries, for the sake of comparison.

This section is not a library guide or a tutorial: only salient
examples are discussed. Furthermore, the set up code (choosing a
back-end, the top-level function causing the code generation and
execution) is omitted; please see the public
repository for the complete, runnable examples.

\subsection{Stateless Pipelines}
\label{s:taste}

This section introduces \strymonas\ and shows several
stateless pipelines as the baseline~-- the least common denominator of
streaming libraries for OCaml, Java, Haskell, etc. \S\ref{s:stateful}
shows examples of stateful streams (stateful filtering and
accumulation), dealing with compression/decompression and nested
grouping/aggregation. The paper \cite{strymonas-PEPM24} describes even
more complicated application: windowing (specialized to different
sizes) and convolutional filtering. All these examples use the
\strymonas\ API as is, without extensions~-- and hence automatically
benefit from the complete fusion guaranteed by it.

We call the pipelines in this section `stateless' because they look and
act as pipelines built with other declarative streaming
libraries, with no hint of 
any state. However, \strymonas\ pipelines have internal state; combinators
such as |take_while| (see \S\ref{s:stateful}) maintain their own
state. This state however is not accessible and
not directly observable.\footnote{One may wonder if the state can be
  deduced from sharing; we show at the end of this section that it cannot.}

It is best to think of \strymonas\ as a sort of Yacc~-- only not as a
standalone tool but embedded in a host language such as OCaml. Yacc
takes the descriptions of a grammar (written in a some sort of BNF)
and of user-actions (written in a different, target language, such as
C). It generates the code in the target language in which the supplied
user-actions are weaved with parsing actions.  The generated code can
be compiled and linked as ordinary target code, put into a library,
etc.

\strymonas\ functions quite similarly~-- only for stream processing
rather than parsing. Unlike (ocaml)yacc, strymonas is a DSL embedded
in OCaml.  Therefore, it integrates as is with the existing OCaml code
and tools. Any typing or pipeline mis-assembling errors are reported
immediately~-- even during editing.

Here is the simplest example (for clarity,
we explicitly write type annotations in all examples, although none of
them are needed since all types are inferred):
\begin{code}
let ex1 : int cstream = iota C.(int 1) |> map C.(fun e -> e * e)
\end{code}
Like Yacc, \strymonas\ uses two languages: one to describe the structure of
the stream pipeline, and the other to specify the user actions such as
filtering conditions, mapping transformations, etc. Since \strymonas\ is an
embedded DSL, both languages are represented by OCaml functions
(combinators), but from two different namespaces (signatures). Stream
structure combinators such as |iota| (producing an infinite stream
of natural numbers starting from the given one) and |map| come from the
signature in Fig.\ref{f:API}. These combinators produce, consume, or
transform values of the type |'a cstream|, representing a
\strymonas\ stream; |'a| is a base type. Recall, a stream is a
pipeline with an output edge, which may still be extended before being
terminated by a consumer. We may now say that a stream is a value of
the type |'a stream|, where |'a| is the type of the data sent along the
output edge. Since most of the time the data is not statically known,
many streams have the type |'a exp stream|, for which we introduce an
abbreviation |'a cstream|. \strymonas\ streams also deal with
structured data such as tuples, records, etc., whose structure, if not
the values themselves, are statically known. In that case, the type
argument of |'a stream| is not of the form |'b exp|~-- as we shall see
in \S\ref{s:tupling}.
The combinators are typically composed
via \lstinline{|>}: the left-associative swapped application operator in
OCaml.\footnote{The OCaml standard library also provides the
  right-associative low-precedence application operator
  \textsf{@@}, which we will see later. Hence, \textsf{x + 1
    $\triangleright$ f} is the same as \textsf{f @@ x + 1} and is the
  same as \textsf{f (x + 1)} but avoids the parentheses.}

User actions (or, semantic actions in Yacc terminology)~-- that is,
the arguments of stream combinators~-- are described via backend
combinators, Fig.~\ref{f:cde}, which build values of the abstract type
|'a exp| representing expressions and |'a stm| representing statements of
the target code. A backend hence is a first-order
statement-oriented imperative language with mutable references and
arrays.  We shall assume one such backend in scope, as the module
named |C| (for `code'), which we will use to explicitly qualify target
code combinators. In OCaml, a module qualification can be attached to
a parenthesized expression, scoping over it. That is, every
unqualified name in |C.(int 1)|, i.e., |int|, should be implicitly
qualified with |C|. In other words, |C.(int 1)| is the abbreviation
for |let open C in int 1|, that is, making |C| operations available
within the expression without qualification.

\begin{figure}
\begin{code}
type 'a stream
type 'a cstream = 'a exp stream

(* Producers *)
val iota : int exp -> int cstream
val from_to : int exp -> int exp -> int cstream
val of_arr : 'a arr -> 'a cstream
val of_int_array : int array -> int cstream       (* Specialization *)

(* Consumers *)
val fold : ('z exp -> 'a exp -> 'z exp) -> 'z exp -> 'a cstream -> 'z stm
val iter : ('a -> unit stm) -> 'a stream -> unit stm
val sum_int : int cstream -> int stm

(* Transformers *)
val map : ('a exp -> 'b exp) -> 'a cstream -> 'b cstream
val flat_map : ('a exp -> 'b stream) -> 'a cstream -> 'b stream
val filter : ('a exp -> bool exp) -> 'a cstream -> 'a cstream
val take : int exp -> 'a stream -> 'a stream
val map_accum : ('z exp -> 'a exp -> ('z exp -> 'b exp -> unit stm) -> unit stm) ->
                $\LS$'z exp -> 'a cstream -> 'b cstream

val drop : int exp -> 'a stream -> 'a stream
val drop_while : ('a exp -> bool exp) -> 'a cstream -> 'a cstream
val take_while : ('a exp -> bool exp) -> 'a cstream -> 'a cstream
val scan : ('z exp -> 'a exp -> 'z exp) -> 'z exp -> 'a cstream -> 'z cstream

val zip_with : ('a exp -> 'b exp -> 'c exp) -> ('a cstream -> 'b cstream -> 'c cstream)
\end{code}
\begin{code}
val (|>) : 'a stream -> ('a stream -> 'b stream) -> 'b stream
\end{code}
\caption{Stream combinators (abbreviated and simplified for
  presentation). Types \textsf{exp}, \textsf{stm}, \textsf{arr}
  and \textsf{mut} refer to expressions, statements, arrays and
  mutable references of the target code,
  see Fig.~\ref{f:cde}.
For ease of reference, the figure also includes the left-associative operator
$\triangleright$. It is actually the function application with swapped
arguments, and provided by the OCaml standard library,
with the more general type. All operations of this interface are
implemented in terms of those in Fig.\ref{f:raw-API}.
}
\label{f:API}
\end{figure}

\begin{figure}
\begin{code}
type 'a exp        $\LS$(* Abstract type of expressions *)
type 'a stm        $\LS$(* Statements *)
type 'a mut        $\LS$(* Mutable references: they are not expressions *)
type 'a arr        $\LS$(* Arrays are not expressions either *)
type 'a tbase      $\LS$(* Type representation for base types *)

val tbool : bool tbase    (* Evidence that the target code has bool type *)
val bool : bool -> bool exp
val not  : bool exp -> bool exp
val (&&) : bool exp -> bool exp -> bool exp   (* shortcut eval *)
val (||) : bool exp -> bool exp -> bool exp

val tint : int tbase    (* Evidence that the target code has int type *)
val int   : int -> int exp
val ( + ) : int exp -> int exp -> int exp
val ( * ) : int exp -> int exp -> int exp
val ( =)  : int exp -> int exp -> bool exp
val logand : int exp -> int exp -> int exp
...
val print_int   : int exp   -> unit stm

val (@.) : unit stm -> 'a stm -> 'a stm        (* sequencing *)

val cond : bool exp -> 'a exp -> 'a exp -> 'a exp
val if_ : bool exp -> unit stm -> unit stm -> unit stm
val if1 : bool exp -> unit stm -> unit stm

val while_ : bool exp -> unit stm -> unit stm

val letl : 'a exp -> (('a exp -> 'w stm) -> 'w stm)      (* let-binding *)

val newref  : 'a exp -> ('a mut -> 'w stm) -> 'w stm
val dref : 'a mut -> 'a exp
val (:=) : 'a mut -> 'a exp -> unit stm
val incr : int mut -> unit stm
val decr : int mut -> unit stm
\end{code}
\caption{Backend code-generating combinators for the target
  language (abbreviated).
  In the present paper,
  they are assumed to be in a module
named \textsf{C}.
}
\end{figure}
\begin{figure}[ht]\ContinuedFloat
\begin{code}
val array_get' : 'a arr -> int exp -> 'a exp
val array_get  : 'a arr -> int exp -> ('a exp -> 'w stm) -> 'w stm
val array_len  : 'a arr -> int exp
val array_set  : 'a arr -> int exp -> 'a exp -> unit stm

val new_array  : 'a tbase -> 'a exp array -> ('a arr -> 'w stm) -> 'w stm
   (* Array of the statically known size with statically known elements,
      allocated in DATA segment *)
val new_static_array  : 'a tbase -> ('b -> 'a exp) -> 'b array -> ('a arr -> 'w stm) -> 'w stm
val new_uarray : 'a tbase -> int -> ('a arr -> 'w stm) -> 'w stm  (* new uninitialized array *)

module F64 : sig    (* long floats *)
  type t
  val tbase  $\LS$: t tbase        $\LS$(* type representation *)
  val lit    $\LS$: float -> t exp
  val of_int $\LS$: int exp -> t exp
  val ( +. ) $\LS$: t exp -> t exp -> t exp
  val ( *. ) $\LS$: t exp -> t exp -> t exp
  val ( /. ) $\LS$: t exp -> t exp -> t exp
  val print  $\LS$: t exp -> unit stm
end
\end{code}
\caption[]{(\emph{continued}) Backend code-generating combinators for the target
  language.
Similarly to \textsf{F64},
strymonas also provides modules \textsf{F32} for short floats
and \textsf{C32} for complex numbers, as well as \textsf{I64} for
64-bit integers.
}
\label{f:cde}
\end{figure}

The above \strymonas\ pipeline is the infinite stream of natural numbers
transformed by squaring each item. It is an
OCaml expression of the type |int cstream|, and can be let-bound, to
give it a name for further reference. The pipeline is completed by
terminating it by an |int cstream| consumer, such as |fold|, or its
instance |sum| below to sum up stream items.
\begin{code}
let ex1 : int cstream = iota C.(int 1) |> map C.(fun e -> e * e)
let sum : int cstream -> int stm = fold C.(+) C.(int 0)
let ex2 : int stm = ex1 |> filter C.(fun e -> e mod (int 17) > int 7) |> take C.(int 10) |> sum
\end{code}
We added to |ex1| (repeated for ease of reference)
two more transformations, to retain only those
items whose remainder mod 17 exceeds 7, and take first 10 such
items. The resulting stream becomes finite and can be meaningfully
summed up. Hence |ex2| is the |int stm| value representing the
integer-producing target code for the complete pipeline, including
stream generation, folding, and the user actions of squaring,
etc.\footnote{\relax The \textsf{C.(\ldots)} notation is hence
  indicative of building target code, and reminds of quasiquotation
  (code template) mechanism of languages like MetaOCaml \cite{metaocaml-10}.}
The
backend that realizes |'a stm| as OCaml code lets us pretty-print this
code:
\begin{code}
let v_1 = ref 0 in
let v_2 = ref 10 in
let v_3 = ref 1 in
while (! v_2) > 0 do
   let t_4 = ! v_3 in
   incr v_3;
   let t_5 = t_4 * t_4 in
   if (t_5 mod 17) > 7 then (decr v_2; v_1 := ! v_1 + t_5)
done;
! v_1
\end{code}
The code can also be saved into a file, compiled,
put into a library~-- or even dynamically linked into the
program that generated it. With the C back-end, the
resulting code is:
\begin{code}
int fn(){
  int x_1 = 0;
  int x_2 = 10;
  int x_3 = 1;
  while (x_2 > 0)
  {
    int const t_4 = x_3;
    x_3++;
    int const t_5 = t_4 * t_4;
    if ((t_5 % 17) > 7)
    {
      x_2--;
      x_1 = x_1 + t_5;
    }
  }
  return x_1;
}
\end{code}
This is what a competent programmer would have written by hand.
Although the pipeline is purely declarative, with first-class (the argument of
|filter|) and higher-order functions, the generated code is
imperative and has no function calls. The main loops runs with no GC,
even in OCaml.

The second example is the pipeline to compute the dot-product of two arrays:
\begin{code}
let ex_dot (arr1:int arr, arr2:int arr) : int stm = zip_with C.( * ) (of_arr arr1) (of_arr arr2) |> sum
\end{code}
The combinator |of_arr| creates a finite stream whose contents is the
given target language array. With the OCaml backend we generate:\footnote{\relax
\textsf{Array.length} and \textsf{Array.get} that appear in the code
are OCaml primitives, which compile directly to the appropriate load
instructions}
\begin{code}
fun (arg1_18, arg2_19) ->
  let v_20 = ref 0 in
  for i_21 = 0 to
    (if (Array.length arg1_18) < (Array.length arg2_19)
     then Array.length arg1_18
     else Array.length arg2_19) - 1
    do
    let el_22 = Array.get arg1_18 i_21 in
    let el_23 = Array.get arg2_19 i_21 in
    v_20 := ! v_20 + el_22 * el_23
  done;
  ! v_20
\end{code}
This is how a C programmer would have written it.

Just as |zip_with| is used for dot-product, |flat_map| can be used for
Cartesian product:
\begin{code}
let ex_cart (arr1:int arr, arr2:int arr) : int stm =
     of_arr arr1 |> flat_map (fun x -> of_arr arr2 |> map C.(fun y -> x * y)) |> sum
\end{code}
The generated code predictably has two nested for-loops.\footnote{See
\nolinkurl{TryFirst/simple.ml} in the
\strymonas\ repository for a more complex version: relational join.}
The argument
of |flat_map| is a function that produces a stream for each element
|x| of the outer stream. Although it may be hard to see from the types,
the |flat_map| interface only supports nested streams of a fixed
structure: the composition of the inner pipeline not
depending on |x|.
% Just like in Yacc, the grammar does not depend on the parsed data.
The limitation
can be gotten around by providing a more
general continuation-passing-style interface, similar to that outlined
in \cite{strymonas-2017}. However, as mentioned there, there does not
seem to be any practical need.

Since the metalanguage (OCaml) has let-expression, we may share
streams, it seems:
\begin{code}
let sts = iota C.(int 1) |> take C.(int 10) in
zip_with C.(+) sts (sts |> filter C.(fun x -> x mod int 3 = int 0)) |> sum
\end{code}
However, if we look at the generated code, we see two copies of the
code responsible for enumerating integers 1 through 10.
Let-expressions in the metalanguage share
stream descriptions, not streams themselves. That subtlety of sharing in
metalanguage vs. target language is expounded in
\cite{multi-stage-monadic-memo06,EDSL-sharing}.
To share streams themselves requires a dedicated primitive, which
\strymonas\ does not provide.

\subsection{Tupling: Abstraction without Guilt}
\label{s:tupling}

Recall that |'a cstream| represents streams of base-type
items such as integers and floating-point numbers~-- after all, the
code-generation interface Fig.\ref{f:cde} lets us build code of base
types only. Nevertheless, we can deal with
streams of tuples, records and other structural types, by using
the so-called raw interface, Fig.\ref{f:raw-API}. This is the real
API provided by \strymonas~-- that is, the library primitives.
What we have been
using so far, Fig.\ref{f:API}, is the `cooked', or `sugared' API, implemented
entirely in terms of the raw one. The raw interface is quite more
general and more difficult to use; the sugared API should be more
familiar and convenient. In full detail the raw API is explained in
the next section, about stateful streams. Here we touch on only one
aspect of it: tuples and structured stream items. We should stress
that the raw API is not an `internal' API: the raw primitives may be
used alongside the cooked ones, as we shall demonstrate; after all,
there is only one stream type, used by both APIs.

So far we have dealt with the streams of the type |'a exp stream|,
abbreviated to |'a cstream|. Generally, in |t stream|, the type |t| is
not necessarily a target code type: it may be a tuple or a record,
etc.\footnote{\relax \textsf{int stream} is also possible: the type of
  streams whose items are a statically known integer. It is hardly
  useful though.} We have already seen, unbeknownst, an example:
|ex_dot|. The combinator |zip_with| used there is syntax sugar over
|zip_raw: 'a stream -> 'b stream -> ('a * 'b) stream| of
Fig.\ref{f:raw-API}:
\begin{code}
let zip_with : ('a exp -> 'b exp -> 'c exp) -> ('a cstream -> 'b cstream -> 'c cstream) =
 fun f s1 s2 -> zip_raw s1 s2 |> map_raw' (fun (x,y) -> f x y)
\end{code}
(where |map_raw'| is |map| with the general type, operating on
|'a stream| rather than |'a cstream|).  That is, |zip_raw| creates a
stream of tuples |'a exp * 'b exp|, which are then mapped to |'c exp|
values. The intermediately produced tuples are the tuples of the host
language; they are not part of the target code. Indeed, the code
generated for |ex_dot| seen earlier has no traces of those tuples, or
other intermediate data structures. That is the key to abstraction
without guilt: by separating the host and the target languages we gain
the freedom of using the full abstraction power of the host language
and yet guarantee that the produced target code has no trace of those
abstractions. To further illustrate this point, here is the
implementation of |take| in terms of |zip_with| and
limited |iota| (viz., |from_to| is described in \S\ref{s:stateful}).
\begin{code}
let take : int exp -> 'a stream -> 'a stream = fun n s -> zip_with (fun _ x -> x) (from_to C.(int 1) n) s
\end{code}
The code is simple and clear. One may worry however if such a tupling
(of an element with its index, produced by |from_to 1 n|) 
followed by the deconstruction of the tuple and discarding that index
would introduce intermediate data structures. The code generated for
|ex2|, see \S\ref{s:taste}, which used |take| alongside other
operators, shows the worry is unfounded: there is no garbage produced;
in fact, the main loop shows no memory allocations at all.

Although the sugared interface suffices for many common examples (such
as most of our benchmarks), sometimes the additional power of the raw
interface is needed, as in the example below, borrowed from
\cite{strymonas-2017}:
\begin{code}
let paper_test : unit stm =
 let square x = C.(x * x) and even x = C.(x mod (int 2) = int 0) in
 Raw.zip_raw
     ([|0;1;2;3|] |> of_int_array |> map square |> take (C.int 12) |> filter even |> map square)
     (iota (C.int 1) |> flat_map (fun x -> iota C.(x+int 1) |> take (C.int 3)) |> filter even)
   |> iter C.(fun (x,y) -> (print_int x) @. (print_int y))
\end{code}
The example shows off the assembling of a complicated pipeline~-- with
flat-mapping (stream nesting) and zipping a finite and an infinite
streams~-- by stringing-up combinators, taking advantages of local
declarations to reduce boilerplate.  |Raw.zip_raw| produces
a |(int exp, int exp) stream| of tuples, to be consumed by |iter|.  Since the
sugared API is just a sugar over the raw one, one may mix-and-match
the two.

The example also highlights the improvement over
\cite{strymonas-2017}. The code produced then contained a local
function declaration (closure creation) and indirect function
invocations in the main loop~-- as well as tuple constructions.  In
contrast, the \strymonas's OCaml code, App.~\ref{a:complex-ex}, is
first-order, with no function invocations of any kind in the main
loop, and also no tuples~-- therefore, it can be also rendered in C.
The code looks nothing like the declarative expression above. Writing
such code by hand, or even reading it, is a challenge.
Benchmarks, \S\ref{s:evaluation}, and the regression test suite offer
many more complex examples.

\strymonas\ distribution has more examples of tupling, such as
accumulating both the minimum and the maximum values of an |int cstream|,
as well as sliding window combinators. The latter, described in
detail in \cite{strymonas-PEPM24}, produce and transform streams of the type
|'a window stream|, where |'a window| is an (OCaml) first-class module
abstracting over
the representation of the window buffer. Again, the generated code, in either
OCaml or C, has no records or closures. These windows are \emph{not}
an extension of \strymonas: they are implemented using the sugared and
raw APIs only, just like the earlier examples~-- thus demonstrating
the expressive power of \strymonas.

\subsection{Stateful Pipelines}
\label{s:stateful}

We have just encountered the raw interface Fig.\ref{f:raw-API}: the
real API of \strymonas, that is, the primitives. The primitives are 
harder to use directly because they are more
general, more abstract, and because some have
side-conditions on their arguments, which the user is obligated to check.
That is why \strymonas\ offers the sugared interface,
Fig.\ref{f:API}, of familiar stateless streams,
implemented entirely in terms of the raw.\footnote{\relax
We have already shown the implementation of \textsf{zip\_with} in
terms of \textsf{zip\_raw}. The combinators \textsf{iter},
\textsf{filter},  \textsf{flat\_map} of the sugared interface are
identical to those of the raw interface, but with more specific, less
confusing to novice types. The implementation of \textsf{from\_to}
and \textsf{take\_while} is
shown in this section. The others are implemented similarly, see the
\strymonas\ repository.}
We call those streams stateless because no
internal state is visible. It exists, as we are about to see, from the 
implementation of the sugared interface combinators. Programmers
likewise may introduce their own state and share it across the
pipeline. After all, the raw primitives apply to streams built with
the cooked API: although the primitives and the sugar are separated into
two APIs, for the convenience of the user, there is only one stream
type.

To demonstrate the introduction and use of the internal state, we take
the example of |from_to| of the sugared interface: |from_to a b|
produces the stream of consecutive integers from |a| to |b|. It is
implemented in terms of the raw API as:
%% \footnote{Fig.\ref{f:API} actually offers a bit more
%% general combinator, with the optional \textsf{step} argument,
%%  defaulted to 1.}
\begin{code}
let from_to : int exp -> int exp -> int cstream = fun a b ->
  let- z = initializing_ref a in
  guard C.(dref z <= b) @@
  infinite @@ fun k -> C.(let- v = letl (dref z) in (incr z) @. k v)
\end{code}

The operator |let-| is a user-definable let-binding--like construct
offered in recent versions of
OCaml\footnote{\url{https://ocaml.org/manual/5.3/bindingops.html}}. We
defined it as
\begin{code}
let (let-) c k = c k
\end{code}
It makes the stream state introduction |initializing_ref| (and also
\textsf{letl} for local binding in the target language, see below)
look like regular OCaml let-expressions.

The |from_to| stream has the internal state: the integer to produce
next. All internal state must be declared, using |initializing_ref|,
whose first argument is the initial value for the state and the second
is the continuation receiving the (code of the) mutable variable for
the state (the |let-| operator hides this continuation).
The argument of |infinite| is a producer of stream items:
upon receiving a continuation it should pass it what should be the
current stream item. The continuation must be invoked exactly
once,\footnote{The exception is the end of the stream, where a
  producer may skip invoking the continuation. For many streams, the
  end is determined by the termination condition. For some streams,
  however, the end cannot be determined before trying to produce an
  item and failing.} which is the side-condition the user has to
check.  In our example, the producer passes to the continuation the
current value of the state, incrementing it; the side-condition of the
linear use of the continuation is satisfied.  The combinator
|infinite|, which is the inverse of |iter|, keeps invoking its
producer argument, indefinitely. The |guard| combinator specifies the
termination condition: (the code of) a boolean expression. It may
access the stream state but must not mutate it~-- the side-condition,
which clearly holds in our example. The stream ends when the
termination expression returns \textsf{false}.

\begin{figure}
\begin{code}
type 'a exp         $\LS$(* Abstract type of code expressions *)
type 'a stm         $\LS$(* Abstract type of statements *)
type 'a mut         $\LS$(* Mutable variables *)
type 'a arr         $\LS$(* Arrays *)
type 'a tbase       $\LS$(* Base types *)
type 'a stream      $\LS$(* Here, 'a is not necessarily the code type! *)

type 'a emit = ('a -> unit stm) -> unit stm

(* Immutable and mutable stream state introduction *)
val initializing : 'z exp -> ('z exp -> 'a stream) -> 'a stream  (* let-binding, essentially *)
val initializing_ref : 'z exp -> ('z mut -> 'a stream) -> 'a stream

(* Producer *)
val infinite : 'a emit -> 'a stream

(* Consumer: the inverse of infinite *)
val iter : ('a -> unit stm) -> 'a stream -> unit stm

(* Transformers *)
val map_raw : ?linear:bool -> ('a -> 'b emit) -> 'a stream -> 'b stream
val map_raw' : ('a -> 'b) -> 'a stream -> 'b stream

val guard : bool exp -> 'a stream -> 'a stream
val filter_raw : ('a -> bool exp) -> 'a stream -> 'a stream
val flat_map_raw : ('a exp -> 'b stream) -> 'a exp stream -> 'b stream
val zip_raw    : 'a stream -> 'b stream -> ('a * 'b) stream
\end{code}
\caption{Raw stream interface, slightly simplified for presentation.
In \textsf{map\_raw}, the notation
  \textsf{?linear:bool} means an optional boolean argument. If not
  specified, the default values (\textsf{true} in this case) is used.
}
\label{f:raw-API}
\end{figure}

More interesting is the difference encoder, which occurs in
JPEG transcoder and Vocoder benchmarks of the StreamIt benchmark
suite, see \cite[Fig. 4: stateful version]{Thies10}.
The example demonstrates that the processing of a stream item may
depend on the earlier seen item(s).
In our case, the first item of the input stream is passed as it is,
but the subsequent output stream items are the differences between the
two consecutive input items.
\begin{code}
let diff : int cstream -> int cstream = fun st ->
  let- z = initializing_ref C.(int 0) in
  st |> map_raw C.(fun e k -> let- v = letl (e - dref z) in (z := e) @. k v)
\end{code}
The general |map_raw| combinator takes as the first argument the item
transformer in the continuation-passing style. The continuation |k|
must be invoked at most once. It does not have to be always invoked~--
in which case the user has to set the optional |linear| argument to
\textsf{false} (unless the stream is about to be ended); |map_raw|
then behaves as a fused map-filter. Thus |map_raw| is harder to use,
but offers extra power for advanced users.  The item transformer is
applied to each stream item in turn, exactly once. If the transformer
maintains state, |map_raw| effectively acts as an accumulating map.

App.~\ref{a:rll} shows an elaborated version of |diff|:
run-length encoder and decoder. Nested grouping/aggregation,
App.~\ref{a:adv22}, is another interesting example of
stateful streams.

The examples have clearly showed that operators in a stateful pipeline may
not only maintain their own private state but also share state. There is
an important side-condition that must be checked whenever a
termination condition (specified in |guard|) shares state with
a mapping or filtering operation~-- that is, when a mutable variable
allocated by |initializing_ref| and accessed in a termination
condition is also used, and assigned to, in a mapping or filtering
operation somewhere in a pipeline. First,
when the termination expression in a |guard| evaluates to false,
no further changes in the state should turn it to
true.\footnote{\relax
This side-condition is actually rather easy to ensure:
introduce a new piece of boolean state (mutable boolean variable)
initialized to \textsf{true}; make \textsf{guard} to dereference this
variable; make sure each assignment to it is monotone:
never assign \textsf{true} if the
current value is \textsf{false}. \strymonas's code use this pattern
extensively. One is reminded of the conversion of flowcharts to
structured programs.} Furthermore,
if a mapping or filtering operation changes the state in such a way that
a termination condition in a |guard|
becomes false, the mapping function must not produce
an item: it must not invoke its continuation.
We shall see in \S\ref{s:core-stateful} how these side-conditions
came about. Here we illustrate with the example of expressing
|take_while| of the sugared interface in terms of the raw interface:
\begin{code}
let take_while : ('a exp -> bool exp) -> 'a cstream -> 'a cstream = fun f st ->
  let- zr = initializing_ref (C.bool true) in
  st |> map_raw C.(fun e k -> if_ (f e) (k e) (zr := bool false)) |> guard C.(dref zr)
\end{code}
When the mapping function decides the stream should be finished, it
`skips' invoking its continuation and sets the termination flag |zr|
to \textsf{false}.

\section{Theory of Stateful Streams: Normal Forms of Stream Pipelines}
\label{s:derivation}

This section presents our theoretical contribution: a theory of
stateful streams. The key results are the notion of a linear stream;
transforming away |zip| when one of the zipped streams is linear; and
linearization. These results then let
us define a normal form of a pipeline and give an algorithm to
compute it~-- normalization-by-evaluation (NbE),~-- proving
its correctness. NbE is the foundation of \strymonas, responsible
for the guaranteed elimination of intermediate data structures in
pipelines. NbE sets \strymonas\ apart from many other stream
systems, which perform optimizations by generally non-confluent
rewriting rules, see \S\ref{s:related} for discussion.

This section presents stateful streams and their semantics, equational
laws, normal forms and their calculation.  \S\ref{s:Agda} describes
the formalization in Agda. Normal forms have the property of being
easily mapped to low-level imperative code; \S\ref{s:implementation}
shows how.

\subsection{Core Stateful Streams}
\label{s:core-stateful}

Our starting point is an idealized stateful stream API, so-called Core
API, Fig.\ref{f:stateful-API}, convenient for theoretical development.
It is quite similar to the raw API in Fig.\ref{f:raw-API}, with
important differences: state is explicit and distinction between host
and target code is erased. In the implementation section
\S\ref{s:implementation} we shall see how Fig.\ref{f:stateful-API} is
turned into the \strymonas\ API in Fig.\ref{f:raw-API}.

The Core API deals with stateful streams |('a,'z) stream|, which carry
values of the type |'a| and have an exposed internal state of the type
|'z|. (We use the same name |stream| as before~-- but with two
arguments, which should disambiguate). The stream producer is
|unroll|, which, by its type looks cunningly similar to the familiar
|unfold|. However, |unroll| may decline to produce an item (`skip'),
while still updating the internal stream state.  It always produces an
infinite stream (even if an infinite skipping stream). Stream
termination is handled separately, by |guard|. A new piece of state is
introduced by |init|.  Strictly speaking, all pipeline
state can be introduced by a producer.  But this is unmodular
and cumbersome~-- in practice and in formalization.
The operation |abstract| hides a part of the state.  The
operation |map_filter| is an accumulating (since it may
also update the state) map-filter. The operation |guard| may examine only the
state of the stream, not the items; it does not update the state.  There
are further side-conditions detailed later. In |flat_map|, the inner
stream may have its own state~-- which should be hidden/abstracted.
What is exposed is the state of the outer stream, which the inner
stream may access.
\begin{figure}
\begin{code}
type ('a,'z) stream

val unroll : ('z -> 'a option * 'z) -> 'z -> ('a,'z) stream
val init : 'z -> ('a, 'z1) stream -> ('a, 'z*'z1) stream
val abstract : ('a, 'z*'z1) stream -> ('a,'z1) stream

val guard : ('z -> bool) -> ('a,'z) stream -> ('a,'z) stream
val map_filter : ('z -> 'a -> 'b option * 'z) -> ('a,'z) stream -> ('b,'z) stream
val flat_map   : ('z -> 'a -> ('b,'z) stream) -> ('a,'z) stream -> ('b,'z) stream
val zip : ('a,'z1) stream -> ('b,'z2) stream -> ('a*'b, 'z1*'z2) stream
\end{code}
\caption{Stateful stream Core API. All pipelines built using this API
  must satisfy side-conditions \ref{d:side-cond}. Just like in full
  \strymonas, nested streams should have fixed structure. That is, the
  argument of \textsf{flat\_map} has to be a function of the form
  $\lam{z}\lam{a}s$ where $s$ is a term build solely by composing the
  operations of the Core API. This restriction is relied upon in
  normalization and linearization, Prop.~\ref{t:linearization}.
}
\label{f:stateful-API}
\end{figure}

\subsubsection*{Notation}
Throughout \S\ref{s:derivation} we use the notational conventions
common in universal algebra \cite[\S4]{jacobs-tutorial}. $X \times Y$
is the cartesian product of sets $X$ and $Y$, viz.,
$X\times Y = \{(x,y) \mid x\in X, y\in Y\}$,
with
\[
\pi: X\times Y\to X = \lam{(x,y)}x \qquad
\pi': X\times Y\to Y = \lam{(x,y)}y
\]
being the projections from it. The product notation
also applies to functions: if $f: X\to X'$ and $g: Y\to Y'$ then
$f \times g: X\times Y \to X'\times Y'$. The tupling, or pairing, of
$h_1: Z\to X$ and $h_2: Z\to Y$ is $\langle h_1, h_2\rangle: Z\to
X\times Y$. A disjoint union (or, coproduct, or sum) of sets $X$ and
$Y$ is $X + Y$; it also applies to functions, such
as $f$ and $g$ above: $f + g: X+Y \to X'+Y'$. Injections into the
union are denoted $\kappa: X\to X+Y$ and $\kappa': Y\to X+Y$.  If
$h'_1: X\to Z$ and $h'_2: Y\to Z$ then $[h'_1, h'_2]: X+Y\to Z$
denotes their `cotupling'.
$\mathbb{B}$ is the set of Booleans; conditional
expressions are written as $(e ? e_1 : e_2)$, where
$e:\mathbb{B}$. The identity function is $\Id$, $\mathsf{K}$ is the
K-combinator, and $\RHD$ is the \emph{left-to-right} functional composition.

A frequently occurring case of coproduct
is $1+X$, corresponding to |$X$ option| in OCaml, for which we
use the special notation $X^\dagger$. Applied to functions, such as
$f: X\to Y$, it means
$f^\dagger: X^\dagger\to Y^\dagger$. In the above, $1$
is the singleton set whose sole element is denoted
$\mathbf{1}$. The same $\mathbf{1}$ also means $\kappa\mathbf{1}: 1 + Y$,
and also the (unique) function
$X\to 1$ and also 
the function $\mathsf{K}(\kappa\mathbf{1}) : X\to 1 + Y$. If
$f: X\to 1+Y$, the frequently occurring
$[\mathbf{1},f]$ means the function $1+X\to 1+Y$.

\subsubsection*{Semantics}

The precise semantics of the core API
Fig.\ref{f:stateful-API}\footnote{\relax Readers who would like to
  view the semantics in a programming-language (OCaml) notation 
  are referred to the
  file \nolinkurl{derivation.ml} (in the \strymonas\ repository
  \url{https://github.com/strymonas/strymonas-ocaml/tree/main/lib/derivation.ml}),
  module \textsf{Stream2Denot}.}
is specified in terms of coinductive stateful
skipping streams $S_{AZ}$~-- $A$ is the set (type) of stream items and
$Z$ is the set of states~-- a stateful version of the
commonly used model of skip streams \cite{coutts_stream_2010}.
$S_{AZ}$ is defined as (the carrier of) the final coalgebra of the
functor $T_{AZ}$:
\[
T_{AZ}(X) = 1\; +\; (1 + A) \times Z \times X^Z
\]
That is, $T_{AZ}(S_{AZ})$ is the observation of the stream $S_{AZ}$:
the stream may be observed as finished or continuing. In the latter
case we observe its current
state $Z$, possibly the current item $1+A$ and the continuation
$S_{AZ}^Z$, which is to receive the state and produce the (remaining)
stream. If the current item is present, we say the stream has produced it;
if absent, we say the stream has \emph{skipped}.

\begin{figure}
\begin{tabular}{McMcMc}
\phantom{{\scriptstyle f'}}U & \xrightarrow{\makebox[5em][c]
  {$\scriptstyle f$}} & S_{AZ}
\\
\left.\begin{gathered}{\scriptstyle f'}\\[1\jot]\end{gathered}
\right\downarrow & & \big\Vert
\\
\phantom{{\scriptstyle f'}}T_{AZ}(U) &
\xrightarrow[{\makebox[5em][c]{$\scriptstyle T_{AZ}(f)$}}]{} & T_{AZ}(S_{AZ})
\end{tabular}
\caption{Final co-algebra}
\label{f:coalgebra}
\end{figure}

Our semantics is set-theoretic. Therefore, we take all functions
appearing in Fig.~\ref{f:stateful-API}
(e.g., |'z -> bool|, the first argument of |guard|) to be
total. Domain-theoretic semantics would be more general
and more realistic~-- and also messier. We use the set-theoretic semantics
for clarity~-- after all, termination is not the main issue and the mapping,
guard etc. functions are often (and recommended, in stream library
guides, to be) total.

Fig.\ref{f:stateful-semantics} shows the
semantic equations: the realization of the Core API of Fig. \ref{f:stateful-API}
when |($A$,$Z$) stream| is $S_{AZ}$. The operations are defined by
coinduction, in the style expounded in \cite{jacobs-tutorial}: to
define an operation $f: U\to S_{AZ}$ where $U$ is some set, we introduce
a mapping $f': U \to T_{AZ}(U)$, thus giving  $U$ a coalgebra structure.
The desired $f$ is then the unique homomorphism from the $U$ coalgebra to
the final coalgebra $S_{AZ}$, see Fig.~\ref{f:coalgebra}.
We write this fact as $f = \coind f'$.
For our functor $T_{AZ}$ it is guaranteed to exist.
In semantic equations we silently
apply isomorphisms, such as commutativity and associativity of the sum
and product, currying, $S_{AZ} = T_{AZ}(S_{AZ})$.

The semantic equations expose a subtlety of |zip|: when one of the
streams to zip skips but the other (call it $s_1$) does not, the result
stream skips and its continuation uses $s_1$ as it was, even though we
have peeked already at its current item. One may say, zipping requires
look-ahead~-- which is not the operation supported by all streams.
Although we certainly support zipping, our API does not provide
look-ahead and so we have to find a work-around~-- what is responsible
for the main complexity of the |zip| implementation, discussed in
\S\ref{s:zip}.

\begin{comment}
An alternative for expressing zip: define an operation
rewind : A^\dagger\times Z\times S_{AZ}^Z\to Z\times S_{AZ}^Z
so that rewind s = (z,t) such that t z = s. In other words, given s
we obtain the state z that has lead to s. For the specification
purposes, we can leave it as thus, non-deterministic and
underspecified. There is a way to express it in our framework, just
representing a stream as a pair (z,t)
type ('a,'z) stream = 'z * ('a,'z) stream_val
and ('a,'z) stream_val = Nil | Cons of 'a * ('a,'z) stream

Then in case s1 has item but s2 skips, we rewind s1. The result of
zipping is a skip that has the current state of s2 but \emph{previous} state
of s1 (that is, state before we examined s1 and found it has the
item).
The advantage: zip with a linear stream is then strongly equivalent to
map_filter, modulo tail.
\end{comment}

%% In |abstract|, we create a closure deliberately, to abstract out a
%% part of state. We could have used some sort of existential, but it is
%% inconvenient.

\begin{figure}
\eqaopenskip=0pt
\begin{eqnarray*}[>{}l]
\O unroll/ (f: Z\to A^\dagger\times Z): Z\to S_{AZ}
 = \coind h \textrm{  where}\\
\quad h: Z\to (1 + (A^\dagger\times Z)\times Z^Z) =
  \lam{z} \kappa' (f\ z,\Id)
\\[1\jot]
\O init/: Z \times S_{AZ_1} \to S_{A(Z\times Z_1)}
 = \coind h \textrm{  where}\\
\quad h: (Z\times 1 + (A^\dagger\times (Z\times Z_1)) \times S_{AZ_1}^{Z_1})\to
  (1 + (A^\dagger\times (Z\times Z_1)) \times (Z \times S_{AZ_1})^{(Z\times Z_1)})\\
\quad = \mathbf{1} + \Id\times(\lam{t}\Id\times t)
\\[1\jot]
\O abstract/: S_{A(Z\times Z_1)}\to S_{AZ_1}
 = \coind h^\dagger \textrm{  where}\\
\quad h: (A^\dagger\times Z_1) \times (Z \times S_{A(Z\times Z_1)}^{(Z\times Z_1)})
  \to (A^\dagger\times Z_1) \times S_{A(Z\times Z_1)}^{Z_1}
= \Id\times \lam{(z,t)} (\lam{z_1} t\ (z,z_1))
\\[1\jot]
\O guard/ (p:Z\to \mathbb{B}): S_{AZ}\to S_{AZ}
 = \coind h \textrm{  where}\\
\quad h: (1+A^\dagger\times Z \times S_{AZ}^Z)\to
 (1+A^\dagger\times Z \times S_{AZ}^Z) =
[\mathbf{1}, \lam{(a',z,t)}(p\ z ? \kappa' (a',z,t) : \mathbf{1})]
\\[1\jot]
\O map\_filter/ (f: Z\times A\to B^\dagger\times Z): S_{AZ}\to S_{BZ}
 = \coind h^\dagger \textrm{  where}\\
\quad h: (Z+Z\times A) \times S_{AZ}^Z\to
         (B^\dagger\times Z) \times S_{AZ}^Z =
[\langle \mathbf{1}, \Id\rangle, f] \times \Id
\\[1\jot]
\O flat\_map/ (f: Z\times A\to S_{BZ}): S_{AZ}\to S_{BZ} =
  \kappa \RHD g  \textrm{  where}\\
\quad g: S_{AZ} + S_{BZ}\times Z\times S_{AZ}^Z \to S_{BZ}
 = \coind [h_o, h_i] \textrm{  where}\\
\quad h_o: (1 + 1\times Z\times S_{AZ}^Z + A\times Z\times S_{AZ}^Z)\to
 (1+B^\dagger\times Z\times (S_{AZ} + S_{BZ}\times Z\times S_{AZ}^Z)^Z) = \\
\quad\quad
[\mathbf{1},\ (\kappa\times\Id\times\lam{t}t\RHD\kappa)\RHD\kappa',\
 \lam{(a,z,t)}h_i (f(z,a),z,t)]\\
\quad h_i: (1\times Z\times S_{AZ}^Z +
   (B^\dagger\times Z)\times S_{BZ}^Z\times Z\times S_{AZ}^Z)\to
 (1+B^\dagger\times Z\times (S_{AZ} + S_{BZ}\times Z\times S_{AZ}^Z)^Z) =\\
\quad\quad
[\kappa\times\Id\times\lam{t}t\RHD\kappa,\
\Id\times\lam{(t',\_,t)}\langle t', \Id, \mathsf{K} t\rangle\RHD\kappa']
\RHD\kappa'
\\[1\jot]
\O zip/: S_{A_1 Z_1} \times S_{A_2 Z_2} \to S_{(A_1\times A_2) (Z_1 \times Z_2)}
 = \coind h \textrm{  where}\\
\quad h:
\left( (1\times S_{A_2 Z_2} +
   A_1^\dagger\times Z_1\times S_{A_1 Z_1}^{Z_1}\times 1) +
   (S_{A_1 Z_1}^{Z_1} \times S_{A_2 Z_2}^{Z_2}) \times
   (A_1^\dagger\times A_2^\dagger\times Z_1\times Z_2)\right)\to\\
\quad\quad
 \left(1 + (A_1\times A_2)^\dagger\times (Z_1\times Z_2) \times
    (S_{A_1 Z_1} \times S_{A_2 Z_2})^{Z_1\times Z_2}\right) =
  \mathbf{1} + g \textrm{  where}\\
\quad g:
(S_{A_1 Z_1}^{Z_1} \times S_{A_2 Z_2}^{Z_2}) \to\\
\quad\quad
\left((Z_1\times Z_2) +
A_1\times Z_1\times Z_2 +
A_2\times Z_1\times Z_2 +
(A_1\times A_2)\times (Z_1\times Z_2)\right)
\to\\
\quad\quad
(A_1\times A_2)^\dagger\times (Z_1\times Z_2) \times
    (S_{A_1 Z_1} \times S_{A_2 Z_2})^{Z_1\times Z_2} = \lam{(t_1,t_2)}\\
\quad\quad
[\lam{z_{12}}(\mathbf{1}, z_{12}, (t_1\times t_2)),
\\
\quad\quad
\lam{(a_1,z_1,z_2)} (\mathbf{1}, (z_1,z_2), (r(\kappa' a_1,z_1,t_1)\times t_2)),
\\
\quad\quad
\lam{(a_2,z_1,z_2)} (\mathbf{1},(z_1,z_2), (t_1 \times r(\kappa' a_2,z_2,t_2))),
\\
\quad\quad
\lam{(a_{12},z_{12})} (a_{12},z_{12},(t_1\times t_2))]
\\[1\jot]
\quad r: A^\dagger \times Z \times S^Z_{AZ} \to S^Z_{AZ} =
  \lam{u}\mathsf{K}(\kappa' u)
\end{eqnarray*}
\caption{Semantics of the Core API}
\label{f:stateful-semantics}
\end{figure}

\begin{definition}[Effectively ended stream]\label{d:ended}
A stream $s$ is called effectively ended if it does not and will never
produce any item. That is, it satisfies the predicate \O Ended/,
(coinductively)
defined as: $\O Ended/ s$ holds just in case either $s=\mathbf{1}$; or
$s=\kappa'(\mathbf{1},z,t) \land \O Ended/ (t\ z)$.
\end{definition}

All pipelines built by applying the operations of the Core
API Fig.\ref{f:stateful-API} must satisfy the following side-conditions:
\begin{definition}[Pipeline side-conditions]\label{d:side-cond}
\leavevmode
\begin{enumerate}
\item[(i)] In a pipeline $s \triangleright \O guard/ g$, if
$T(s) = \kappa'(a,z,t)$ and $g\ z = \O false/$ then $s$ must be
  effectively ended (and hence it must be that $a=\mathbf{1}$).
That is: a guard may terminate only an effectively ended stream.
\item[(ii)] If a pipeline contains |guard $g$| followed by
|map_filter $f$| then $g\, z = g\, (\pi' (f\, z\, a))$ for all $z$ and $a$.
That is: any state
change effected by |map_filter|s should not affect the terminating
condition of the preceding guards.
\item[(iii)] If a pipeline contains |guard $g$| followed by
|flat_map $f$| then $g\,z=\mathsf{true}$ implies
$f (z,a) \triangleright \O guard/ g = f(z,a)$, for all $a$.
That is: any state
change effected by nested streams should not affect the terminating
condition of the preceding guards.
\end{enumerate}
\end{definition}
% actually, the side-condition (ii) may be relaxed: map_filter may
% affect g of the preceding guard such that g z becomes false in that
% new state, provided that there is another guard g' following map_filter
% such that g' z is also false.
%
\noindent
The side-condition (i) is not an actual restriction: we may
always precompose any |guard g| with
$\O map\_filter/ \lam{(a,z)}(g\ z ? (\kappa' a,z) : (\mathbf{1},z))$
that ensures (i). It does simplify the
formalization and implementation.

\subsection{Equational Laws}
\label{s:eq-laws}

The semantics of the core API lets us state and verify equational laws
of streams, which, in turn, help us define normal forms of stream
pipelines in \S\ref{s:NF}.
Fig.\ref{f:equations} collects the equations, in the form
of strong or weak (see \S\ref{s:zip}) equivalences.

\begin{definition}[Strong Stream Equivalence]
Two streams $s_1: S_{AZ}$ and $s_2: S_{AZ}$
are strongly equivalent, written $s_1 \cong s_2$, just in case they
are bisimilar: there exists a relation $R \subset S_{AZ}\times S_{AZ}$
such that $(s_1,s_2)\in R$ and for all $s$ and $s'$ such that
$(s,s')\in R$ it follows $(T_{AZ}(s),T_{AZ}(s'))\in T_{AZ}(R)$.
\end{definition}
\noindent
In plain English, two streams are strongly equivalent when they cannot
be distinguished by observation. Being terminated, the current
item or its absence (skip) and the state are all observable. Strong
equivalence is a congruence.
%% We extend the relations $\cong$ and $\approx$ from streams to stream
%% transformers in the obvious way.

The laws \eqref{l:unroll-init}-\eqref{l:zip-init} justify that |init|
is indeed strictly speaking unnecessary and all state for the pipeline can
be introduced by the producer.  Law \eqref{l:init-abstract} shows that
|init| and |abstract| are sort of inverses.  In
\eqref{l:abstract-init}, the abstracted part $Z$ of the overall state
and the state $Z'$ introduced by |init| are generally different;
$iso$ in that law denotes the application of the state isomorphism
$Z'\times (Z\times Z_1) \leftrightarrow Z\times(Z'\times Z_1)$.

\begin{figure}
\begin{eqnarray}[rcl]
\O unroll/ u\ z_1 \triangleright \O init/ z &\cong&
\O unroll/ (\Id\times u)\ (z,z_1)
\label{l:unroll-init}
\\
\O guard/ g \RHD \O init/ z &\cong&
\O init/ z \RHD \O guard/ (\pi'\RHD g)
\label{l:guard-init}
\\
\O map\_filter/ f \RHD \O init/ z &\cong&
\O init/ z \RHD \O map\_filter/ (\Id\times f)
\label{l:map-init}
\\
\O flat\_map/ f \RHD \O init/ z &\cong&
\O init/ z \RHD
\O flat\_map/ (\lam{((z,z_1),a)} f (z_1,a) \triangleright \O init/ z)
\label{l:flatmap-init}
\\
\O zip/ (s_1 \RHD \O init/ z, s_2) &\cong&
\O zip/ (s_1,s_2) \RHD \O init/ z
\label{l:zip-init}
\\
\O zip/ (s_1 \RHD \O abstract/, s_2) &\cong&
\O zip/ (s_1,s_2) \RHD \O abstract/
\label{l:zip-abstract}
\\
\O init/ z \RHD \O abstract/ &\cong& \Id
\label{l:init-abstract}
\\
\O abstract/ \RHD \O init/ z &\cong&
\O init/ z \RHD iso \RHD \O abstract/
\quad \textrm{see text for $iso$}
\label{l:abstract-init}
\\
\O abstract/ \RHD \O guard/ g &\cong&
\O guard/ (\pi'\RHD g) \RHD \O abstract/
\\
\O abstract/ \RHD \O map\_filter/ f &\cong&
\O map\_filter/ (\Id\times f) \RHD \O abstract/
\\
\O abstract/ \RHD \O flat\_map/ f &\cong&
\O flat\_map/ (\lam{((z,z_1),a)} f (z_1,a) \triangleright \O init/ z)
   \RHD \O abstract/
\label{l:abstract-flatmap}
\\
\O unroll/ g\ z \triangleright \O map\_filter/ f &\cong&
\O unroll/ (g\RHD [\langle \mathbf{1},\Id\rangle,f])\ z
\label{l:unroll-map}
\\
\O guard/ f_1 \RHD \O guard/ f_2 &\cong& \O guard/ (f_1 \land f_2)
\label{l:guard-guard}
\\
\O guard/ g \RHD \O map\_filter/ f &\cong&
\O map\_filter/ f \RHD \O guard/ g
\quad\textrm{ (*)}
\label{l:guard-map}
\\
\O guard/ g \RHD \O flat\_map/ f &\cong&
\O flat\_map/ f \RHD \O guard/ g
\quad\textrm{ (*)}
\label{l:guard-flatmap}
\\
\O flat\_map/ f_1 \RHD \O map\_filter/ f_2 &\cong&
\O flat\_map/ (f_1 \RHD \O map\_filter/ f_2)
\label{l:flatmap-map}
\\
\O flat\_map/ f_1 \RHD \O flat\_map/ f_2 &\cong&
\O flat\_map/ (f_1 \RHD \O flat\_map/ f_2)
\label{l:flatmap-flatmap}
\\
\O zip/ (s_1 \RHD \O guard/ g, s_2) &\cong&
\O zip/ (s_1,s_2) \RHD \O guard/ (\pi\RHD g)
\label{l:zip-guard}
\\
\O zip/ (\O unroll/ u_1\ z_1, \O unroll/ u_2\ z_2) &\approx&
\O unroll/ ((u_1\times u_2)\RHD (\textit{pair}^\dagger\times\Id))\ (z_1,z_2)
\quad\textrm{(**)}
\label{l:zip-bothlinear}
\\
\O zip/ (\O unroll/ u_1\,z_1, s_2) &\approx&
\O init/ (z_1,s_2) \triangleright
\O map\_filter/ (f\,u_1)
\quad\textrm{(**)}
\label{l:zip-linear}
\end{eqnarray}
\caption{Equational laws of stateful streams. (*) marks the laws that
require side-conditions \ref{d:side-cond}.
(**) The weak equivalence $\approx$ is defined in \S\ref{s:zip}; see
that section for details on these laws. In particular,
$f$ appearing in \eqref{l:zip-linear} is defined in
Prop.~\ref{t:zip-linear}.
The operator $\RHD$ is the left-to-right composition:
$(f \RHD g)\ x = g (f\ x)$. As before, $\triangleright$ is the
left-to-right application: $x \triangleright f = f\ x$. Both emphasize
left-to-right flow of information.
}
\label{f:equations}
\end{figure}

The proofs are either the straightforward calculation (applying
isomorphisms: isomorphic streams are clearly strongly equivalent),
or following the definition of the strong or weak equivalence.
The most important among the isomorphisms are the ones that follow
from the definition of coalgebra, Fig.~\ref{f:coalgebra}.
Concretely, if $h: U\to T_{AZ}(U)$ and $u\in U$, then
\[
u \triangleright \coind h =
u \triangleright h \triangleright T_{AZ}(\coind h) =
u \triangleright h \;\triangleright\; \mathbf{1} + \Id \times \Id \times
\lam{t} t \RHD \coind h
\]
For example, for $\O unroll/ f$ we have
\[
\O unroll/ f z
= z \triangleright \O unroll/ f
= z \triangleright
  \lam{z} \kappa' (f\ z,\Id) \triangleright
  T_{AZ}(\O unroll/ f)
= \kappa' (f\ z,\O unroll/ f)
\]
\begin{comment}
For example, for $\O init/$ we have
\begin{eqnarray*}
\O init/ (z,s) =
(z,s) \triangleright \O init/ =
(z,s) \triangleright \mathbf{1} + \Id\times(\lam{t}\Id\times t)
\triangleright T_{A(Z\times Z_1)}(\O init/)
\\
=
\begin{cases}
\mathbf{1} \triangleright T_{A(Z\times Z_1)}(\O init/)
& if $s=\mathbf{1}$
\\
\kappa' (a,(z,z_1),\lam{(z,z_1)}(z,t\,z_1)) \triangleright T_{A(Z\times Z_1)}(\O init/)
& if $s = ((a,z_1),t)$
\end{cases}
\\
=
\begin{cases}
\mathbf{1}
& if $s=\mathbf{1}$
\\
\kappa' (a,(z,z_1),\lam{(z,z_1)}(z,t\,z_1) \triangleright \O init/)
& if $s = ((a,z_1),t)$
\end{cases}
\end{eqnarray*}
\end{comment}
Therefore, for the equational law \eqref{l:unroll-init} we have:
\begin{eqnarray*}
\O init/ (z,\O unroll/f \ z_1)\\
% unroll the definition of init, from the commutative diagram
= (z,\O unroll/f \ z_1)
  \triangleright
  \mathbf{1} + \Id\times(\lam{t}\Id\times t)
  \triangleright
   T_{A(Z\times Z_1)}(\O init/)\\
% unroll the definition of unroll, from the commutative diagram
= (z,z_1 \triangleright \lam{z} \kappa' (f\ z,\Id)
         \triangleright T_{AZ}(\O unroll/ f))
  \triangleright
  \mathbf{1} + \Id\times(\lam{t}\Id\times t)
  \triangleright
   T_{A(Z\times Z_1)}(\O init/)\\
% inner application
= (z,\kappa' (f\ z_1,\Id\RHD\O unroll/ f))
  \triangleright
  \mathbf{1} + \Id\times(\lam{t}\Id\times t)
  \triangleright
   T_{A(Z\times Z_1)}(\O init/)\\
% pairs/sums distributivity, dropping composition with \Id
%% = \kappa' (z, (f\ z_1,\O unroll/ f))
%%   \triangleright
%%   \mathbf{1} + \Id\times(\lam{t}\Id\times t)
%%   \triangleright
%%    T_{A(Z\times Z_1)}(\O init/)\\
% reshuffling
= \kappa' ((z, f\ z_1),\O unroll/ f)
  \triangleright
  \mathbf{1} + \Id\times(\lam{t}\Id\times t)
  \triangleright
   T_{A(Z\times Z_1)}(\O init/)\\
% application
= \kappa' ((z, f\ z_1),\Id\times \O unroll/ f)
  \triangleright
   T_{A(Z\times Z_1)}(\O init/)\\
% expansion of \Id\times \O unroll/ f
= \kappa' (z, f\ z_1,\lam{(z,z_1)} (z, \O unroll/ f\ z_1))
  \triangleright
   T_{A(Z\times Z_1)}(\O init/)\\
% application of T
= \kappa' (z, f\ z_1,\lam{(z,z_1)} \O init/ (z, \O unroll/ f\ z_1))\\
% beta-expansion
= \kappa' ((\Id\times f) (z,z_1),
\lam{(z,z_1)} \O init/ (z, \O unroll/ f\ z_1))
\\
% beta-expansion
(z,z_1) \triangleright
  \lam{z} \kappa' ((\Id\times f)\ z,\Id) \triangleright
 T_{A(Z\times Z_1)}(\lam{(z,z_1)} \O init/ (z, \O unroll/ f\ z_1))
%% % coinductive hypothesis
%% = \kappa' (z, f\ z_1,\O unroll/ (\Id\times f))\\
%% % definition of unroll
%% = \O unroll/ (\Id\times f) (z,z_1)
\end{eqnarray*}
We have therefore established that
$\lam{(z,z_1)} \O init/ (z, \O unroll/ f\ z_1)$ is a coalgebra
morphism to $S_{A(Z\times Z_1)}$
from the coalgebra with the carrier
$Z\times Z_1$ and the operation
$\lam{z} \kappa' ((\Id\times f)\ z,\Id)$. From the |unroll| definition,
$\O unroll/(\Id\times f)$ is also such morphism.  Since
$S_{A(Z\times Z_1)}$ is final and a morphism to it is unique,
$\lam{(z,z_1)} \O init/ (z, \O unroll/ f\ z_1)$ must be the same as
$\O unroll/(\Id\times f)$ (modulo isomorphisms).

% sometimes, h \triangleright T_{AZ}(op) \triangleright h' \triangleright
% T_{BZ'}(op')
% can be represented as
% h \triangleright h' \triangleright T_{BZ'} (op \RHD op')
% if h : U -> T_{AZ}(U) and h' are `sufficiently polymorphic'

As an example of bisimulation, we show the proof of \eqref{l:guard-map}.
Let $R$ be the relation
$(s \triangleright p_1, s \triangleright p_2)$,
where $p_1 = \O guard/ g \RHD \O map\_filter/ f$ and
$p_2 = \O map\_filter/ f \RHD \O guard/ g$, for all streams $s$ and
functions $f$ and $g$ of appropriate types.
We aim to prove that $R$ is a bisimulation: if
$(s_1,s_2)\in R$ then either: (i) $s_1=s_2=\mathbf{1}$, or
(ii) $s_1=\kappa'(a,z,t_1),s_2=\kappa'(a,z,t_2), (t_1,t_2)\in R$.
The proof is by (long) case-analysis on $s$. The case $s=\mathbf{1}$ is
trivial. If $s=\kappa'(\mathbf{1},z,t)$ ($s$ skips), then
|map_filter $f$ $s$| skips as well. There are two subcases:
If $g\ z = \O false/$, then $p_1\ s = p_2\ s = \mathbf{1}$; otherwise
$p_1\ s = \kappa'(\mathbf{1},z,t\RHD p_1)$,
$p_2\ s = \kappa'(\mathbf{1},z,t\RHD p_2)$ and
$(t\RHD p_1,t\RHD p_2)\in R$ by premise.
In the remaining case $s=\kappa'(\kappa' a,z,t)$. The subcase of
$g\ z$ being \textsf{false} is ruled out by side-condition
\ref{d:side-cond}(i). Then
$p_1\ s = \kappa'(f(a,z),t\RHD p_1)$,
$p_2\ s = \kappa'(f(a,z),t\RHD p_2)$ keeping in mind
side-condition \ref{d:side-cond}(ii).

\begin{proposition}
The equational laws in Fig.\ref{f:equations} preserve
side-conditions \ref{d:side-cond}.
\end{proposition}
We only need to consider the laws involving |map_filter| and
|flat_map| (since these are the only transformers that may update the
state).

Formal proofs of the laws are discussed in \S\ref{s:Agda}.

\subsection{The Problem of zip}
\label{s:zip}

As already noted, |zip|~-- specifically, zipping of streams with
different production rates~-- poses a complication, which motivated the
theoretical development and responsible for most of its subtle points.
Here we describe the problem and its solutions in detail.

Let's consider a stream of natural numbers $0,1,\ldots$
\[
\mathrm{nat} \Def \O unroll/ (\lam{z} (\kappa' z,z+1))\ 0
\]
Each time the state (the next number to produce) is advanced, a new
item is produced. Each observation produces a new state, and a new item.
The stream
\[
\mathrm{nat10} \Def \O map/ (+ 10) \ \mathrm{nat}
\]
likewise has the strict unit production rate,
producing an item (natural numbers starting from $10$) with
each advancement of its state. Zipping such lockstep-matching 
streams is simple: each observation advances the state of both
streams and produces a pair of their items. Formally,
\[
\O zip/ \mathrm{nat}\ \mathrm{nat10} =
\O unroll/ (\lam{(z_1,z_2)} (\kappa' (z_1,z_2), (z_1+1,z_2+1))\ (0,10)
\]
In other words, zip and unroll fuse, in this case (see 
Prop.~\ref{t:zip-unroll-unroll} below for the precise statement).

Skipping or flat-mapping, however, may prevent the lockstep processing. Consider
\begin{eqnarray*}
\mathrm{evens} \Def \O filter/ \mathrm{even} \ \mathrm{nat}\\
\mathrm{natsevens} \Def \O zip/ \mathrm{nats}\ \mathrm{evens}
\end{eqnarray*}
Here, evens filters out the odd numbers from nat: whenever nats
produces an odd number, evens skips. Determining an observation of
natsevens is no longer simple. Suppose zip observes nats first,
obtaining its current item $a_1$ and the new state $z_1$. Then 
evens is observed; suppose it skipped: produced no item but gave the new
state $z_2$. The natsevens observation in this case should
be $(z_1,z_2)$ as the new state, and no item: there is nothing to pair
$a_1$ with. That item $a_1$ is left in limbo, with nowhere to put it.

Generally speaking, |zip| maps the observation of a pair of streams to
a pair of observations:
\[ \textit{pair}^\dagger: \left(A_1^\dagger\times
A_2^\dagger\to (A_1\times A_2)^\dagger = (1 + A_1 + A_2 + A_1\times
A_2)\to (A_1\times A_2)^\dagger\right) = [\mathbf{1},?,?,\kappa']
\]
The problematic cases, marked with $?$, correspond to the mappings
$A_1\to (A_1\times A_2)^\dagger$ and $A_2\to (A_1\times A_2)^\dagger$.
Clearly they have to be $\mathbf{1}$: say, if we only have $a_1:A_1$ we
cannot return a pair $\kappa'(a_1,a_2)$ since $a_2$ is not
available; we can only return $\mathbf{1}$, which means $a_1$ is
thrown out since there is nowhere to put it.\footnote{Formally, the
  problem is that a mapping $(1+A_1)\times(1+A_2) \to
  (1+A_1\times A_2)$ is not injective.}
To recover the thrown
out item, |zip| has to effectively backtrack the stream
(or remember the original stream before the observation, which is what
the semantics in Fig.~\ref{f:stateful-semantics} does: see $r$). This
backtracking not only complicates the semantics: it demands a
complicated implementation. It is worth then try to avoid it.

Throwing away an already observed item is harmless if occurs
at the end, so to speak, for example:
\[
\O zip/ \mathrm{nats}\ (\O filter/ (< 10)\ \mathrm{nats})
\]
After ten observations, nats produces $10$ and the filtered stream
skips (and goes on skipping forever). Therefore, the zipped stream
will not produce any more items; disregarding the observed $10$ is
harmless since it never gets paired up.

In the case of zipping of nats and evens, the observed item of nats
cannot be thrown out, however: it will be paired up. The problem
described earlier remains. There is a simple way around, however, in
this case: the stream evens should be observed first.  Indeed,
if zip first observes evens and notices it skips, natsevens should naturally
skip; the new state will be the pair of old state of nats and the new
state of the evens. The stream nats hence gets observed only when 
evens produced an item. Since
nats always produces an item, we will never end up holding an observed item with
nothing to pair it with. Formally, this case is described in
Prop.~\ref{t:zip-linear} below.

We thus notice the importance of distinguishing streams that always produce
an item when observed~-- except at the very end. We call such
streams linear (unrollings).

\begin{definition}[Linear stream and linear unrolling]
\label{d:linear-unrolling}
A stream $s$ is called linear if it skips only at the end: if $s$
skips then it is effectively ended.\\
A function $f: Z\to A^\dagger\times Z$ is called a linear unrolling
whenever $f z\triangleright \pi = \mathbf{1}$ implies $z\in \delta_f$.
Here, $\delta_f$ is the largest subset of $Z$ with the property
that for all $z \in \delta_f$ we have $f z = (\mathbf{1},z')$
and $z' \in \delta_f$.\\
A function $f: Z\times A\to B^\dagger\times Z$ is called a linear
mapping just in case $\lam{z}f(z,a)$ is a linear unrolling for all
$a\in A$.
\end{definition}

\begin{proposition}\label{t:linear-trans} If $u$ is a linear unrolling, then
$\O unroll/ u\ z$ is a linear stream. If $s$ is a linear stream then
$s \triangleright \O init/ z$,
$s \triangleright \O abstract/$,
$s \triangleright \O guard/ g$,
$s \triangleright \O map\_filter/ f$ ($f$ is a linear mapping) are all
  linear streams. If $s_1$ and $s_2$ are linear, so is
$\O zip/ (s_1,s_2)$.
\end{proposition}
The proposition follows from the observation:
if $s$ is linear, then it can be represented as
$s=\overline s\ccat \underline s$, where
$\overline s$ is a stream with no skips and $\underline s$ is
effectively ended, and $\ccat$ is stream concatenation. Operations
|init|, |abstract|, |guard| and |map_filter| distribute over the
concatenation. In particular, we have
$(\overline s\ccat \underline s)\triangleright \O guard/ g =
\overline s\ccat (\underline s\triangleright \O guard/ g)$
(keeping in mind side-condition \ref{d:side-cond}(i))
and
$\O zip/ (\overline s_1\ccat \underline s_1,\overline s_2\ccat \underline s_2)=
\O zip/ (\overline s_1,\overline s_2) \ccat \underline s$ for some
effectively ended $\underline s$.

\subsubsection*{Coarser Stream Equivalence}

By definition, a linear stream does not skip, except perhaps at the
end~-- because skipping forever is essentially the end of the stream.
To formalize this `essentially' we have to refine the notion of the
stream equivalence.

\begin{definition}[Weak Stream Equivalence] Two streams $s_1$ and $s_2$
are weakly equivalent, written $s_1 \approx s_2$, just in case
$\O NoSkip/ s_1 \cong \O NoSkip/ s_2$, where
\O NoSkip/ is coinductively defined as
\[
\O NoSkip/ s =
\begin{cases}
\mathbf{1} & if $s$ is effectively ended\\
\kappa'(\kappa' a, z, t\RHD \O NoSkip/) & if $s = \kappa'(\kappa' a, z, t)$\\
\O NoSkip/ (t\, z) & if $s = \kappa'(\mathbf{1}, z, t)$
and not effectively ended.
\end{cases}
\]
Recall, that if a stream is not effectively ended, it will
eventually produce a value. \O NoSkip/ is hence well-defined.
\end{definition}
\noindent
In other words, two streams are weakly equivalent when they cannot be
distinguished by the observation of the current item.
Skipping is not observable. It is immediate from the definition that
two effectively ended streams are weakly
equivalent; conversely, if two streams are weakly equivalent and one is
effectively ended, so is the other.  Clearly strong equivalence
implies weak one. Since \O NoSkip/ is idempotent, 
$s \approx \O NoSkip/ s$  
for any stream $s$.

\begin{proposition}\label{t:weak-cong}
Weak stream equivalence is congruence.
\end{proposition}
In the proof, the only interesting case is $s_1\approx s_2$ implying
$s_1\triangleright \O guard/ g \approx s_2\triangleright \O guard/ g$.
By side-condition \ref{d:side-cond}(i), in order for a guard to have
any effect, the stream has to be effectively ended. On the other hand,
if $s$ is effectively ended then so is $s\triangleright \O guard/ g$
for any $g$, and effectively ended streams are weakly equivalent.

\begin{comment}
For Prop.\ref{t:zip-unroll-unroll}, we could have used a stronger
equivalence: strong modulo the end: $s_1 \approx s_2$ just in case
$s_1 = s_1' ++ s_1'', s_2 = s_2'++s_2'', s_1' \cong s_2'$ and $s_1''$ and
  $s_2''$ are effectively ended.
\end{comment}

\subsubsection*{Zip-elimination for Linear Unrollings}

The notions of linear stream and weak stream equivalence lets us formally
describe zip-unroll-unroll fusion, exemplified in zipping of nats and nats10
at the beginning of the section.
\begin{proposition}\label{t:zip-unroll-unroll}
If $u_1: Z_1\to A_1^\dagger\times Z_1$ and $u_2: Z_2\to A_2^\dagger\times Z_2$ are
linear unrollings, then
$\O zip/ (\O unroll/ u_1\ z_1,\ \O unroll/ u_2\ z_2) \approx
\O unroll/ u\ (z_1,z_2)$
where $u = (u_1\times u_2)\RHD (\textit{pair}^\dagger\times\Id)$, which
is a linear unrolling.
\end{proposition}
\begin{comment}
The proof follows from the observation
$\O zip/ (\overline s_1\ccat \underline s_1,\overline s_2\ccat \underline s_2)=
\O zip/ (\overline s_1,\overline s_2) \ccat \underline s$.
and the fact that for streams with no skips, the proposition trivially
holds.
\end{comment}

Prop. \ref{t:zip-unroll-unroll} indicates that in the case of two
linear unrollings, zipping can be transformed away. In fact, a more
general result holds: if only one stream is a linear unrolling,
zipping can be transformed away.

\begin{proposition}\label{t:zip-linear}
If $u_1: Z_1\to A_1^\dagger\times Z_1$ is a linear unrolling, then
$\O zip/ (\O unroll/ u_1\,z_1,\ s_2) \approx
\O init/ (z_1,s_2) \triangleright
\O map\_filter/ (f\,u_1)$ where
$f\,u_1 = \lam{((z_1,z_2),y)} (u_1\,z_1) \triangleright
((\lam{x}(x,y))^\dagger \times \lam{z_1}(z_1,z_2))$
\end{proposition}
The proof is showing that the relation
$R=(\O NoSkip/ (\O zip/ (\O unroll/ u_1\,z_1,\ s_2)),\
\O NoSkip/ (\O init/ (z_1,s_2) \triangleright \O map\_filter/
(f\,u_1)))$ is a bisimulation. The only interesting case is
\[
u_1\,z_1=(\kappa' a,z_1') \qquad s_2=\kappa'(\mathbf{1},z_2,t_2)
\]
whereupon
\begin{eqnarray*}[rl]
\O zip/ (\O unroll/ u_1\,z_1,\ s_2) &=
\kappa'(\mathbf{1},(z_1',z_2),(\mathsf{K}(\O unroll/
u_1\,z_1)\times t_2)\RHD\O zip/)\\
\O init/ (z_1,s_2) \triangleright \O map\_filter/ (f\,u_1) &=
\kappa'(\mathbf{1},(z_1,z_2),
\lam{(z_1,z_2)}\O init/ (z_1,t_2\,z_2)\triangleright
\O map\_filter/ (f\,u_1))
\end{eqnarray*}
However, \O NoSkip/ skips those states.

\subsubsection*{Linearization}

We have seen that if one of the streams to zip is a linear unrolling,
|zip| can be transformed away, and hence no longer pose any
complication. There still remains the general case: of zipping two
non-linear streams, i.e., streams which are nested or filtered (or
both).  We now show that any skipping non-nested, or nested stream can
always be transformed to a (guarded) linear unrolling. The idea is
simple: apply \O NoSkip/, i.e., keep observing the stream until it
produces an item or (effectively) ends. Recall, \O NoSkip/ is
basically a loop that consumes (or skips over) skips. The result
surely has no skips until the end.

\begin{proposition}\label{t:linearization-flat}
If $\O unroll/ u\,z \triangleright \O guard/ g$ is not a linear
stream, it can be converted to the weakly equivalent linear
$\O unroll/ u'\,z \triangleright \O guard/ g$
where
\[
u' = \O fix/\lam{w}\lam{z}u\,z\triangleright
[\lam{z'}(z' \in \delta_u\: ?\: (\mathbf{1},z'):
(g\,z'\:?\: w\,z': (\mathbf{1},z))),\ \kappa']
\]
is a linear unrolling.
\end{proposition}
\noindent
(To remind, the term in square brackets is the function
$Z+A\times Z \to Z+A\times Z$, which is isomorphic to
$A^\dagger\times Z \to A^\dagger\times Z$). That is,
$\O unroll/ u'\,z \triangleright \O guard/ g$ is obtained from
$\O NoSkip/ (\O unroll/ u\,z \triangleright \O guard/ g)$ by 
inlining the \O NoSkip/ definition. 
In words: if the unrolling function $u$, given the state $z$, produced
an item and the new state, they are returned as is. If it produced a
new state $z'$ but no item, it is applied again, to $z'$~-- until either
an item is produced, or the termination guard triggers, or the state is
found to be the infinite skipping state for $u$.
Here, \O fix/ is the fixpoint combinator and $\delta_u$ is defined as
in Defn.~\ref{d:linear-unrolling}. We have to stress that checking
$z' \in \delta_u$ is generally not effectively computable.  This is
the price to pay for the set-theoretic semantics, in which all
functions (including $u'$) are total. Had we used domain-theoretic
semantics, we could have skipped the check $z' \in \delta_u$ and let
the fixpoint diverge. The weak equivalence follows from the already noted
fact that
$s \approx \O NoSkip/ s$  
for any stream $s$.

\begin{proposition}\label{t:linearization-nested} The pipeline
\begin{eqnarray*}[l]
\O unroll/ u_1\,z_1 \triangleright \O guard/ g_1
\triangleright \O flat\_map/ f \triangleright \O guard/ g_3\\
\quad
\textrm{where } f =
\lam{z}\lam{x} \O unroll/ u_i (x,z)\, z_i (x,z)
\triangleright \O guard/ g_i (x,z)
\triangleright \O abstract/
\end{eqnarray*}
(where $u_i$ is the linear unrolling for any $z$ and $x$)
is weakly equivalent to the linear stream
$\O unroll/ u_0\,z_0 \triangleright \O guard/ g_0 \triangleright
\O abstract/$
where the linear unrolling $u_0$ and $z_0,g_0$ are given as below.
\end{proposition}

\noindent
The structure
of the inner pipeline
is fixed: does not depend on the outer stream~-- but its
parameters do. That is, the inner unrolling function, its initial
state and the inner guard are all generally functions of $x$ and $z$:
the current element of the outer stream and its state.  The
state of the inner pipeline is a pair $(z_p,z)$ where $z_p$ is a
private inner state, which is abstracted away at the end. The 
linear unrolling $u_0$ and $z_0,g_0$ are as follows.

\begin{eqnarray*}[rl]
u_0 &= \O fix/ \lam{w}\\
&
[\lam{z}
 u_1 z \triangleright [
\lam{z} (z \in \delta_{u_1}) \:?\: (\mathbf{1},(\mathbf{1},z)) :
  w\, (\mathbf{1},z),
\\
&
\hskip 4.5em
  \lam{(x,z)}
  (g_1 z \:?\: (\mathbf{1},(\mathbf{1},z)) :
  w\, (\kappa' (u_i (x,z), g_i (x,z), \pi(z_i (x,z))), \pi' (z_i (x,z))))
]
,
\\
&\;\lam{((u',g',z_p),z)}
u' (z_p,z) \triangleright [
\lam{(z_p,z)}
(g' (z_p,z) ? w\, (\kappa'(u',g',z_p),z) : w\, (\mathbf{1},z)),
\\
&\hskip 11em
\lam{(y,(z_p,z))}
(\kappa' y, (\kappa'(u',g',z_p),z))
]
]
\\
z_0 &= (\mathbf{1}, z_1)
\\
g_0 &= \pi' \RHD (g_1 \land g_3)
\end{eqnarray*}
% we use the equational laws to pull in guard g3 and merge it

Although it might be hard to see, the idea is the same as in
Prop.~\ref{t:linearization-flat}: applying \O NoSkip/ and inlining,
using the semantics of \O flat\_map/ from
Fig.~\ref{f:stateful-semantics}.  The state of the resulting linear
stream is a pair, whose first component describes the inner stream and
the second component is the state of the outer stream. The first
component is $\mathbf{1}$ when the the inner stream has not been
started or already finished. The unrolling in this case has to observe
the outer stream, (eventually) obtain its item $x$ and state $z$ and
hence determine the unrolling function $u'=u_i(x,z)$, the initial
state $z_p=z_i(x,z)$ and the guard $g'=g_i(x,z)$ of the inner
stream. This triple $(u',g',z_p)$ characterizes the inner stream and
lets us observe it, skipping over the skip states as we did in
Prop.~\ref{t:linearization-flat}. When the inner guard triggers, we
recur to observing the outer stream. The algorithm hence ping-pongs
between the inner and the outer streams. How this algorithm looks in code,
as a state machine, is discussed in depth in \S\ref{s:closure-conv}.

The two propositions concern rather specific streams.  We shall see in
\S\ref{s:NF} that any other stream pipeline can be converted to one of
the two cases.

\subsection{Normal Forms of Stream Pipelines}
\label{s:NF}

The equational laws in Fig.\ref{f:equations} help us distill normal
forms of stream pipelines: the basis of the stream space, so to
speak. This section defines the normal forms and presents an algorithm
to compute them, by applying the laws in a particular direction and
sequence.

Given the operations of the Core API, Fig.\ref{f:stateful-API}, we
construct the term algebra, i.e., initial algebra,
by considering all terms (pipelines) built using this interface, for
example,
\begin{code}
unroll |> map_filter,    unroll |> map_filter |> map_filter,    unroll |> map_filter |> guard, ...
\end{code}
(The arguments of the API functions are elided. Of
interest is the structure of the pipeline: the sequence of API calls.)
By equational laws, many of the terms are equivalent. The goal
thus is to determine the classes of equivalent terms
and choose their representative~-- the one that is straightforward
to implement in low-level code. Such an equivalence class
representative will be called a normal form, or, normal
pipeline.
%% Recall, in normalization-by-evaluation (NbE), normal forms
%% are defined upfront.

\begin{definition}[Normal form]
A normal form pipeline $\mathit{NF}$ is defined by the grammar
\def\cceq{{:}{:}\!{=}}
\begin{eqnarray*}[r@{\;\cceq\;}l]
\mathit{NF} & \mathit{Flat}\triangleright \mathit{Abstracts}
\mid \mathit{Nested} \triangleright \mathit{Abstracts}
\\
\mathit{Flat} & \O unroll/ \triangleright \O guard/
\\
\mathit{Nested} & \mathit{Flat} \triangleright
\O flat\_map/ (\lam{z}\lam{x} \mathit{NF}) \triangleright \O guard/
\\
\mathit{Abstracts} & \Id \mid \mathit{Abstracts} \RHD \O abstract/
\end{eqnarray*}
Non-terminals are in the upper-case and italics. Terminals are the Core API
operations with appropriate arguments.
\label{d:nf}
\end{definition}

Figure~\ref{f:NF} presents the algorithm to convert a pipeline to
$\mathit{NF}$. For clarity and to save space, the Figure shows only
the case analysis and which equational laws to apply in which case.
When normalizing \O flat\_map/ pipelines, we rely on the fact that the
structure of an inner pipeline is statically known (does not depend
on the items or the state of the outer stream), as stated in
Fig.\ref{f:stateful-API}.

\begin{figure}
\begin{flushleft}
$\mathcal{N}(s)$ converts stream $s$ to $\mathit{NF}$\\
($p$ below stands for the stream
transformer \O init/, \O guard/, \O map\_filter/, or \O flat\_map/)
\end{flushleft}
\begin{eqnarray*}[rl]
\mathcal{N}(\O unroll/) &= \O unroll/ \triangleright \O guard/ (\O K/ \O true/)
\\
\mathcal{N}(s\triangleright \O abstract/) &=
\mathcal{N}(s) \triangleright \O abstract/
\\
\mathcal{N}(s\triangleright p) &= \mathcal{A}(\mathcal{N}(s), p)
\qquad \textrm{except for } p=\O flat\_map/
\\
\mathcal{N}(s\triangleright \O flat\_map/(\lam{z}\lam{x}s')) &=
\mathcal{A}(\mathcal{N}(s), \O flat\_map/(\lam{z}\lam{x}\mathcal{N}(s')))
\\
\mathcal{N}(\O zip/(s_1,s_2)) &= \mathcal{Z}(\mathcal{N}(s_1),\mathcal{N}(s_2))
\end{eqnarray*}

\bigskip
\begin{flushleft}
$\mathcal{A}(\mathit{NF},p)$ normalizes $\mathit{NF}\triangleright p$
by pulling all \O abstract/ out, using using
\eqref{l:abstract-init}-\eqref{l:abstract-flatmap}
\end{flushleft}
\begin{eqnarray*}[rl]
\mathcal{A}(\mathit{Flat}\triangleright \mathit{Abstract},p) &=
\mathcal{T}_f(\mathit{Flat},p) \triangleright \mathit{Abstract}
\\
\mathcal{A}(\mathit{Nested}\triangleright \mathit{Abstract},p) &=
\mathcal{T}_n(\mathit{Nested},p) \triangleright \mathit{Abstract}
\end{eqnarray*}

\bigskip
\begin{flushleft}
$\mathcal{T}_f(\mathit{Flat},p)$ converts $\mathit{Flat}\triangleright
p$ to either $\mathit{Flat}$ or $\mathit{Nested}$ pipeline
\end{flushleft}
\begin{eqnarray*}[rl]
\mathcal{T}_f(\mathit{Flat},\O init/) &=
\mathit{Flat}
\textrm{ using \eqref{l:guard-init}, \eqref{l:unroll-init}}
\\
\mathcal{T}_f(\mathit{Flat},\O guard/) &=
\mathit{Flat}
\textrm{ using \eqref{l:guard-guard}}
\\
\mathcal{T}_f(\mathit{Flat},\O map\_filter/) &=
\mathit{Flat}
\textrm{ using {\eqref{l:guard-map}, \eqref{l:unroll-map}}}
\\
\mathcal{T}_f(\mathit{Flat},\O flat\_map/(\lam{z}\lam{x}\mathit{NF})) &=
\mathit{Nested}
\textrm{ by definition}
\end{eqnarray*}

\bigskip
\begin{flushleft}
$\mathcal{T}_n(\mathit{Nested},p)$ converts $\mathit{Nested}\triangleright
p$ to $\mathit{Nested}$ pipeline
\end{flushleft}
\begin{eqnarray*}[rl]
\hbox to 0.35\textwidth{$
\mathcal{T}_n(\mathit{Flat}\,\triangleright\,
\O flat\_map/(\lam{z}\lam{x}\mathit{NF})\,\triangleright\,
\O guard/\, \triangleright\, \O init/) =$}\\
&\hbox to 0.45\textwidth{$
\mathit{Flat}\, \triangleright\,
\O flat\_map/(\lam{z}\lam{x}\mathcal{A}(\mathit{NF}, \O init/))
\,\triangleright\, \O guard/$}
\\
&
\hfil
\textrm{ using \eqref{l:flatmap-init}, \eqref{l:guard-init},
\eqref{l:unroll-init}}
\\
\mathcal{T}_n(\mathit{Nested},\O guard/) &=
\mathit{Nested}
\textrm{ using {\eqref{l:guard-guard}}}
\\
\hbox to 0.35\textwidth{$
\mathcal{T}_n(\mathit{Flat} \,\triangleright\,
\O flat\_map/(\lam{z}\lam{x}\mathit{NF}) \,\triangleright\,
\O guard/ \,\triangleright\, \O map\_filter/) =$}\\
&\hbox to 0.45\textwidth{$
\mathit{Flat} \,\triangleright\,
\O flat\_map/(\lam{z}\lam{x}\mathcal{A}(\mathit{NF}, \O map\_filter/))
\,\triangleright\, \O guard/$}
\\
&
\hfil
\textrm{ using \eqref{l:guard-map}, \eqref{l:flatmap-map}}
\\
\hbox to 0.35\textwidth{$
\mathcal{T}_n(\mathit{Flat} \,\triangleright\,
\O flat\_map/(\lam{z}\lam{x}\mathit{NF}) \,\triangleright\,
\O guard/\,\triangleright\,
\O flat\_map/(\lam{z}\lam{x}\mathit{NF})) =$}\\
&\hbox to 0.55\textwidth{$
\mathit{Flat} \,\triangleright\,
\O flat\_map/(\lam{z}\lam{x}\mathcal{A}(\mathit{NF},
   \O flat\_map/(\lam{z}\lam{x}\mathit{NF})))
\,\triangleright\, \O guard/$}
\\
&
\hfil
\textrm{ using \eqref{l:guard-flatmap}, \eqref{l:flatmap-flatmap}}
\end{eqnarray*}

\bigskip
\begin{flushleft}
$\mathcal{Z}(\mathit{NF},\mathit{NF})$ converts \O zip/ of two
$\mathit{NF}$ to $\mathit{NF}$
\end{flushleft}
\begin{eqnarray*}[rl]
\mathcal{Z}(\mathit{Flat}_1\triangleright \mathit{Abstracts}_1,
\mathit{Flat}_2\triangleright \mathit{Abstracts}_2) &=
\mathit{Flat}\triangleright \mathit{Abstracts}_1 \triangleright
\mathit{Abstracts}_2
\textrm{ using \eqref{l:zip-abstract}, \eqref{l:zip-guard},
\eqref{l:zip-bothlinear}}
\\
&\hfil
\textrm{ provided $\mathit{Flat}_1$ and $\mathit{Flat}_2$ are both linear}
\\
\mathcal{Z}(\mathit{NF}_1,\mathit{NF}_2) &=
\mathcal{N}(\mathit{NF}_2 \triangleright \O init/ \triangleright \O map\_filter/
\triangleright \mathit{Abstracts})
\\
&
\hfil
\textrm{ using \eqref{l:zip-abstract},
    \eqref{l:zip-linear}, \eqref{l:zip-guard}
    and Prop.~\ref{t:linearization} for $\mathit{NF}_1$
}
\end{eqnarray*}

\caption{Normalization algorithm.
The meta-variable $s$ stands for any stream, $p$ stands for the stream
transformer \O init/, \O guard/, \O map\_filter/, or \O flat\_map/.
We only show the structure and refer to equational laws for detail.
}
\label{f:NF}
\end{figure}

The general case of normalizing zip pipelines relies on the fact that
any normal form can be converted to a weakly equivalent so-called linear
normal form.

\begin{definition}[Linear Normal form]
A pipeline of the form
$\O unroll/ f\ z \triangleright \O guard/ g \triangleright \mathit{Abstracts}$
where $f$ is a linear unrolling is called linear normal form
$\mathit{LNF}$.
\label{d:nf-linear}
\end{definition}

\begin{comment}
LNF itself can be interpreted as a $T_{AZ}$ co-algebra:
if $g\, z$ is false, we observe the stream as finished. Otherwise,
applying $f\, z$ gives the new state and possibly an stream item.
\end{comment}

\begin{proposition}[Linearization]\label{t:linearization}
Every $\mathit{NF}$ can be converted to a weakly equivalent $\mathit{LNF}$.
\end{proposition}
A $\mathit{NF}$ is either  $\mathit{Flat}$ or  $\mathit{Nested}$:
and these are exactly the cases covered by
Prop.~\ref{t:linearization-flat} and Prop.~\ref{t:linearization-nested}.

\begin{theorem}[Normalization]
\leavevmode
\begin{itemize}
\item[(i)] Algorithm $\mathcal{N}(s)$, Fig.~\ref{f:NF},
  applies to every pipeline $s$ built using
  the operations of the Core API, Fig.\ref{f:stateful-API}, satisfying
  side-conditions \ref{d:side-cond}. The algorithm terminates after
  a finite number of
  steps.
\item[(ii)] Algorithm is idempotent:
$\mathcal{N}(\mathcal{N}(s)) = \mathcal{N}(s)$
\item[(iii)] $\mathcal{N}(s) \approx s$
\end{itemize}
\end{theorem}
For part (i) we notice that the stream producer |unroll| can be
converted to $\mathit{NF}$ by adding the trivial guard, and
an application of any stream transformer to a
$\mathit{NF}$ can be
converted to a normal form, in a finite number of steps.
Thus the normal form exists and is computable. The normalization is
meaning-preserving, part (iii), which is
justified by the fact that the stream equivalence (weak and strong) is
congruence. One may notice from the algorithm that if
the pipeline $s$ does not contain |zip| then
$\mathcal{N}(s) \cong s$.

One may notice that the normalization algorithm Fig.~\ref{f:NF}
looks similar to call-by-value evaluation. As we see in
\S\ref{s:implementation}, the
normalization is indeed performed as evaluation.
One may also wonder why not to use the linear normal form $\mathit{LNF}$
always. As we have seen in Prop.~\ref{t:linearization-flat} and
\ref{t:linearization-nested},
linearization requires a fixpoint (or loop in
the generated code), which adds overhead however small. Therefore,
linearization should be performed only when necessary.

As an interesting example of normalization, consider the pipeline
\begin{code}
iota 1 |> flat_map (fun x -> from_to x (x+5)) |> take 10
\end{code}
expressed in the sugared API (Fig.\ref{f:API}).
In the Core API, it has the form
\begin{eqnarray*}[l]
\O zip/ (\O unroll/ u_{nm} (1,10,\mathsf{true})
   \triangleright \O guard/ grd,\\
\qquad
  \O unroll/ u_n 1 \triangleright
  \O flat\_map/ \lam{(z,x)}
    \O unroll/ u_{nm}\, (x,x+5,\mathsf{true})
   \triangleright \O guard/ grd \triangleright
   \O init/ z \triangleright iso \triangleright \O abstract/)\\
\quad
 \triangleright
 \O map\_filter/ (\lam{(z,(\_,x))} (\kappa' x,z)) \triangleright \O abstract/
\end{eqnarray*}
where
\begin{eqnarray*}[rl]
grd &= \pi'\RHD\pi'
\\
u_n &= \lam{z}(\kappa'z,z+1)
\\
u_{nm} &= \lam{(x,y,w)} (x\le y \:?\: (\kappa'x,(x+1,y,w)) :
 (\mathbf{1},(x,y,\mathsf{false})))
\end{eqnarray*}
and $iso$ is the $(x,y) \leftrightarrow (y,x)$ state isomorphism.
(In \strymonas, |take n| is expressed as zipping with
|from_to 1 n|, as was described in \S\ref{s:taste}.)
Applying the normalization algorithm gives the following normal
form:
\begin{eqnarray*}[l]
\O unroll/ (\lam{(z_0,z)} u_n\,z \triangleright \lam{(x',z)}(x',(z_0,z))\:
((1,10,true),1) \triangleright
\O guard/ (\mathsf{K}\mathsf{true})\\
\triangleright \O flat\_map/ (\lam{((z_0,z),x)}\\
\quad
        \O unroll/
(\lam{((z_0,z),z_p)} u_{nm}\,z_p \triangleright
[\langle \mathbf{1},\Id\rangle,
 \lam{(y,((z_0,z),z_p))}
 u_{nm} z_0 \triangleright
 ((\mathsf{K}y)^\dagger\times
           \lam{z_0}((z_0,z),z_p))]\\
\qquad\qquad
          ((z_0,z),(x,x+5,\mathsf{true})))\\
\quad
        \triangleright \O guard/ (\pi'\RHD grd)
    \triangleright  iso \triangleright \O abstract/)\\
\triangleright \O guard/ (\pi\RHD grd)
\triangleright \O abstract/
\end{eqnarray*}
The file \nolinkurl{derivation.ml} in the \strymonas\ repository
source code\footnote{\relax
\url{https://github.com/strymonas/strymonas-ocaml/tree/main/lib/derivation.ml}}
shows the normalization
in detailed steps.

\subsection{Formalization}
\label{s:Agda}

The equational laws~-- the main theoretical contribution~-- have been
mechanically proven in Agda, as described in this section.

First, however, we have to mention a `lightweight' formalization: the
OCaml file \nolinkurl{derivation.ml} in the \strymonas\ repository. It
demonstrates that all of Fig.\ref{f:API} API is indeed implementable
in terms of the Core API, Fig.~\ref{f:stateful-API}, and the latter in
terms of $S_{AZ}$, Fig.~\ref{f:stateful-semantics}~-- spot-checking
them on sample inputs. The file also does sanity checking of
equational laws (type preservation). In our formal development in
\S\ref{s:derivation} we take isomorphic states to be equal, as
common in mathematics. In programming and mechanization, however,
isomorphisms must be marked explicitly. Therefore, the Core API has an
extra operation
\begin{code}
val adjust : (('z1 -> 'z2) * ('z2 -> 'z1)) -> ('a,'z1) stream -> ('a,'z2) stream
\end{code}
where the first argument is a bijection: a pair of a mapping and its
inverse.

The file \nolinkurl{Derivation.agda} in the repository\footnote{\relax
\url{https://github.com/strymonas/strymonas-ocaml/tree/main/lib/Derivation.agda}}
recasts
\nolinkurl{derivation.ml} in Agda, proving that equational laws are not just
type preserving but are in fact equivalences. It begins by formalizing
the basic interface of stream operations Fig.~\ref{f:stateful-API}
(denoted as |stream2| in \nolinkurl{derivation.ml} and |Stream$_2$| in
\nolinkurl{Derivation.agda}), co-inductively defining $S_{AZ}$ (called
|StStream|) and demonstrating it implements the core API. In effect,
the Agda code transcribes Fig.\ref{f:stateful-semantics}~--
verifying, at the same time, that the definitions are productive and
well-founded. We rely on Agda's support for inductive families,
coinduction, and guarded recursion (the |--guardedness| flag).

% In defining weak bisimulation, the Agda code inline NoSkip
% Also, even in strong bisimulation, the Agda code ignores state
% One may say that Agda bisimulation is stream equality after
% abstracting state

The Agda development then defines Strong and Weak stream equivalences
and proves they are indeed equivalence relations and form a
Setoid. The largest part of Agda development is demonstrating that
equational laws in Fig.\ref{f:equations} (including linearization:
Prop.\ref{t:linearization}) are stream
equivalences.

\section{Implementation}
\label{s:implementation}

This section ties the just developed theory~-- fusion by
normalization~-- to the practice, the actual implementation of the
\strymonas\ API in Fig.\ref{f:raw-API} delivering the complete fusion.
On one hand, the implementation \emph{just} realizes the algorithm in
Fig.\ref{f:NF}. However, there are a few subtle points, related to the
actual representation of the theoretical constructs and, in particular,
to staging. The central idea, to re-iterate, is that normal forms
\emph{straightforwardly}, directly correspond to an output state machine.

The complete code of \strymonas, which spells out all details, is
freely available from the \strymonas\ repository. Therefore, in this
section we concentrate on the main ideas and subtle points, which are
best illustrated by examples. The complete code corresponding to these
examples in in \nolinkurl{intuition.ml} in the \strymonas\
repository.\footnote{\relax
\url{https://github.com/strymonas/strymonas-ocaml/tree/main/lib/intuition.ml}}

%% Along the way we will see how
%% Fig.\ref{f:stateful-API} turns into the \strymonas\ API in
%% Fig.\ref{f:raw-API}.

\subsection{From Pipelines to Normal Forms to State Machines}
\label{s:impl-informally}

This section demonstrates how a pipeline turns into an imperative
state machine. The goal here is to give the intuition, highlight
the normalization and explain the connection between the theoretical development
in \S\ref{s:derivation} and the \strymonas\ implementation. The actual
implementation~-- the computation of normal forms and code
generation~-- is the topic of \S\ref{s:computing-NF}.

We will be using the introductory example, |ex1|-|ex2| from
\S\ref{s:taste}, repeated, after inlining auxiliary definition, below:
\begin{code}
let ex2 : int stm =
  iota C.(int 1) 
  |> map C.(fun e -> e * e) |> filter C.(fun e -> e mod (int 17) > int 7) |> take C.(int 10)
  |> fold C.(+) C.(int 0)
\end{code}
Recall, the module |C| is an implementation of the Fig.\ref{f:cde}
interface~-- an embedding of a target language~-- assumed in scope
throughout the paper. We will talk about it in more detail in
\S\ref{s:computing-NF}.
The operations |iota|, |take|, etc. of \strymonas\ are implemented in
terms of the Raw API, Fig.\ref{f:raw-API}. Inlining the implementation
gives:
\begin{code}
let u1 = C.(fun e -> e * e)
let u2 = C.(fun e -> e mod (int 17) > int 7)
let ex2raw = zip_raw
     (initializing_ref C.(int 10) @@ fun i ->
        let ud = C.(fun k -> decr i @. k ()) in let g1 = C.(dref i > int 0) in
        infinite ud |> guard g1)
     (initializing_ref C.(int 1) @@ fun z ->
        let un = C.(fun k -> letl (dref z) @@ fun v ->  incr z @. k v) in
        infinite un |> map_raw (fun e -> C.letl (u1 e)) |> filter_raw u2)
  |> map_raw' (fun (_,x) -> x) |> fun st -> let open C in
  let- zs = newref (int 0) in iter (fun a -> zs := dref zs + a) st @. ret (dref zs)
\end{code}
As noted several times earlier, |take| is implemented in terms of |zip|.
Written using the stateful API
(\S\ref{s:derivation}, Fig.\ref{f:stateful-API}), the example becomes
\begin{eqnarray*}[l]
\mathit{ex}_2 = \O zip/\\
\quad
 \left(\O unroll/ u_d\, 10 \triangleright \O guard/ g_1\right)\\
\quad
 \left(\O unroll/ u_n\, 1 \triangleright
  \O map\_filter/ \lam{zx}(\kappa' (u_1\, x),z) \triangleright
 \O map\_filter/ \lam{zx} ((u_2\, ? \kappa' x : \mathbf{1}), z)\right)\\
 \triangleright
 \O map\_filter/ (\lam{(z,(\_,x))} (\kappa' x,z)) \triangleright \O
 abstract/
\\
\textrm{where }
u_n = \lam{z}(\kappa'z,z+1)
\qquad
u_d = \lam{z} ((z>0 \:?\: \kappa'z : \mathbf{1}),z-1)
\quad
g_1 = \lam{z}z\ge 0
\end{eqnarray*}

The comparison $\mathit{ex}_2$ and |ex2raw| well demonstrates the
relation between theory and practice. The two expressions
clearly have the same structure and, generally, are quite alike. There are
also several differences. First jumps out is the handling of
the stream state. In our theoretical development in
\S\ref{s:derivation} the stream state, generally a tuple, is
explicitly passed around all throughout the stream operations. It is
hence, a \emph{heap}, treated explicitly, which suits theoretical
development. In practical programming, the program heap, which is
likewise conceptually
passed from one operation to the next, is implicit; its contents
is modified by mutation. This is the basis of
stateful streams of \strymonas. Therefore, |'a stream| of \strymonas\ does
not carry the state explicitly, and the stream type is not indexed by
the state. The operations |abstract| and |adjust| of
the stateful API, Fig.\ref{f:stateful-API}, become trivial and omitted
from the \strymonas\ Raw API.

\begin{comment}
One may say that the correspondence of explicit state passing and
mutable state holds only if the state is used linearly. Actually, it
is OK to reference state many times. In the implicit-heap case,
referencing the state corresponds to the dereferencing the mutable
state. Therefore, when the state is updated, the previously
dereferenced value remains as is.  Anyway, in our theoretical
implementation, $S_{AZ}^Z$ is pure and not closure. This is always
true except for zip. But NF eliminates zip.
\end{comment}

The second difference between $\mathit{ex}_2$ and |ex2raw| is that
the latter is completed: it is terminated by the observation function
|fold| (or, |iter|, in the raw form). As the name |iter| implies, it
observes the stream by passing each produced item to a consumer
function, until the stream ends, if ever. In the case of |ex2raw|, the
consumer accumulates the received items (integers) in a mutable cell
|zs|, whose contents is returned when the stream ends.

The normal form of $\mathit{ex}_2$, obtained by applying the
normalization algorithm in Fig.\ref{f:NF}, is as follows.
\begin{eqnarray*}[l]
\mathit{ex}_2^{norm} =\\
\;
\O unroll/ \lam{(z_1,z)} (u_2 (u_1\, z) \:?\:
   ((z_1 > 0 \,?\, \kappa' (u_1\, z) : \mathbf{1}), (z_1-1,z+1)) :
   (\mathbf{1},(z_1,z+1)))\: (10,1)
\\
\;\triangleright \O guard/ (\pi \RHD g_1)
    \triangleright \O abstract/
\end{eqnarray*}
The central idea of \strymonas\ is that normal forms
\emph{straightforwardly}, directly correspond to an output state machine.
Indeed, recall the $\mathit{Flat}$ normal form:
$\O unroll/ f\,z  \triangleright \O guard/ g
\triangleright \mathit{Abstracts}$. Here $z$ is the initial state,
$f$ is the state transition cum output function; the final states are
those at which $g$ is $\mathit{false}$.
The observation |iter| then `drives' this machine,
repeatedly obtaining its output items and passing it the the consumer
function. With implicit heap, the machine is imperative,
and driving it corresponds to the \textsf{while} loop iterated while
the guard $g$ is true, with the transition function $f$ being the loop
body. Let's see how it works on the $\mathit{ex}_2^{norm}$ example,
recast into the implicit heap model,
with the consumer function being the stateful accumulation (like in
|ex2raw|).
The state is the pair of integers with the initial
values of |10| and |1| correspondingly, for which we allocate
reference cells. While the guard remains true, we evaluate the
transition function that may also produce an output item, which is
then passed to the consumer. In the implicit heap realization,
$\O abstract/$ is dummy. All in all, we obtain (handwavingly, for now)
the following
imperative code (in OCaml notation):
%% let u1 = fun e -> e * e;;
%% let u2 = fun e -> e mod 17 > 7;;
\begin{code}
let zs = ref 0 in (* accumulator *)
let consumer = fun a -> zs := !zs + a in
let z1 = ref 10 and z = ref 1 in (* stream state *)
while (! z1 > 0) (* guard *) do
   let zcurr = ! z in incr z;
   if u2 (u1 zcurr) then
   (let z1curr = ! z1 in decr z1; if z1curr > 0 then consumer (u1 zcurr))
done;
! zs
\end{code}
%% - : int = 853
or, after the obvious optimization:
\begin{code}
let zs = ref 0 in (* accumulator *)
let consumer = fun a -> zs := !zs + a in
let z1 = ref 10 and z = ref 1 in (* stream state *)
while (! z1 > 0) (* guard *) do
   let zcurr = ! z in incr z;
   if u2 (u1 zcurr) then (decr z1; consumer (u1 zcurr))
done;
! zs
\end{code}
%% - : int = 853

This code is very much like the one obtained by \strymonas, shown in
\S\ref{s:taste}. Compared to the original |ex2raw| pipeline, the
fusion is clear, even across |zip|. The fusion is not complete,
however: we see function invocations (|consumer|, |u1|, |u2|) and
the repeated code evaluation: |u1 zcurr|. The
first problem is solved by staging, and the second by
let-insertion~-- as we now demonstrate.

\subsection{Normalization as Computation}
\label{s:computing-NF}

To show how the complete fusion, of |ex2raw|, is concretely
accomplished, we turn it to code~-- this time, not handwavingly but
rigorously, mechanically: as an evaluation, that is, as normalization
by evaluation (NbE). We come to see how the \emph{complete} fusion is
guaranteed, by the very design.

One problem with the state machine obtained in the previous section is
the function calls, to user-defined mapping and filtering actions such
as |u1| and |u2|. The calls are eliminated if we \emph{inline} these
functions. Formally, inlining is an instance of partial evaluation, or
staging
\cite{sorensen-unifying,jones_introduction_1996,taha_gentle_2004}.
Staging splits evaluation into several (often two) stages: the first,
or, present, stage performs the reductions that are possible to do
without knowing the input data, such as inlining calls to
statically-known functions, constant-propagation and simplifications
(applications of $\eta$-laws, etc.). The result of the first stage is
the (more performant) code, executed on input data in the second, or
future, stage.  The first stage is where the stream fusion,
normalization, is performed in \strymonas.

The first stage hence act as a code generator. To compute on, or
generate, code~-- build simple expressions and combine them~-- the
first stage has to represent the code as values.  In \strymonas, the
target code is represented in so-called tagless-final style, as
explained in detail in \cite[\S5]{generating-C}:
as values of an abstract type, such as |exp| for the code
of expressions or |stm| for the code of statements, parameterized by
the type of the expression or statement. The abstract types are
associated with the operations to build and combine their values (i.e.,
algebra), collected in a signature,
Fig.~\ref{f:cde}.  A stream operator such as mapping\footnote{The actual
  operation in the Raw API, Fig.\ref{f:API}, has a more general
  signature to accommodate tupling.}
\begin{code}
val map_raw' : ('a exp -> 'b exp) -> 'a exp stream -> 'b exp stream
\end{code}
takes as the mapping function a code transformer, such as
|u1| of our running example:
\begin{code}
let u1 : int exp -> int exp = C.(fun e -> e * e)
\end{code}
Recall that throughout the paper we have assumed one
implementation of the Fig.~\ref{f:cde} signature, as the module named |C| (for
`code'). Given the code of an integer expression, |u1| produces the
code of the product of that expression with itself. 

We should stress
it again: the mapping is specified \emph{not} as |('a -> 'b) exp|~--
that is, not as a function in the target language. After all, the
target language is strictly first order and, deliberately, 
has no facilities to define functions. The mapping is a metalanguage function
|'a exp -> 'b exp|: a target code transformer. Therefore, it is
inlinable; in fact, it cannot be left unlinable.

The running example pipeline |ex2raw : int stm| is hence a
generator~-- the computation over expressions and statements of the
target language, eventually producing an |int|-returning statement.
We now explain how exactly this computation is performed.  We thus
present an implementation of \strymonas: the realization of the type
|'a stream| and the implementation of the Raw API operations such as
|infinite|, |map_raw|, |zip_raw|. For the sake of explanation and
clarity, the shown here implementation is simplified. We also omit the
parts not relevant to the running example, in particular, nested
streams.

As befits the normalization-by-evaluation approach, the realization of
|'a stream| is the stream normal form. Stream constructors such as |infinite|
produce an instance of the normal form, and stream transformers such
as |map_raw| and |zip_raw| take a normal form as arguments and return
a normal form as the result~-- effectively implementing the
normalization algorithm in Fig.\ref{f:NF}.

Since we elide nesting for the sake of explanation, we will be dealing
with a $\mathit{Flat}$ normal form, which is, recall:
\[
\O unroll/ f\, z \triangleright \O guard/ g \triangleright \mathit{Abstract}
\]
where $z:Z$ is the initial state, $f: Z\to A^\dagger\times Z$ is an
unrolling and $g:Z\to \mathbb{B}$ is the guard. In OCaml code, it is
realized as the following data structure.
\begin{code}
type 'a emit = ('a -> unit stm) -> unit stm
type 'a stream =
          | Flat : 'a flat -> 'a stream
          | Init : ('z exp * ('z mut -> 'a stream)) -> 'a stream        (* $\zeta$ is existentially quantified *)
and 'a flat = {unr: 'a emit;  grd:bool exp; linear:bool}
\end{code}
The type |'a| is actually
an expression code type; therefore, |'a stream| represents a code
generator. Since we are the using the 
implicit, mutable heap, the stream state is
implicit and $\mathit{Abstract}$ is trivial and elided.  The stream state
$Z$ hence is the collection of mutable variables,
introduced/initialized by |Init| and used (dereferenced and mutated)
in the unrolling and the guard. The unrolling function, given the state,
updates it and possibly produces an item.  We represent it in the
continuation-passing style, as |'a emit| (to facilitate let- and
if-insertion, as we see soon). That is, the unrolling |unr| takes the
continuation |'a -> unit stm| (where |'a| is generally an |exp| type:
the code of an expression) and produces the code of a
statement, which may dereference and mutate the state. Producing an
item corresponds to invoking the continuation, passing to it the code
of the producing expression. Skipping corresponds to not invoking
the continuation. An unrolling is assumed to invoke the
item continuation at most once. A guard also operates on the
implicit state: it is the code of an expression that accesses the heap
and returns a boolean.  We also keep track if the unrolling is linear
(that is, if the normal form is a LNF).

\begin{comment}
The side-condition (i) says that the
guards apply to a skipping stream. So, we may as well pull the guards
forward. Note, when stream skips, it essentially `skips through the
end' (because nothing acts on a skipping stream except guard, and
nothing can recover from skipping.)
\end{comment}

The Raw API operations |initializing_ref| and |infinite| are the
constructors of the |'a stream| data type:
\begin{code}
let initializing_ref  : 'z exp -> ('z mut -> 'a stream) -> 'a stream = fun i k -> Init (i,k)
let infinite : 'a emit -> 'a stream = fun unr -> Flat {unr; grd = C.bool true; linear=true}
\end{code}
The argument of |infinite| is assumed to be a linear unrolling. The
implementation of |infinite| is literally the $\mathcal{N}(\O unroll/)$
stanza of the normalization algorithm, Fig.\ref{f:NF}.
As an example, the infinite stream that counts from 1~--
|iota (C.int 1)|,~-- or
\begin{code}
let un z = C.(fun k -> letl (dref z) @@ fun v ->  incr z @. k v) in
initializing_ref C.(int 1) @@ fun z -> infinite (un z)
\end{code}
hence evaluates to the following (Linear) NF:
\begin{code}
Init (C.int 1, fun z -> Flat {unr=un z; grd=C.boot true; linear=true})
\end{code}

The stream transformer |guard| receives the guard expression and the
NF and returns the updated NF:
\begin{code}
let guard : bool C.exp -> 'a stream -> 'a stream = fun g ->
  map_flat (function {unr;grd;linear} -> {unr;linear;grd= C.(g && grd)})
\end{code}
where
\begin{code}
let rec map_flat : type a b. (a flat -> b flat) -> (a stream -> b stream) = fun f -> function
    | Init (i,k) -> Init (i, k >> map_flat f)
    | Flat x -> Flat (f x)
\end{code}
The code directly corresponds to $\mathcal{T}_f(\mathit{Flat},\O
guard/)$ of the normalization algorithm, relying on the equational law
\eqref{l:guard-guard}, Fig.\ref{f:equations}.
Likewise, |map_raw| (which corresponds to $\O map\_filter/$ of
the theoretical Core API, Fig.\ref{f:stateful-API}, but expressed in the
continuation-passing style) implements
$\mathcal{T}_f(\mathit{Flat},\O map\_filter/)$,
relying on laws \eqref{l:guard-map} and \eqref{l:unroll-map}.
\begin{code}
let map_raw : ?linear:bool -> ('a -> 'b emit) -> 'a stream -> 'b stream = fun ?(linear=true) f ->
    map_flat (function {unr;grd;linear=l} ->
      {grd; linear=l && linear; unr=fun k -> unr (fun x -> f x k)})
\end{code}
If the mapping transformer is known to be linear~--
that is, if the continuation |'b -> unit stm|
of |'b emit| is invoked exactly once (except, perhaps, at the very
end)~-- the linearity of the NF is preserved. If the mapping
transformer is not known to be linear, the user has to specify
the optional argument |?linear| as false.  The use of the
continuation-passing style in |map_raw| allows us to do let-insertion: see
|map_raw (fun e -> C.letl (u1 e))| in the code of our running example
|ex2raw|. When the mapping function is applied, to a code value |e|,
it is passed to the user-action |u1|, which generates the value
representing the product |e * e|. Effectively, the squaring user
action becomes inlined. Furthermore, this code value is passed to
|letl|, which generates a let-expression binding the squaring
expression to a fresh variable, whose code is eventually passed to the
mapping's continuation. If that code is later spliced several times, only
the variable reference (rather than the squaring computation) is being
replicated~-- as we will see in the final generated code below.

The operations |map_raw'| and |filter_raw| are the particular cases of
|map_raw|. The former is a linear mapping by construction
\begin{code}
let map_raw' : ('a -> 'b) -> 'a stream -> 'b stream = fun f -> map_raw (fun e k -> k (f e))
\end{code}
whereas the latter is, in general, non-linear (unless the filtering
predicate is constant):
\begin{code}
let filter_raw : ('a -> bool exp) -> 'a stream -> 'a stream = fun f ->
  map_raw ~linear:false C.(fun x k -> if1 (f x) (k x))
\end{code}

The operation |zip_raw| implements
$\mathcal{Z}(\mathit{NF},\mathit{NF})$ of the normalization algorithm:
If two streams to zip are linear, the result is also linear. If one is
linear, Prop. $\ref{t:zip-linear}$ is applied. Otherwise, one of the
streams in linearized and then Prop. $\ref{t:zip-linear}$ is
applied. In \strymonas, to chose which of the two streams to linearize
we estimate their complexity (that is, how deeply nested they are) and
linearize the less complex stream.
\begin{code}
let rec zip_raw : type a b. a stream -> b stream -> (a * b) stream = fun st1 st2 ->
  match (st1,st2) with
  | (Init (i,k),st2)  -> Init (i, fun x -> zip_raw (k x) st2)
  | (st1,Init (i,k))  -> Init (i, fun x -> zip_raw st1 (k x))
  | (Flat {linear=true;unr;grd},st2) ->
         map_raw (fun b k -> unr (fun a -> k (a,b))) st2 |> guard grd
  | (st1,(Flat {linear=true} as st2)) -> zip_raw st2 st1 |> map_raw' (fun (x,y) -> (y,x))
  | (st1,st2) -> (* linearize one of st1 or st2  and recur *)
\end{code}

The stream observation function |iter| completes the stream.
It converts the stream NF to the driven state machine code: allocating the
machine state and generating the \textsf{while}-loop whose body
is produced by the unrolling function upon receiving the item |consumer|.
\begin{code}
let rec iter : type a. (a -> unit stm) -> a stream -> unit stm = fun consumer -> function
    | Init (i,k) -> C.newref i (k >> iter consumer)
    | Flat {unr;grd} -> C.while_ grd (unr consumer)
\end{code}

With the just presented implementation of the Raw API the running example
|ex2raw| evaluates to the following code
(assuming the module |C| to be the OCaml backend):
\begin{code}
let v_1 = Stdlib.ref 0 in
(let v_2 = Stdlib.ref 10 in
 let v_3 = Stdlib.ref 1 in
 while (! v_2) > 0 do
   let t_4 = ! v_3 in
   Stdlib.incr v_3;
   (let t_5 = t_4 * t_4 in
    if (t_5 mod 17) > 7 then (Stdlib.decr v_2; v_1 := ((! v_1) + t_5)))
   done);
! v_1
\end{code}
This is \emph{exactly} the code generated for |ex2| seen in
\S\ref{s:taste}. Compared to the code at the end of
\S\ref{s:impl-informally}, the user actions |u1| and |u2| are inlined
and the duplication of |u1| is eliminated via the let-binding,
|t5|. The fusion is now complete.

Inlining (and closure conversion, in the next section) sound like
ordinary optimizations to improve code quality. They are more than
that: without them no code is produced. Looking back at the target
language, Fig.~\ref{f:cde}, one notices that it has no facilities to
define functions (named or anonymous), to represent closures, or to
define and use data structures. Were we unable or unwilling to eliminate
function definitions via inlining, we would not have been able to
produce any code. The complete fusion~-- the absence of data structures
and closures~-- is ensured statically by the very design of the target
language. Such a restricted design poses the question if the target 
language can express
all the desired pipelines. As we have demonstrated, every expressible
(|flat_map|-free, in this section) pipeline is convertible to 
|'a stream| representation, turned into a while-loop by
|iter|. Thus every expressible pipeline is compilable into 
in the target language, and hence completely fused. One may say that
the optimizations are complete.

\begin{comment}
\subsubsection{Pull array}

An attentive reader may notice an extra operation
produces a finite stream.
It is a particular form of unroll, and convertible to it.
In contrast, |pull_array z n| always produces
a finite stream, of |n| items.

\begin{code}
val pull_array : 'z -> int -> ('z -> int -> ('a*'z)) -> ('a,'z) stream
\end{code}
Technically it is unnecessary (since it can clearly can be expressed
via |unroll| and |guard|~-- which we show in mechanization: Agda code). It is
convenient for later implementation, since it corresponds to for-loop.
Some laws  are valid for |pull_array|; some don't and so
|pull_array| has to be converted to unroll to proceed with
normalization. We keep |pull_array| distinction as far as possible
during normalization.

Although |pull_array| is expressible in terms of unroll, it is worth
distinguishing it since it naturally maps to for-loop, and also easier
to zip.
\end{comment}

Nested streams generalize straightforwardly: the corresponding state
machine has nested loops. There is one subtlety: linearizing
nested loops, described next.

\subsection{Nested Streams and Closure Conversion}
\label{s:closure-conv}

As we have just seen, |zip_raw| when given two non-linear streams has
to linearize one of them. Linearization, Prop.\ref{t:linearization-nested},
is most complex for nested streams. The proposition presents and
proves the linearization algorithm (recalled below);
\strymonas\ `merely' needs to follow it. There is a subtlety, however,
described in this section. Staging proves indispensable, effectively
implementing closure conversion, in a manifestly type-preserving way.
This closure conversion has been the stumbling block, preventing
fusion of zipped flatmapped streams in streaming libraries in Java,
Haskell, OCaml, etc.~-- including the original \strymonas.

We describe how closure conversion arises in linearization and how
it is performed on the example of linearizing the following nested
pipeline:
\begin{code}
let ex_nested = from_to 1 5  |> flat_map (fun x -> from_to x (x+3))
\end{code}
Desugared and in the stream realization of \S\ref{s:computing-NF} it
takes the form:
\begin{code}
let un n z = C.(fun k -> let- v = letl (dref z) in (incr z) @. if1 (v <= n) (k v))
let g1 n z = C.(dref z <= n)
let ex_nested =
 Init (C.(int 1), fun z -> Flat {unr=un C.(int 5) z; grd=g1 C.(int 5) z; linear=true}
 |> flat_map_raw (fun x ->
        Init (x, fun zi -> Flat {unr=un C.(x+int 3) zi; grd=g1 C.(x+int 3) zi; linear=true}))
\end{code}
which has just the right form to apply the algorithm of
Prop.\ref{t:linearization-nested}. Recall, the proposition shows that the
pipeline
\begin{eqnarray*}[l]
\O unroll/ u_1\,z_1 \triangleright \O guard/ g_1
\triangleright \O flat\_map/
  (\lam{z}\lam{x} \O unroll/ u_i (x,z)\, z_i (x,z)
\triangleright \O guard/ g_i (x,z)
\triangleright \O abstract/)
\end{eqnarray*}
is weakly equivalent to the linear stream
$\O unroll/ u_0\,z_0 \triangleright \O guard/ g_0 \triangleright
\O abstract/$
where
\begin{eqnarray*}[rl]
u_0 &= \O fix/ \lam{w}\\
&
[\lam{z}
 u_1 z \triangleright [
\lam{z} (z \in \delta_{u_1}) \:?\: (\mathbf{1},(\mathbf{1},z)) :
  w\, (\mathbf{1},z),
\\
&
\hskip 4.5em
  \lam{(x,z)}
  (g_1 z \:?\: (\mathbf{1},(\mathbf{1},z)) :
  w\, (\kappa' (u_i (x,z), g_i (x,z), \pi(z_i (x,z))), \pi' (z_i (x,z))))
]
,
\\
&\;\lam{((u',g',z_p),z)}
u' (z_p,z) \triangleright [
\lam{(z_p,z)}
(g' (z_p,z) ? w\, (\kappa'(u',g',z_p),z) : w\, (\mathbf{1},z)),
\\
&\hskip 11em
\lam{(y,(z_p,z))}
(\kappa' y, (\kappa'(u',g',z_p),z))
]
]
\\
z_0 &= (\mathbf{1}, z_1)
\\
g_0 &= \pi' \RHD g_1
\end{eqnarray*}
Thus the implementation of |flat_map_raw| merely needs to compute
$g_0$ and $u_0$ as described above, transcribing into the
implicit heap model (in
which the state $z$ etc. is represented as mutable cells). There are
two problems, however. First, we notice that the state may store the
tuple $(u_i (x,z), g_i (x,z), \pi(z_i (x,z)))$ whose elements $u_i
(x,z)$ and $g_i (x,z)$ are closures: as the unrolling and the guard of the
inner stream they depend on, or `close over', the current item of the
outer stream $x$.
This dependence of the state, unrolling and guard of the nested stream
on  |x| is clearly visible in
|ex_nested|. The need for something like closures is also easy to see.
The first few items of |ex_nested| are
|1, 2, 3, 4, 2, 3, 4, 5, ...|: when the outer stream produced |1|, the inner
stream produces |1| through |4|; when the outer stream produced |2|,
the inner stream can be observed to produce |2|, then |3|, etc.,
through |5|. During these observations, the inner stream has to somehow
remember the outer stream item on which its unrolling depends on.
A closure over the outer stream item is one way.
\strymonas, however, takes great pains to avoid closures.

There is a more literal way of saving the outer stream item: in a
mutable cell, call it |xr| for our example. The linearized stream also
has to maintain the state of the outer and the inner streams: the
integer mutable cells |zo| (initialized with 1) and |zi| (initial
value is arbitrary). The linear unrolling for |ex_nested| computed by the
linearization algorithm adjusted for the mutable heap then corresponds
to the following state machine:
\begin{code}
q3: if g1 (int 5) zo = false then goto q0;
      un (int 5) zo (fun x -> xr := x; zi := x; goto q7);
      goto q3;                     (* skipped *)
q7:  if g1 (dref xr + int 3) zi = false then goto q3;
      un (dref xr+int 3) zi (fun y -> k y; goto q5);  (* exit *)
      goto q7;                     (* skipped *)
\end{code}
The states are |q0|, |q3|, |q5| and |q7| (the numbering is chosen for
easy switching, see below). States |q3| and |q7| are the entrance
states: |q3| at the first stream observation, |q7| at the subsequent ones.
The final states are |q0| and |q5|. The former is the terminal:
the stream has finished and no further observations should be performed.
The continuation |k| is the continuation to pass the produced
item to. Instead of closures we see the expressions like
|un (dref xr+int 3) zi|, where, recall, |fun x z -> un (x+int 3) z|
is the unrolling of the inner
stream as the function of the outer state item and the inner stream state.

The remaining problem is how to actually extract the unrolling and the
guard of the inner stream as functions of the outer-stream item. In
our example we found that the inner stream unrolling is
|fun x -> un (x+int 3)| by looking into the lambda-expression passed
to |flat_map_raw|. Although \emph{we} can look inside lambda-expressions,
|flat_map_raw| cannot: In most programming languages, functions are
opaque and cannot be examined. Surprisingly, staging does let us
examine functions and even extract parts of their bodies, as functions
of the argument. The trick obviously does not apply to all functions,
only to code transformers~-- the functions receiving a code value. The
first argument of |flat_map_raw|, of the type |'a exp -> 'b stream|
is, fortunately, one of such functions.  The key idea is that a
function |'a -> 'b| (where |'b| is a non-function type)
is equivalent to a pair of a mutable cell |'a ref|
and a thunk |unit -> 'b|. Applying a function corresponds to writing
the argument to the reference cell and then forcing the thunk. This
alternative representation can be obtained by allocating a fresh
uninitialized mutable cell and substituting the bound variable in the
function's body with the dereferencing of that cell.\footnote{Strictly
  speaking, we have to let-bind the dereferencing.}
For functions
like |'a exp -> 'b stream| such substitution can be mechanically
performed: just by applying the function to the \emph{code} of the
cell dereferencing. The result is a |'b stream| value, which can be
patterned-matched. Effectively, staging lets us do closure
conversion.

Concretely, this closure conversion is performed by the following
|closure_convert|. It receives an inner-stream function
|int exp -> 'b stream| (for simplicity, we restrict ourselves here to
integer argument) and returns the allocated mutable cell |xr|,
the |'b flat| record data structure (see \S\ref{s:computing-NF})
whose components, unrolling and the guard, depend on |xr|,
and a sequence of statements to initialize the inner stream state,
again, depending on |xr|.
\begin{code}
let closure_convert :
    (int exp -> 'b stream) -> (int mut * 'b flat * unit C.stm -> 'w stream) -> 'w stream =
  fun stf k ->
   initializing_ref C.(int 0) @@ fun xr ->
     let rec loop acc = function
       | Init (i,sk) -> initializing_ref i @@ fun z -> loop C.(acc @. (z := i)) (sk z)
       | Flat fl -> k (xr,fl,acc)
    in loop C.unit (stf C.(dref xr))
\end{code}
%% Crucial restriction: nested streams of a fixed structure.
%% Also, staging lets us normalize flat-map body (and examine, for
%% linearization).

We are now in position to mechanically compute the linearized stream
state machine.  A minor problem is that our code representation,
module |C|, has no provisions for |goto|: not every backend supports
it (OCaml does not). Representing arbitrary control flow via
structured control flow (iteration and conditions) has, however, been
long-solved, at the birth of structured programming. All in all, we
obtain the following implementation for |flat_map_raw|, specialized to
integer outer streams for simplicity.
\begin{code}
let flat_map_raw : (int exp -> 'b stream) -> int exp stream -> 'b stream = fun inner_fn st ->
   initializing_ref C.(int 1) @@ fun q ->
   closure_convert inner_fn @@ fun (xr,{unr=unri; grd=grdi},init_state) ->
   let linearize {unr;grd} = let open C in
     let unr' k =
     (q := dref q + int 2) @.
     while_ (logand (dref q) (int 2) <> int 0) @@ begin
     (if1 (dref q = int 3)
        (if_ grd (* advancing outer *)
           (unr (fun x -> (xr := x) @. init_state @. (q := int 7)))
           (q := int 0))) @.
     (if1 (dref q = int 7)
        (if_ grdi        (* advancing inner *)
             (unri (fun y -> k y @. (q := int 5)))
             (q := int 3)))
   end
   in {unr=unr'; grd =C.(dref q <> int 0); linear=true}
   in
   map_flat linearize st
\end{code}

Of course if |ex_nested| is terminated by |iter|, it can easily, and
simply, be represented as a nested loop without the need for
linearization. It is only when it comes to zipping that the
linearization becomes necessary.

\subsection{Push and Pull Processing and \strymonas}
\label{s:push-pull}

The terms `push' and `pull' are often used to characterize stream
processing~-- and also database queries
\cite{push-pull-databases,Amir-query-hll}
and even array processing \cite{Amir-DestPassing}.
One cannot avoid discussing the connection of \strymonas\ to the
two models~-- especially because \strymonas\ is neither, or both.

The original \strymonas\ paper has had a detailed
introduction to push and pull streams \cite{strymonas-2017}. 
Here we recall them on a different, simpler example tailored to
signal processing (software-defined radio, etc.): infinite imperative
streams, whose state is maintained in mutable variables. Assume the
function |get_data: unit -> int| that obtains a sample from a sensor,
blocking if necessary.

A push stream is represented by the fold; in the context of our example:
\begin{code}
type 'a push = ('a -> unit) -> unit
let push_data : int push = fun k -> 
  while true do
    get_data () |> k
  done
\end{code}
It is a function that takes a continuation |'a -> unit| and \emph{repeatedly}
applies it
to each stream element in turn, indefinitely. Characteristically for
push streams, it is the producer that drives the processing (has the loop).
If a sample is processed within the sampling period, the
sampling rate, set by the sensor, is maintained.
Push stream transformers are continuation transformers; they are not recursive:
\begin{code}
let map : ('a -> 'b) -> ('a push -> 'b push) = fun f st -> fun k -> st (f >> k)
let downsample : int -> 'a push -> 'a push = fun n st -> fun k -> 
    let r = ref n in st (fun x -> decr r; if !r = 0 then begin r := n; k x end)
\end{code}
Downsampling is stateful; the skipped samples are just dropped and not
passed to the further continuation. Zipping is hard to express with
push streams: getting two producers, two loops, to advance in parallel
requires some sort of suspension. Although workarounds exists, none of
them support constant space- and time processing and can maintain the
sampling rate. To write a consumer we assume 
|put_data : int -> unit|. 
Push streams do not support early termination.  If only a
certain number of samples is required, some sort of out-of-band
signaling such as exceptions are needed.
\begin{code}
exception End
let consume : int -> int push -> unit = fun n st -> 
  let r = ref n in
  st (fun x -> decr r; if !r >= 0 then put_data x else raise End)
\end{code}
The complete processing pipeline is then
\begin{code}
try push_data |> map ((+) 10) |> downsample 4 |> consume 10 with End -> ()
\end{code}
Recall, the processing loop is in the producer |push_data|; each iteration
starts by calling |get_data|. The input sampling
rate is hence easy to control.

Pull streams are unfolds; in our case of an infinite imperative stream:
\begin{code}
type 'a pull = unit -> 'a
let pull_data : int pull = get_data
\end{code}
A pull stream is a thunk that produces a new stream item every time it
is invoked.
\begin{code}
let map : ('a -> 'b) -> ('a pull -> 'b pull) = fun f st -> st >> f
let downsample : int -> 'a pull -> 'a pull = fun n st -> fun () -> 
  for i = 1 to n-1 do ignore (st ()) done; st ()
\end{code}
Downsampling requires iteration, to affirmatively skip. Zipping is 
easily expressible with pull streams: there are no infinite loops to interleave.
\begin{code}
let rec zip_with : ('a -> 'b -> 'c) -> ('a pull -> 'b pull -> 'c pull) = 
  fun f st1 st2 -> fun () -> f (st1 ()) (st2 ())
\end{code}
However, if |st1| takes more time than the sampling period of |st2|,
we lose data in st2. Input sampling rate, on both inputs, is hard to
maintain. With pull streams, it is the consumer that drives the processing:
\begin{code}
let rec consume : int -> int pull -> unit = fun n st -> 
  for i = 1 to n do put_data (st ()) done
\end{code}
Therefore, it can easily terminate the sampling.
The complete pipeline looks similar to that for the push streams; there is
no longer any need for exceptions, however.
\begin{code}
pull_data |> map ((+) 10) |> downsample 4 |> consume 10
\end{code}
Each iteration of the loop (now, in the consumer) outputs one
sample. How many input samples this corresponds to, is, generally,
hard to tell. With pull streams, therefore, it is easy to control the
output sampling rate but not the input one.  Our example is designed
to be simple, therefore, one may work out that each iteration in the
complete pipeline pulls four input samples.

%% The semantics is via a pull stream, but the Core API is not a pull
%% stream, as defined.
We now come to \strymonas, which represents stream as
\begin{code}
type 'a strym = ('a -> unit) -> unit
let strym_data : int strym = fun k -> get_data () |> k
\end{code}
(cf. |'a emit| in Fig.~\ref{f:raw-API} and its discussion in
\S\ref{s:computing-NF}). The type |'a strym|
looks exactly like |'a push|. The meaning
is different however: it is a function that accepts continuation and
possibly passes a sample to it (if there is)~-- and then
returns. Whereas push streams invoke the received continuation
repeatedly, \strymonas\ streams invoke it at most once (exactly once,
for linear streams). Although the representation type looks like that of a
push stream, it acts as a pull stream. The \strymonas\
transformers are (almost) exactly like the push transformers:
\begin{code}
let map : ('a -> 'b) -> ('a strym -> 'b strym) = fun f st -> fun k -> st (f >> k)
let downsample : int -> 'a strym -> 'a strym = fun n st -> 
  let r = ref n in
  fun k -> st (fun x -> decr r; if !r = 0 then begin r := n; k x end)
\end{code}
(There is a slight difference in |downsample|, however.) In
particular, there are no recursion or iteration. Unlike push streams,
zip is expressible: in exactly the same way as in the pull stream.
\begin{code}
let rec consume : int -> int strym -> unit = fun n st -> 
  let r = ref n in
  while !r > 0 do st (fun x -> put_data x; decr r) done
\end{code}
Like in the pull stream, the consumer drives the iteration. There is a
difference however: each iteration corresponds to one \emph{input}
sample~-- just as it was the case with the push streams. One may say
that in \strymonas, the loop controls both the input and the output.
The complete pipeline looks exactly the same as in the pull stream,
but each iteration does exactly one input sample.

\section{Evaluation}
\label{s:evaluation}

We evaluated \strymonas\ on a set of benchmarks, chosen deliberately
from published literature, with a couple of additions. Since we
position \strymonas\ as a Java8-stream--like library (or, OCaml's
|Seq| or Haskell list and similar libraries), our benchmarks are
mostly chosen from this area. The earlier \strymonas\
\cite{strymonas-2017} paper collected the benchmark suite (from a
combination of Murray et al.~\cite{murray_steno:_2011} and Coutts et
al.~\cite{coutts_stream_2007}), which we deliberately re-use as is,
to demonstrate the improvement in the current version
of \strymonas.  To that suite we add
\textsf{zipFilterFilter} and \textsf{zipFlatMapFlatMap} from
\cite{staged-rewriting}, as well as the run-length decoding benchmark.
The benchmarks range from micro-benchmarks (targeting one or two
stream operators) to more realistic; in particular,
the decoding benchmark is representative of stream
run-length-decompression and processing.

In contrast, the
paper \cite{strymonas-PEPM24} presents \emph{macro}-benchmarks of \strymonas\
on a realistic application of software-defined radio (FM Radio
reception) in comparison with GNU Radio.

Below we describe the suite in more detail, including the input data
set. The overall pattern is typical of Java8-stream--like processing:
taking input from a large array (or arrays),
transforming and accumulating (summing up). Each
benchmark program first allocates and initializes input arrays, runs
the benchmark code as a warm-up, and then runs it tens of times
measuring the execution time of each run. We report the averaged
measured execution time and its standard deviation.
Knowing the array sizes, it can be converted
to throughput.

The benchmark suite contains the following benchmarks:
\begin{bullets}
 \item \textbf{sum}: the simplest
\lstinline{of_arr arr |> sum} pipeline, summing up an array;
 \item \textbf{sumOfSquares}: summing the squared array elements;
 \item \textbf{sumOfSquaresEven}: the sumOfSquares benchmark with an added
   filter, summing the squares of only the even array elements;
 \item \textbf{cart}: $\sum x_iy_j$, with |flat_map| for the
   outer-product
   (viz., |ex_cart| from \S\ref{s:taste});
 \item \textbf{mapsMegamorphic}: seven consecutive map operations with integer
   multiplication;
 \item \textbf{filtersMegamorphic}: seven consecutive filter operations using
   integer comparison;
 \item \textbf{dotProduct}: a simple |zip_with| operation
   (i.e.,|ex_dot| from \S\ref{s:taste})
 \item \textbf{flatMapAfterZip}: compute $\sum (x_i+x_i)y_j$,
   like cart above, doubling the |x| array via |zip_with (+)| with itself;
 \item \textbf{zipAfterFlatMap}: |zip_with| of two streams one
   of which is the result of |flat_map|;
 \item \textbf{flatMapTake}: |flat_map| followed by |take|;
 \item \textbf{zipFilterFilter}: |zip_with| with two nested filtered streams;
 \item \textbf{zipFlatMapFlatMap}: |zip_with| with two nested |flat_map|,
 producing effectively two nested streams traversed in parallel for
 zipping;
 \item \textbf{decode}: run-length-decoding and processing, see
  App.~\ref{a:rll}. The benchmark is a
  realistic version of \textsf{zipFlatMapFlatMap}. The benchmark
  represents a common pattern in video and
audio processing: overlaying two video streams, combining video and
audio tracks, or mixing two audio streams. For example, the popular video/audio
processing application |ffmpeg| provides operations (called filters in
ffmpeg) `amix' and `amultiply'. The latter multiplies 
``each sample from first stream with sample at same position from second
stream'', to ``create amplitude fades and amplitude modulations.''
(see |ffmpeg-filters(1)| manual page). The audio inputs to ffmpeg are
typically compressed/encoded (e.g., |.mp3|); 
ffmpeg hence has to decode the streams
and down- or up-sample them before applying the filter. The \textbf{decode}
benchmark is hence a slightly simplified version of 
|ffmpeg -filter_complex amultiply|. 
\end{bullets}
The complete source code of the benchmarking
suite is available in the \strymonas\
repository.\footnote{\url{http://www.github.com/strymonas}}

The input data for all benchmarks are synthetic,
prepared as follows (shown for OCaml; the initialization code in
C is the same, modulo syntax).
\begin{code}
let v     $\LS$= Array.init 100_000_000 (fun i -> i mod 10)
let vHi   $\LS$= Array.init 10_000_000 (fun i -> i mod 10)
let vLo   $\LS$= Array.init 10 (fun i -> i mod 10)
let vFaZ  $\LS$= Array.init 10_000 (fun i -> i)
let vZaF  $\LS$= Array.init 10_000_000 (fun i -> i)
\end{code}
These arrays are used as inputs to benchmarks, as follows.
\begin{tabular}[C]{ll @{\hskip 3em} ll}
\textbf{sum} & \textsf{v} & \textbf{sumOfSquares} & \textsf{v}
\\
\textbf{sumOfSquaresEven} & v
&
\textbf{mapsMegamorphic} & v
\\
\textbf{filtersMegamorphic} & v
&
\textbf{cart} & \textsf{vHi}, \textsf{vLo}
\\
\textbf{dotProduct} & \textsf{vHi}, \textsf{vHi}
&
\textbf{flatMapAfterZip} & \textsf{vFaZ}, \textsf{vFaZ}
\\
\textbf{zipAfterFlatMap} & \textsf{vZaF}, \textsf{vZaF}
&
\textbf{flatMapTake} & \textsf{vHi}, \textsf{vLo}
\\
\textbf{zipFilterFilter} & \textsf{v}, \textsf{vHi}
&
\textbf{zipFlatMapFlatMap} & \textsf{v}, \textsf{vLo}
\\
\textbf{decode} & \textsf{v}, \textsf{v}
\end{tabular}

The benchmarks are essentially sums over very long integer
arrays. OCaml's version of \strymonas\ uses the |int| type (which, on
x86-64 provides 63 bits: 1 bit is tag); the generated OCaml
code also uses |int|. The C backend uses |int| all throughout,
which is likewise used in the handwritten C baseline; the summation
accumulator however is |int64_t|.
\begin{comment}
On JVM,
an |int| is always a 32-bit signed two's complement integer and a
long data type is a 64-bit two's complement integer.
As a result, we used long integers for our benchmarks on the JVM. For the Scala version
all pipelines operate on |long|. For indexing of arrays we still use
integers.
\end{comment}
We have verified that each benchmark produces the same result across
all platforms and across each variation (baseline,
generated by \strymonas\ or using the other benchmarked libraries).
When presenting results, we eschew the log scale in order to clearly compare
with the baselines, and the original strymonas which we improve upon
(therefore, results other than the baseline and \strymonas\ may be
truncated if the other libraries turn out too slow for a particular
benchmark.)

\paragraph*{Setup:}
All benchmarks are run on AMD Ryzen 7 under Ubuntu 24.04.1 LTS.
The system is equipped with a 2 GHz 8-Core AMD Ryzen 7 7730U CPU
(eight independent CPU cores on a single socket,
with two (hyper)threads per core), 32 KiB L1 instruction and data
cache per core, 512 KiB L2 cache per core and 16 MiB L3 cache shared
across all cores.
The total memory of the system is 16 GB.
When running the benchmarks, we take care to disable Turbo-Boost and to
pin the benchmark process to a single CPU thread
(thread 4). For each C benchmark,
after sleeping for 0.1sec (to let the CPU cool down) and warmups, and
right before and after running the main loop,
we estimate the CPU frequency from the
execution time of a calibrated loop, and confirmed it stays within
$2362 \pm 3$ MHz.
%%We use version build 14.0.2 of the OpenJDK.
The
compiler versions of our setup are presented in the table below:

\begin{center}
\begin{tabular}{ c c c }
  \toprule
  Language & Compiler & Compiler Flags \\
  \midrule
%%  Java & Java 14 (14.0.2+12)  & \textemdash \\
%%  Scala 3 & 0.27.0-RC1 & (vanilla Scala 3) \\
  OCaml & 4.14.1 & \textsf{-O2 -nodynlink} \\
  C & GNU GCC 13.2.0 & \textsf{-W -Wall -O2 -fomit-frame-pointer -lm} \\
  \bottomrule
\end{tabular}
\end{center}

\begin{figure*}
\centering
\begin{subfigure}{\textwidth}
 \centering
 \includegraphics[width=\linewidth]{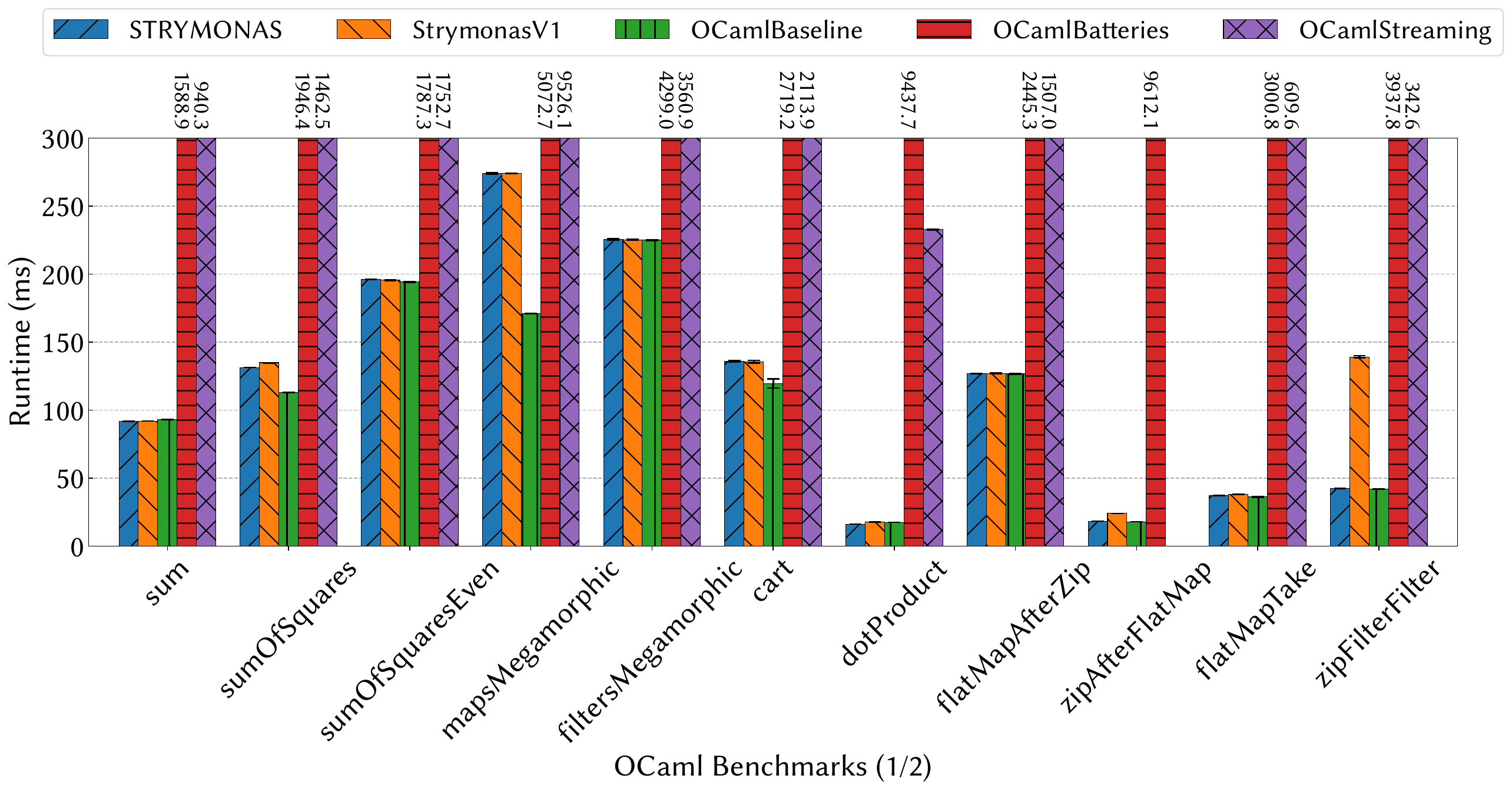}
 \includegraphics[width=\linewidth]{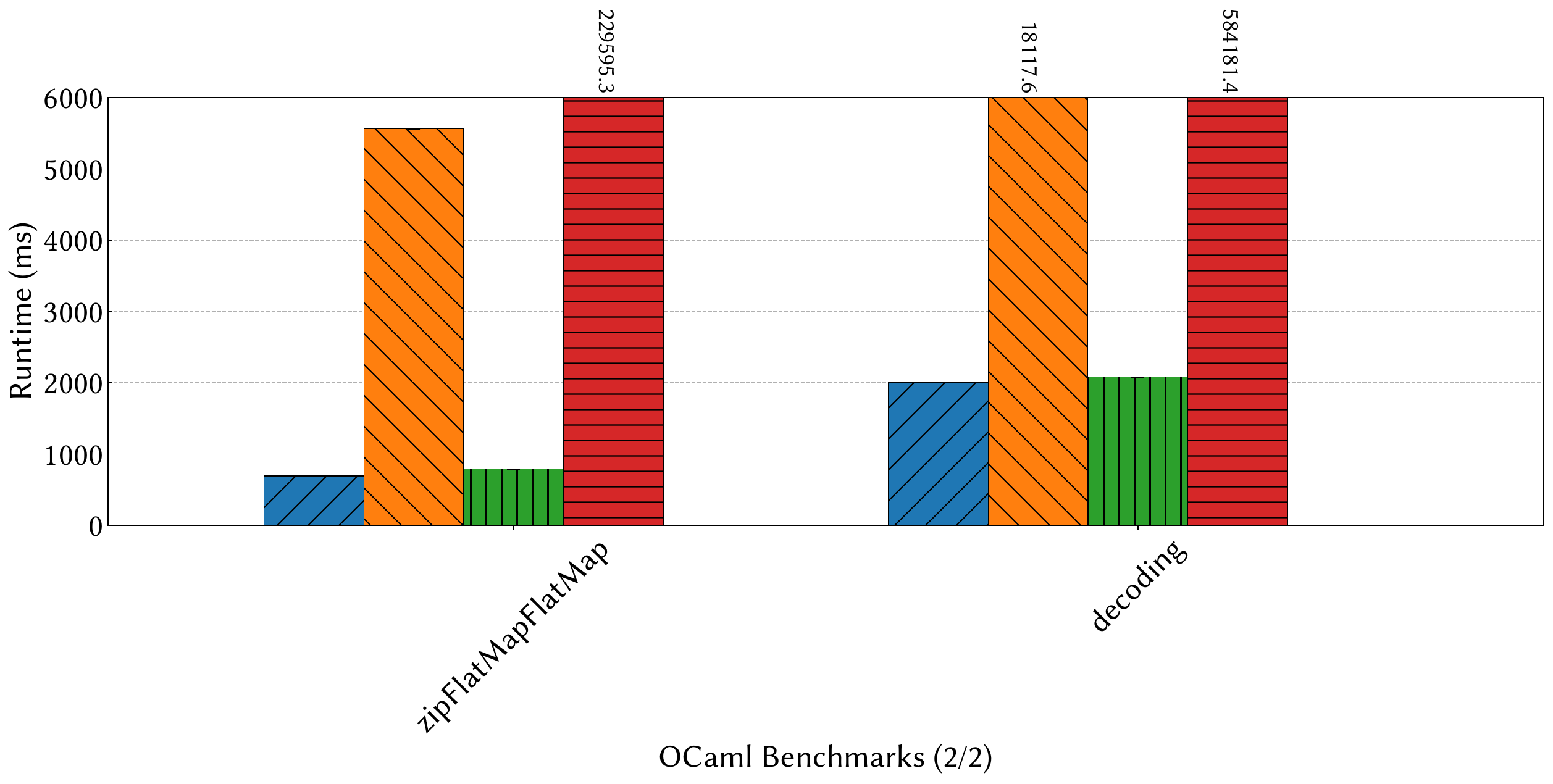}
\end{subfigure}
\caption{OCaml microbenchmarks in msec / iteration (avg. of 20, with
  mean-error bars shown). OCamlBaseline
  contains handwritten loops, OCaml's Batteries do not have a unifying
  module that implements all the combinators we employ and as a result we
  use data type conversions where possible. OCaml's Streaming does not
  offer a \texttt{flat\_map} combinator.
  \strymonas\ original version is annotated
  as StrymonasV1. The figure is truncated: OCaml batteries take
  up to 580sec (per iteration!), for the decode benchmark. }
\label{f:microbenchmarks1}
\end{figure*}

\begin{comment}
\begin{figure*}
\centering
%\begin{subfigure}{0.9\textwidth}
%  \centering
%  \includegraphics[width=.98\linewidth]{scala.eps}
%\end{subfigure}
\caption{JVM microbenchmarks (both Java and Scala) in msec / iteration
  (avg. of 10 after 10 warmup iterations, for 2 forks of the VM to
  reduce run-to-run variance, with mean-error bars shown).
  Java Streams do not offer a zip combinator. We complement them
  by Google's Guava to complete the benchmarks. However Guava does not
  offer a primitive specialization for longs and that introduces
  boxing effects. JAYield is a library offered for the Java
  space. JavaBaseline contains handwritten loops. \strymonas\ original
  version is annotated as Strymonas1. The figure is also
  truncated. }
\label{f:microbenchmarks2}
\end{figure*}
\end{comment}

\begin{figure*}
\centering
\begin{subfigure}{\textwidth}
 \centering
 \includegraphics[width=\linewidth]{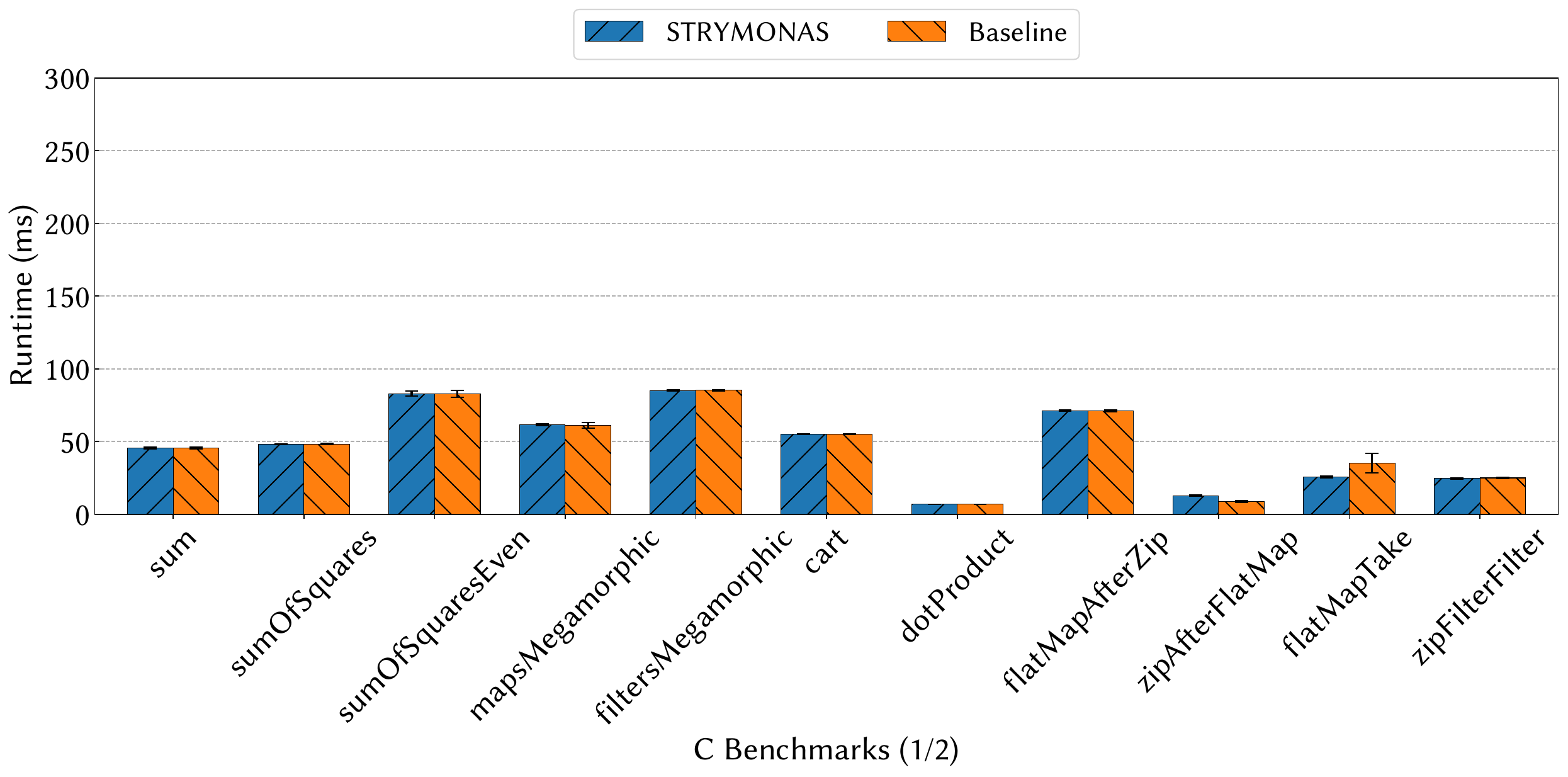}
 \includegraphics[width=\linewidth]{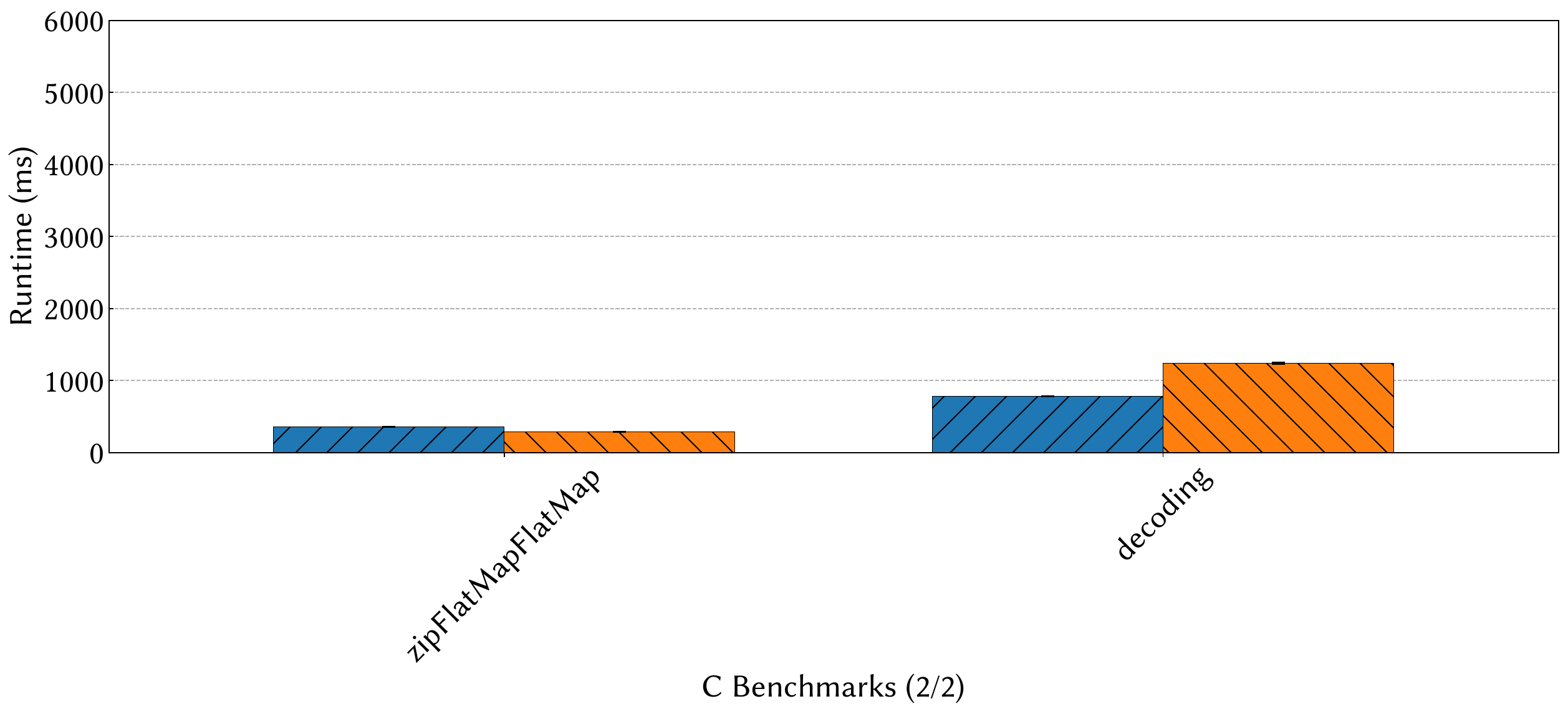}
\end{subfigure}
\caption{C microbenchmarks generated by the OCaml version, in msec / iteration
  (avg. of 20 iterations, with mean-error bars shown).
  Baseline is handwritten.}
\label{f:microbenchmarks3}
\end{figure*}

\paragraph*{Automation:}
\begin{comment}
For Java and \scala{} benchmarks we used the Java Microbenchmark
Harness (JMH)~\cite{aleksey_shipilev_openjdk} tool: a benchmarking
tool for JVM-based languages that is part of the OpenJDK.
We employed 30 warm-up iterations and 30 proper
iterations. We also force garbage collection before benchmark
execution and between runs.
\end{comment}
For OCaml, we used a custom harness, which
compiles a benchmark into a separate executable
(using |ocamlopt|) and runs it.
The harness calculates the average execution time, computing
the mean error and standard deviation using the Student-T
distribution.
We also force garbage collection before benchmark
execution and between runs.
We use the similar harness for C benchmarks, also estimating the
effective CPU frequency before and after the main repetition loop.
%%The same method is employed in JMH.
In all benchmarks, we do
not measure the time needed to initialize data-structures (filling
arrays), nor the run-time for code generation. These costs are
constant (i.e., they become proportionally insignificant for larger
inputs or more iterations) and they were small, between 5 and 10ms,
for all our runs.

\paragraph*{Results:}

Figures~\ref{f:microbenchmarks1}
%% \ref{f:microbenchmarks2}
and \ref{f:microbenchmarks3} present the results from the OCaml
%% Scala 3 ports
and for the generated C code respectively.  We emphasize that
the baselines for
all diagrams are hand-written code.
\strymonas\ matches closely all handwritten benchmarks.

%% \strymonas\ is faster up to $22\times$ ($5\times$-$17\times$ for the
%% zipped filters and zipped |flat_map|s resp. with the JAYield library,
%% see Fig.\ref{f:microbenchmarks2}).
Compared to the state-of-the art
OCaml streaming library we offer orders of magnitude speed-up (see
Fig.\ref{f:microbenchmarks1}) on benchmarks supported by streaming
(streaming does not support |flat_map|),
and up to two-three orders of magnitude speed-up compared to
OCamlBatteries (on the most demanding \textbf{zipFlatMapFlatMap} and
\textbf{decode} benchmarks).
Regarding the original strymonas, in OCaml,
the current version exhibits $4\times$ speed-up
over the zip-based pipelines.
%% On the JVM \strymonas\ achieves $3\times$ speed-up over the same
%% pipelines.
%% We observe that in OCaml the
%% baseline is roughly $1.5x-2x$ faster than \strymonas\ (however, on C with
%% optimizations, \strymonas\ is on par with the baseline, see below).

More instructive is comparison with the handwritten code (the
baseline). In OCaml, on average the generated \strymonas\ code matches
the hand-written, sometimes slower, sometimes quite faster. The
notable slowdown is in \textbf{mapsMegamorphic}:
in the the generated code, the loop body is
\begin{code}
let el_328 = Stdlib.Array.get t_240 i_327 in
let t_329 = el_328 * 2 in
let t_330 = t_329 * 3 in
let t_331 = t_330 * 4 in
let t_332 = t_331 * 5 in
let t_333 = t_332 * 6 in
let t_334 = t_333 * 7 in v_326 := ((! v_326) + t_334)
\end{code}
whereas the handwritten code uses
\begin{code}
let item1 = arr1.(counter1) in
sum := !sum + item1*1*2*3*4*5*6*7
\end{code}
The used version of OCaml (4.14.1) apparently does not optimize-out
even trivial let-bindings. GCC does optimize: \strymonas-generated
and handwritten code for \textbf{mapsMegamorphic} compile to the
identical assembly.

In C, on the simple benchmarks, the difference between the
\strymonas-generated and the handwritten (baseline) disappears
completely: for \textbf{sum}, \textbf{sumOfSquares},
\textbf{sumOfSquaresEven},
\textbf{mapsMegamorphic},
\textbf{filtersMegamorphic},
\textbf{cart},
\textbf{dotProduct},
\textbf{flatMapAfterZip} benchmarks,
the assembly code is \emph{identical}.

\section{Related Work}
\label{s:related}

Our inspiration has been the original \strymonas\ library
\cite{strymonas-2017}, which demonstrated the potential of code
generation for eliminating abstraction overhead in almost all stream
processing pipelines. However, the original library still left closures in the
most complex case of zipping of two flat-mapped streams. It was
conjectured that this problem was unavoidable due to inherent and
statically unpredictable rate differences in zipping streams. We show
the conjecture to be wrong.

Not only the current \strymonas\ significantly improves the original
library performance in complex cases, it also enriches the API,
supporting |take_while| and |drop_while|, map accumulation and
map-filtering, and sliding windows. Its most notable feature, however,
is normalization-by-evaluation, responsible for
the guaranteed elimination of intermediate data structures and
closures in all pipelines, no matter how
complex. Normalization-by-evaluation sets \strymonas\ apart from other
streaming libraries, which perform optimizations by generally
non-confluent rewriting rules.

%% The original strymonas also showed that it is quite unlikely to expect the
%% highest performance just by throwing in a PE at the stream pipeline.

In hindsight, the original \strymonas\ also had the idea of
normalization-by-evaluation~-- although it was not mentioned as such in
\cite{strymonas-2017} and was not (consciously)
applied to the development of the library:
there was no derivation for the intermediate language
and no concrete correctness argument. Because the intermediate language
was ad hoc, extensions were difficult. Code generation (using
MetaOCaml and Scala LMS) was not separated into a backend; therefore,
portability and extensibility was limited.

Our design of normal forms (\S\ref{s:NF}) evokes the `algebraic
approach' of identifying essential operations and their laws, so well
illustrated in Hughes' pretty-printing \cite{hughes-design}. One also sees
differences: for example, although |zip| is an essential operation in
\strymonas, it is absent in the normal form. On the other hand, a
prominent normal form operation |guard| hardly ever used directly in
user-written pipelines. Our normal form is the result of deep (and
iterated) conceptual analysis, identifying operations that are not
immediately apparent.

Our API matches those of popular libraries like Akka Streams. The
semantics of latter, however, is markedly different: Akka Actors are
based on an asynchronous non-blocking back-pressure protocol.

\strymonas\ ensures the constant-space execution of the processing
loop by the careful design of its API. A different approach is to
provide a very general and expressive API, of dataflow programming~--
and select the pipelines that can be compiled into a bounded space-
and time low-level code using synchronicity constraints.  The
constraints are imposed via types, or clocks
\cite{Caspi-TCS,halbwachs_synchronous_1991,Paspi-Pouzet-SyncKahn}.
This so-called synchronous programming is widely used in
embedded systems practice \cite{Caspi-synchronous-programming}.

Parreaux et al.~\cite{staged-rewriting} advance the state of the art
of staging systems. Their library offers rewrite rules which can
execute arbitrary code during pattern matching and code reconstruction
and the original \strymonas\ was similarly an inspiration. They
demonstrate that by constructing an internal IR as well (e.g.,
|doWhile|, |pull|, |doFlatMap| marker methods). To deal with the
complex |flat_map| pipelines they inspect open code, taking advantage
of Scala 2.12.x quasiquotes.  While Scala 3 offers the ability to
pattern match over code with native language
features~\cite{Stucki:277946} we eschew code inspection since it is
not portable, and requires a significantly complex type system if
implemented assuredly.

\begin{comment}
Be sure to mention various restrictions introduced in other
  libraries. Look at Ben's write-up, take a few things from
  strymonas-2017, and then refer to it for the remaining details.
\end{comment}

Optimizing stream pipelines with static analysis was the topic of
M\o{}ller et al.~\cite{streams_staticanalysis}, who demonstrated the
optimization and transformation of pipelines prior their loading on
VM, using interprocedural flow- and context-sensitive type and
pointer analyses.  They use a similar to us set of benchmarks
and show promising results. Most of them, however,
are within a fair margin slower than the baseline. Furthermore,
the analyses could not deal with
|flatMapTake|.  Zip-based pipelines were also not included since |zip|
is missing from Java Streams.

The paper \cite{semantics-streams} noting that ``Most work on
algorithms and architectures for data stream management, however,
never defines what a stream means'' proposes the semantics of streams
based on so-called reconstruction functions, mapping prefixes of a
stream to an element in denotation domain. Alas, as the paper itself
admits in the conclusions, ``it is not yet tested whether they
[reconstruction functions] can deal with the range of data streams
seen in practice, or are helpful in proving properties of stream
operators.''

%Robert Soulé, Martin Hirzel, Buğra Gedik, and Robert Grimm. 2016.
%River: an intermediate language for stream processing.
%Software: Practice and Experience 46, 7 (2016), 891–929.

The paper \cite{catalog-optimizations} presents the catalog of stream
processing optimizations proposed for and used in the whole area of
stream processing. The paper hence gives an overview of the entire
area, from digital signal processing to MapReduce, databases and
complex event processing. It sets up the common terminology, which we
used all throughout.

Since we restrict ourselves to (single-threaded) fusion, when
operators run sequentially, most of the
optimizations described in \cite{catalog-optimizations} do not apply.
Within fusion, we use operator reordering and algorithm selection.
For the former we use algebra (see \S\ref{s:derivation}), as common in
CQL-like streaming systems coming from the database world; in their case,
algebra is relational algebra. Our algebra however is not.
Fusion, however, enables many compiler optimizations such as
inlining, constant propagation, instruction scheduling etc.
% example:  StreamIt: Gordon et al 2002
However, as we showed in
\cite{strymonas-2017}, the most profitable optimizations are algorithmic
and cannot be obtained this way. They require rethinking of the
representation of streams.

StreamIt\footnote{\url{https://github.com/bthies/streamit}}
\cite{streamit-thesis,streamit-ref,Thies10} is a domain-specific
language for high-performance stream programming based on synchronous
dataflow. A StreamIt program is, generally, a cyclic graph built of
user-defined operators (called `filters' in StreamIt) with one input
and one output. The operators are wired together into a graph using
three built-in combinators: pipeline (sequential composition),
split-join (splitters can be either duplicating or round-robin, and
joiners are always round-robin), and feedback loop.  The number of
produced and consumed data items during an execution of an operator
has to be declared.  An operator may `peek' into its input stream at a
given index; this is how sliding window computations are
programmed. The maximum amount of peeking is also declared.  All
operators are thought to run concurrently, limited only by the
availability of data in input channels or the capacity of output
channels.  However, since the rate of consumption, peeking and
production is statically known for each filter, the execution can be
and is statically scheduled and the necessary buffer space computed
and pre-allocated.\footnote{There is support, without much optimization though
\cite{Thies10}, for dynamic scheduling, in case of complex operators
whose data rate cannot be determined statically. Dynamic data rates
occurred only in 9 out of 65 StreamIt benchmarks~-- but are said to be
absolutely necessary for these benchmarks \cite{Thies10}.} StreamIt
also uses fusion and fission to regroup operators for
better load balancing.

Both StreamIt and \strymonas\ are hence built around constructing
pipelines by plugging primitive operators together. The set of
operators and plugs differ: while \strymonas\ supports only sequential
composition, StreamIt also offers split-join and feedback loop. On the
other hand, \strymonas\ is not limited to one-input operators and
offers |zip|.
%% In StreamIt, since each filter has only one input and one output, and
%% joiners have to be used in connection with a spliiter, |zip| is
%% formally out of scope.
StreamIt optimizes using both fusion and fission, and does standard
low-level optimizations on fused operators, up to and including
register allocation and instruction scheduling. Strymonas considers
only fusion; it also leaves low-level optimizations to the backend
compiler (GCC or clang), which is well-maintained and is being constantly
improved. \strymonas\ concentrates instead on pipeline transformation
optimizations: for example, eliminating zip, essentially fusing across
split~-- which has no analogue in StreamIt.

In short, StreamIt and \strymonas\ are based on complementary designs
and make different trade-offs: StreamIt is designed for parallelism
and static scheduling; \strymonas\ targets single-core, single-thread
processing and dynamic scheduling. As noted in \cite{strymonas-PEPM24}
both systems accomplish, using different means, high-performance
processing needed for the FM Radio reception.

\begin{comment}
StreamIT does support flatmap: it is a pipeline whose worker pops 1
item and pushes several.
When a worker pops 1 and pushes 0 or 1, it is a classic filter.
Popping n and pushing m (n /= m) corresponds to the general,
stateful flatmap.

Although StreamIt has a sort of a flatmap: an operator whose
pops 1 item and pushes several, the number of pushed items must be
statically known (for the static scheduling). In realistic flatmaps
(used in our benchmarks and examples) the number of items produced by
the inner stream is not statically known. (Although later version of
StreamIt did support some sort of dynamic scheduling, many
optimizations did apply in that case. Also, it required a special
run-time system, meaning performance degradation due to calls to the
dynamic scheduler and its overhead.)
\end{comment}

To understand why \strymonas, despite its significantly simple
streaming model and absence of any static scheduling and related
optimizations manages nevertheless to accomplish the tried
applications with high performance, we point to
\cite{Thies10}. It introduces the StreamIt benchmark suite, consisting of
65 programs and 33,600 lines of code.
%developed by 22 programmers, mostly MIT students, over 8 years
The suite contains whole applications (MPEG2 encoder/decoder, 802.11a
transmitter, raytracer), graphics pipelines, sorting as well as kernels
(DCT and FFT, CRC, matrix-matrix multiplication), mostly from
the area of signal processing, multimedia and graphics.
The paper retrospectively looks at the implementations of benchmark
programs in StreamIt and describes their salient features and lessons
learned implementing them.
%Some of the lessons: stateful filters are needed in 1/4 of
%benchmarks; but that state is often the running counter. We support
%it via zipping with iota.
%Feedback loop are quite uncommon, but when they do occur they are
%really needed by the algorithm, such as MPEG decoding.
%Sliding windows are very common. Common types: FIR-like filtering
%(pop 1, push 1, peek N) and peek exactly 1. The latter filter
% accounts for 1/3 of all peeking declarations in the benchmark.
The most unexpected and interesting lesson is that the neighboring
operators often have matched data rates.
``The common case [89\%] is that the entire benchmark
is operating on a logical frame of data which is passed through the
entire application.'' Many filters execute only once in steady state:
that means there is no need for buffering, and no opportunity for
pipeline parallelism. Advanced scheduling does not seem to be needed
for most of the realistic benchmarks.
%\cite{Thies10} contains pointers to several other streaming
%languages, mostly for multimedia and graphics.

%% Most of the meat of compiling StreamIt is in the thesis:
%% [19] M. I. Gordon. Compiler Techniques for Scalable Performance of
%% Stream Programs on Multicore Architectures. PhD thesis, MIT, 2010
%% The following thesis is mainly about StreamIt as a language
%% [52] W. Thies. Language and Compiler Support for Stream Programs.
%% PhD thesis, MIT, 2009.

%https://github.com/bthies/streamit/tree/master/streams/apps/benchmarks
%https://github.com/LaminarIR/framework

%https://github.com/bthies/streamit/tree/master/streams/apps/benchmarks/pldi-03
% Has C reference implementation
%https://github.com/bthies/streamit/tree/master/streams/apps/benchmarks/fm
%https://github.com/bthies/streamit/tree/master/streams/apps/benchmarks/asplos06-superset

%https://github.com/bthies/streamit/tree/master/streams/apps/sorts/MergeSort
%https://github.com/bthies/streamit/tree/master/streams/apps/applications/crc
%https://github.com/bthies/streamit/tree/master/streams/apps/applications/802.11a

% FMradio, vocoder, OFDM, Filterbank, 802.11a, mergesort, trellis,
% It looks like beamformer, filterbank2 and fmradio is where the full fusion
% is relatively less beneficial, at least according to
% \cite{Sermulins05} benchmarks. Also, ofdm does not seem to benefit
% from optimizations. These are benchmarks we should try.

The design choices of \strymonas~-- single-thread processing and
fusion~-- are further justified by the following two studies.
\cite{Sermulins05} presents three cache aware optimizations for
StreamIt programs. Remarkably, its benchmarks show (Fig 16 and 17 of
that paper) that on modern CPUs (even of those times) simply fusing
all operators~-- and removing any pipeline parallelism~-- achieves most
and on some platforms all of the sophisticated optimizations presented
in that paper.
\begin{comment}
This paper isn't at all remarkable: its value is in a good
introduction to StreamIt and making a point for us. The caching model
proposed in the paper is quite simplistic, especially for modern
CPUs. No account for prefetching, branch prediction.

Incidentally, execution scaling is subsumed by vectorization.

``Fusion also reduces the overhead of switching between work
functions. In our infrastructure, the steady state is a loop that in-
vokes the work functions via method calls. Thus, every pair of
fused actors eliminates a method call (per invocation of the
actors).'' (\S4.2)/
No wonder they report significant gains in optimization: the
unoptimized model is so simplistic: a function call for each
operation, which often is so simple.

The modulation buffer management is just a circular buffer, which is
what I do.
\end{comment}
\citet{Mendes-TPC} benchmarks three complex event processing (CEP) engines
(two of which are commercial) and notes that ``It was also quite
surprising and disappointing to realize that CEP engines were not able
to automatically benefit from the multi-core hardware used in our
tests.'' That is, sequential processing seems rather to be the norm in
CEP.

%Sanjeev Kulkarni, Nikunj Bhagat, Maosong Fu, Vikas Kedigehalli,
%Christopher Kellogg, Sailesh Mittal, Jignesh M. Patel, Karthik Ra-
%masamy, and Siddarth Taneja. 2015. Twitter Heron: Stream Processing
%at Scale. In ACM SIGMOD.

\strymonas\ is based on operator fusion: replacing all calls to
individual primitive operators
with one big state-machine ``operator''. We hence share the goals with
database systems that employ code generation \citet{neumann_2011},
and modern big-data distributed computing infrastructures such as
Spark. Spark's Tungsten engine has introduced whole-state code
generation, to remove virtual method calls and memory allocations.
One may say that \strymonas\ is using the techniques that make
database engines thrive, as was called for in \cite{thriving}.

Weld~\cite{weld} enables powerful compiler optimizations and automatic
parallelization across functions by expressing the core computations
via a small common intermediate representation.
Weld does not have |flat_map|
%% a short circuiting or a streaming one like unroll (for good reasons
%% in its
%% design\footnote{\url{https://web.archive.org/web/20201120234957/https://github.com/weld-project/weld/blob/master/docs/language.md}})
but has |zip| and performs horizontal fusion (fusion of parallel
loops).  \strymonas' fusion of zipped pipelines can also be
considered a form of horizontal fusion.

Machine fusion \cite{merge_fusion} is a technique for producing fused
imperative code from pipelines (actually, networks) of functional
stream combinators, and which, like us, aims to overcome the simple
push/pull stream processing. Unlike us, it is quite more powerful:
able to fuse networks with both splits and joins, with multiple
combinators reading data from the same input stream. It does hence
provide for sharing of the same stream. The primitive fusion operator
is not associative (nor commutative), however. Therefore, there may be
many possible ways to fuse a network, not all of which are valid, with
no sure way to determine the valid ones ahead of time.

Of the most recent close work we have to mention indexed streams
\cite{indexed-streams}: like us, the authors stress correctness, code
generation and operator fusion. Indexed streams, the intermediate
representation proposed in \cite{indexed-streams}, is akin to our
normal forms, in that it is formally specified with a formal algebra,
proven correct with respect to a functional specification, and can be
efficiently implemented in a low-level language, effecting fusion.
The paper \cite{indexed-streams} targets sparse tensor (linear)
algebra and relational algebra, and hence their set-up is
higher-dimensional streams augmented with an increased
(multi-dimensional) index value. High-dimensional indexing imposes
structure and constraints absent in our stream model.  The indexing
also presupposes that the the user may observe the current state (the
current index), perhaps repeatedly, and explicitly request to advance
to the next index. In our model, an observation \emph{implicitly} and
irrevocably advances the state of the stream: to paraphrase
Heraclitus, one cannot enter into (observe) the same stream twice.

As a large-scale real-life application of \strymonas,
\cite{strymonas-PEPM24} develops Software-Defined Radio~-- FM Radio
reception~-- contrasting and benchmarking it against StreamIt, and the
state-of-the art: GNU Radio.  \strymonas\ turns out to offer portable
high performance, well enough for real-time FM Radio reception. It is
on par with (or, on Raspberry Pi Zero, outstripping) GNU Radio, while
providing static guarantees of complete fusion and type safety.

FM Radio reception is fundamentally digital signal filtering~-- that
is, window stream processing. Although \strymonas\ API,
Fig.~\ref{f:API}, does not provide for windowing, it turns
implementable entirely in terms of the raw API Fig.~\ref{f:raw-API}.
The paper \cite{strymonas-PEPM24} also presents the optimizations
related to windowing: in \cite{Sermulins05} these optimizations were
implemented within a compiler; we, on the other hand,
support them using the \strymonas\ API
as is, and hence automatically preserve typeability and performance
guarantees of \strymonas, in particular, complete fusion. The paper
\cite{strymonas-PEPM24} 
hence highlights the benefit of the embedded DSL nature of \strymonas:
whereas introducing a new optimization in a standalone DSL like
StreamIt requires changing the compiler, we can do it at a `user
level' so to speak, which encourages experimentation (no need to
recompile the library) and maintains guarantees by
construction. Furthermore, our transformations and optimizations
(relating user API to the raw API) are deterministic and
type-preserving by construction~-- in contrast to typical compiler
re-writing passes (or Haskell RULES).

Designing compilers by normalization is rather old approach, as
detailed in \cite{Hoare-NF}. However, the approach in \cite{Hoare-NF}
is not NbE; rather, it is based on re-writing: applying re-writing
laws justified by a set of axioms, justified only by a few
brief references to Dijkstra calculus of predicate transformers. No
confluence is investigated.  Furthermore, re-writing is based not in
equality but on inequalities (refinements).

\section{Conclusions}
\label{s:conc}

We have presented a theory of stateful streams and its implementation
as a stream processing library \strymonas. The library supports freely
assembling pipelines from combinators including |flat_map| and |zip|
and compiles them to state machines introducing no intermediate data
structures or function calls, thus achieving complete fusion.  The
theory and the library were developed in tandem. 

The first main result is the normal form of stream pipelines, which
represents stream processing in a wide domain (much wider than Java
Streams) and still is efficiently mappable to low-level state machine
code.  Notably, although normal form represents pipelines with |zip|,
it does not include |zip|: |zip| is eliminatable.

The second main result is the equational theory, which lets us
transform a higher-level pipeline description to its normal form using
a deterministic and terminating process with the guaranteed result:
normalization-by-evaluation~-- quite unlike typical compiler
optimizations based on generally best-effort, non-confluent rewriting
rules. The theory lets us prove that our optimizations are correct:
meaning-preserving.

The approach is general enough: \strymonas\ is implemented in two
different languages with three distinct target code backends: C, Scala
and OCaml. It turns out expressive enough to implement efficient
windowing processing without needing extensions.  The generated code
is assuredly well-typed (that is, should compile without errors) and
assuredly completely fused.

All in all, we offer a way to intuitively assemble pipeline
descriptions, using higher-order functions, records, modules, etc.~--
and assuredly and entirely avoid the overhead of the abstractions in
the resulting code.  In the domain of stream processing as well, one
attains abstraction without guilt.

In the future work we would like to extend \strymonas\
to support more or all
pipelines in the domain of Weld. We also want to look into the domain of
ETL pipelines, specifically, to support various forms of merging and
stream correlation. It is also interesting to look into a way to
(semi-) automatically verify or enforce the side-conditions
\ref{d:side-cond} on raw stream pipelines, perhaps via a type system.

As StreamIt has clearly demonstrated, the start-up behavior (e.g., filling
in the window) differs significantly from
steady-state. StreamIt has therefore special support for it: so-called
`prework'. Another common use of prework
is to introduce initial delay. It seems worth implementing a similar
feature in \strymonas.

% drop/dropWhile could be implemented via it. I guess
% it should be implemented via a switch. It is best not to duplicate
% code.
% Also add EOF processing: EOF reflection.
% See stream_raw_fn.ml for more optimizations
%
% See also
% https://stevana.github.io/parallel_stream_processing_with_zero-copy_fan-out_and_sharding.html
% Say that parallelization is future work.

\begin{comment}
I would have focused on the improvement of data
pipelines as a whole. From Extraction, to Transformation to Loading the
data back. Stream Fusion can be exhibited in *all of these* steps and
currently it doesn't show up. Kafka doesn't, Batch Processing Systems do
not, Spark just started but RDBMS systems do it for a long time. So I would
have tried to bring in the insight that heterogeneity of sources of data
needs a data engine able to fuses streams on all of these steps.

It is not clear to me what this "data engine" is, but I am sure that this
data engine can be powered by \strymonas\ in some way. The more I see our
code I see that \strymonas\ can be used for the implementation of a physical
query plan scheduler. It can be used for a Kafka Connector which cleans and
combines data before a pipeline starts and certainly when the Spark job, to
run efficient code....
\end{comment}

\begin{acks}

We are deeply indebted to Aggelos Biboudis for his many comments,
very many discussions and encouragement.  
We thank Ben Lippmeier for great many
suggestions, which have lead to the considerable improvement in the
presentation. We gratefully acknowledge Yannis Smaragdakis and Marc Pouzet
for their comments and encouragement, and anonymous referees for many
comments and suggestions.

This work was partially supported by
\grantsponsor{KAKENHI}{JSPS KAKENHI}{} Grants Numbers
\grantnum{KAKENHI}{17K12662},
\grantnum{KAKENHI}{18H03218},
\grantnum{KAKENHI}{21K11821} and
\grantnum{KAKENHI}{22H03563}.
\end{acks}

\bibliographystyle{ACM-Reference-Format}
\bibliography{../streams.bib}

\newpage
\appendix
\section{Generated code for the complex example in \S\ref{s:tupling}}
\label{a:complex-ex}

For the last example of
\S\ref{s:overview}, repeated below for reference:
\begin{code}
let square x = C.(x * x) and even x = C.(x mod (int 2) = int 0) in
Raw.zip_raw
    ([|0;1;2;3|] |> of_int_array |> map square |> take (C.int 12) |> filter even |> map square)
    (iota (C.int 1) |> flat_map (fun x -> iota C.(x+int 1) |> take (C.int 3)) |> filter even)
  |> iter C.(fun (x,y) -> (print_int x) @. (print_int y))
\end{code}
the generated OCaml code is as follows.
\begin{code}
let t_77 = [|0;1;2;3|] in
let v_76 = ref 12 in
let v_78 = ref 1 in
let v_82 = ref 0 in
while ((! v_76) > 0) && ((! v_82) < 4) do
  let t_83 = ! v_78 in
  incr v_78;
  let v_84 = ref 3 in
  let v_85 = ref (t_83 + 1) in
  while ((! v_84) > 0) && (((! v_76) > 0) && ((! v_82) < 4)) do
    decr v_84;
    let t_86 = ! v_85 in
    incr v_85;
    if (t_86 mod 2) = 0
    then
      while
         let v_87 = ref true in
         ((let el_88 = Array.get t_77 (! v_82) in
           let t_89 = el_88 * el_88 in
           decr v_76;
           if (t_89 mod 2) = 0
           then
              let t_90 = t_89 * t_89 in
              v_87 := false;
              Format.print_int t_90;
              Format.force_newline ();
              Format.print_int t_86;
              Format.force_newline ());
          incr v_82);
         (! v_87) && (((! v_76) > 0) && ((! v_82) < 4))
        do () done
    done
  done
\end{code}

\section{Run-length encoding and decoding}
\label{a:rll}

Run-length encoding and decoding is a real-life example of using
\strymonas, showing off the raw API, Fig.~\ref{f:raw-API}, and
flat-mapping.\footnote{It is the real-life example indeed: three and a half
  decades ago the
  first author applied almost the same run-length scheme to
  encode the stream of neuron firings, when helping to acquire and
  then correlated neuron firing streams in a neuro-physiological experiment.}
We see that flat-mapping arises naturally in decoding
and decompression.

The input for the run-length encoder is a stream of |1| and |0| (or
\textsf{true} and \textsf{false}), often with long stretches of zeros
(i.e., \textsf{false}). Such a stream is typical in black/white scanning or
sensor data. The encoded stream is a stream of 8-bit unsigned
integers: An element $0\le n < 255$ represents a stretch of $n$
zeros followed by |1| in the original stream.  An element $255$
represented a stretch of 255 zeros in the original, \emph{not}
followed by one.

\begin{code}
type byte = int
let byte_max = 255
let encode : bool cstream -> byte cstream = fun st ->
  let- zeros_count = Raw.initializing_ref C.(int 0) in
  st |> Raw.map_raw ~linear:false @@ fun el k -> let open C in
    let- zeros = letl (dref zeros_count) in
    if_ el ((zeros_count := int 0) @. k zeros)
     ((zeros_count := zeros + int 1) @.
      if1 (dref zeros_count = int byte_max) (
        (zeros_count := int 0) @. k (int byte_max)))
\end{code}

The decoding shows off an interesting use of |flat_map|:
\begin{code}
let decode : byte cstream -> bool cstream =
    flat_map @@ fun el ->
      let- el1 = Raw.initializing C.(el + int_of_bool(el = int byte_max)) in
      Raw.pull_array el1 @@ fun i k -> (k C.(i = el))
\end{code}

Here is an example of using the decoder: ORing of two compressed
binary streams. It is the \textbf{decode} benchmark:
\begin{code}
fun (arr1,arr2) ->
  zip_with C.(||) (of_arr arr1 |> decode) (of_arr arr2 |> decode)
  |> map C.int_of_bool
  |> sum_int_long
\end{code}

\section{Nested Grouping-Aggregation}
\label{a:adv22}

As a less-than-trivial example of using \strymonas\ and building abstractions
on the top of it, we present nested grouping-aggregation: solving the
problem from Advent of Code 2022.\footnote{See
  \url{http://okmij.org/ftp/Algorithms/nested-aggregation.html} for
  further discussion and background.} The solution is the composition
of three Mealey machines.  The input is a sequence of numbers
separated by `\lstinline{|}' into chunks where each chunk contains
multiple comma-separated numbers: for example:
\begin{code}
   100,200,300|400|500,600|700,800,900|1000
\end{code}
The task is to compute the sum of each chunk
and find the maximum across all chunks.

Recall, Mealey machine is an input-output state automaton where the
output depends on the input and the current state. It is generically
realized in \strymonas\ as follows, where |z:'s exp| is the initial
state, and |tr| is the user-provided transition function, which, given
the current state |'s exp| and the current input |'a| invokes the
continuation |k:'b option -> 's exp -> unit stm| passing it the
possible output and the new state.
\begin{code}
let map_accum_filter (z: 's exp) (tr: 's exp -> 'a -> ('b option -> 's exp -> unit stm) -> unit stm) :
  'a seq -> 'b seq = fun st ->
 let open C in
 let- s = initializing_ref z in
 st |> map_raw ~linear:false @@ fun c k ->
 let- os = letl (dref s) in
 tr os c (function
 | None   -> fun ns -> s := ns
 | Some c -> fun ns -> (s := ns) @. k c)
\end{code}

Parsing of sequences of numbers to an integer is one instance of
Mealey machine:
\begin{code}
let parse_ints : int exp seq -> (int exp * int exp) seq = let open C in
  map_accum_filter (int 0) @@ begin fun s c k ->
  if_ (c >= int (Char.code '0') && c <= int (Char.code '9'))
    (k None (int 10 * s + (c - int (Char.code '0'))))
    (k (Some (s,c)) (int 0))
  end
\end{code}
Monoid accumulation is another instance:
\begin{code}
type 'a monoid = {unit: 'a; op: 'a -> 'a -> 'a}
let group_by_aggregate : int exp -> 'a exp monoid -> ('a exp * int exp) seq -> ('a exp * int exp) seq =
  fun sep m -> let open C in
  map_accum_filter m.unit @@ fun s (x,c) k ->
  let- ns = letl (m.op s x) in
  if_ (c = sep)
    (k None ns)
    (k (Some (ns,c)) m.unit)
\end{code}

The complete solution is hence as follows (where |of_stdin| makes a
stream of characters from the program's standard input):
\begin{code}
of_stdin ~terminator:0
|> parse_ints
|> group_by_aggregate (int (Char.code ',')) {unit=int 0;op=(+)}
|> group_by_aggregate (int (Char.code '|')) {unit=int Int32.(to_int min_int);op=imax}
|> iter (fun (x,_) -> print_int x)
\end{code}

The generated C code is clearly strictly sequential,
processing the input character-by-character with no buffering, no
re-scanning, and no allocations in the main loop:
\begin{code}
void main(){
  int x_1 = -2147483648;
  int x_2 = 0;
  int x_3 = 0;
  bool x_4 = true;
  while (x_4)
  {
    int const t_5 = getc_term(0);
    if (t_5 == 0)
      x_4 = false;
    int const t_6 = x_3;
    if ((t_5 >= 48) && (t_5 <= 57))
      x_3 = (10 * t_6) + (t_5 - 48);
    else {
      x_3 = 0;
      int const t_7 = x_2;
      int const t_8 = t_7 + t_6;
      if (t_5 == 44)
        x_2 = t_8;
      else {
        x_2 = 0;
        int const t_9 = x_1;
        int const t_10 = (t_9 < t_8 ? t_8 : t_9);
        if (t_5 == 124)
          x_1 = t_10;
        else {
          x_1 = -2147483648;
          printf("%d\n",t_10);
        }
      }
    }
  }
}
\end{code}
\end{document}